\RequirePackage{fix-cm}
\documentclass{article}
\usepackage{CJKutf8}

\usepackage[OT1]{fontenc}
\usepackage{textcomp}
\usepackage[utf8]{inputenc}
\usepackage[a4paper]{geometry}
\usepackage{algorithm2e}
\usepackage{amsmath}
\usepackage{amssymb}
\usepackage{graphicx}
\usepackage[pdfusetitle,
 bookmarks=true,bookmarksnumbered=true,bookmarksopen=true,bookmarksopenlevel=2,
 breaklinks=false,pdfborder={0 0 1},backref=false,colorlinks=false]
 {hyperref}

\makeatletter
\numberwithin{equation}{section}

\makeatother

\begin{document}
\begin{CJK}{UTF8}{}%
\sloppy
\title{Physical Modeling of Piano Sound}
\author{Haifan Xie\thanks{Third-year master student, School of Accounting, Guangdong University
of Foreign Studies, Guangzhou, China. Email: fan455@outlook.com.}}
\maketitle
\begin{abstract}
This paper aims to develop a comprehensive physical model and numerical
simulation schemes for a grand piano. The model encompasses various
subsystems, including hammer felt, hammer shank, string, soundboard,
air and room barriers, each modeled in three dimensions to approach
their realistic dynamics. A general framework for 3D elastic solids
accounting for prestress and prestrain is introduced, particularly
addressing the the nonlinearities arising from the large deformation
of piano strings and the one-sided nature of hammer felt-string contact.
The study also examines coupling between subsystem through mechanisms
of surface force transmission and displacement/velocity continuity.
To facilitate numerical simulations, strong PDEs are translated into
weak ODEs via a flexible space discretization approach. Modal transformation
of system ODEs is then employed to decouple and reduce DOFs, and an
explicit time discretization scheme is customized for generating digital
audio in the time domain. The study concludes with a discussion of
the piano model’s capabilities, limitations, and potential future
enhancements.

\tableofcontents{}
\end{abstract}

\section{Introduction}

This study aims to present a detailed physical model of a grand piano
and its numeric implementation schemes. The main features of our model
and numeric schemes include:
\begin{itemize}
\item As 3D as possible. Subsystems including hammer felt, hammer shank,
soundboard, air and room barriers are all fully 3D geometrically modeled.
This also means unless for rigid bodies (hammer shank), their 3-directional
displacements fully vary with their 3D positions. Our geometric model
of piano strings is semi-3D, viz. a cylinder geometry described by
coordinates of one axis, but accounts for the strings' 3-directional
displacements. 
\item A general framework for 3D elastic solid, from both Newtonian and
Lagrangian perspectives. Rooted in the 3D elasticity theory, this
framework not only follows the classic stress-strain relationship,
but also introduces a simple and intuitive model for prestrain and
prestress. For significantly prestressed structures like piano strings,
we offer a more ``naturally linear'' prestress model compared with
the elaborated geometrically exact nonlinear stiff string model in
\cite{morse1986theoretical}. For nonlinearity due to large deformation,
we present a time discretization scheme based on the scalar auxiliary
variable (SAV) method \cite{shen2018scalar}. For nonlinearity due
to collision, we present a time discretization scheme based on the
master-slace approach.
\item Consideration of coupling between subsystems of the piano. Two mechanisms
are considered here: surface force transmission and displacement/velocity
continuity. Furthermore, nonlinear coupling due to collision rather
than fixation is considered in the interaction between hammer felt
and string, and discussed (but failed to implement) in the interaction
between string and two bridge pins.
\item A straightford framework for weak forms and system ordinary differential
equations (ODE). For the sake of numeric simulation via the finite
element method (FEM), we transform the strong form partial differential
equations (PDE) into first or second order system ODEs for each subsystem
of the piano. This is achieved via a flexible space discretization
framework, which can easily incoporate Dirichlet boundary conditions.
\item Modal transformation for solving system ODEs. By solving generalized
eigenvalue problems of the mass and stiffness matrices, we decouple
the coupled system ODEs and significantly reduce the large vector
of degrees of freedom (DOF) into a much smaller vector of modal DOFs.
\item An explicit time discretization scheme customized for solving modal
ODEs exhibiting source term coupling. This scheme is relatively efficient
in that no mid steps are evaluated and no inversion of large non-diagonal
matrices is needed. This scheme is relatively accurate in that the
rhs source terms of the next time step are used for approximate integration
whenever possible, and more than one iteration of numeric integration
at each time step may be run to improve convergence.
\end{itemize}
The rest of this paper starts from a basic 3D elastic prestressed
solid model in section 2, which will be applied in several cases later.
Subsequently sections 3 to 5 introduce models for subsystems of the
piano: soundboard, string, air and room barriers. Section 6 investigates
the mechanisms for coupling between these subsystems, where hammer
felt and shank models are introduced. Section 7 presents the modal
superposition method and explicit time stepping schemes to solve the
derived system ODEs numerically, and a comprehensive computation procedure
for running the simulation. Section 8 gives a brief summary of the
whole model and discusses current limitations and future outlooks.

Here we explain some notation patterns that will appear throughout
this paper. If not explicitly specified, symbols defined only apply
to the current section. In cases where avoiding symbol conflict is
necessary, superscripts $^{(a)},{}^{(b)},{}^{(c)},{}^{(d)},{}^{(e)}$,$^{(f)}$
refer to the string, soundboard, air, hammer shank, hammer felt parts,
room barriers of a piano physical system respectively. For the convenience
of notation, $(x,y,z)$ and $(x_{1},x_{2},x_{3})$, $(u,v,w)$ and
$(u_{1},u_{2},u_{3})$ are used interchangeably. Bold letters represent
vectors or matrices while non-bold letters represent scalars. Definitions
for scalars, vectors and matrices with subscript $_{0}$, if not explicited
stated, are automatically inferred from their counterparts without
subscript $_{0}$. 

\section{Preliminary: 3D linear elastic solid with prestress}\label{sec:Preliminary:-3D-linear}

\subsection{Strong form (PDEs)}

\begin{figure}
\begin{centering}
\includegraphics{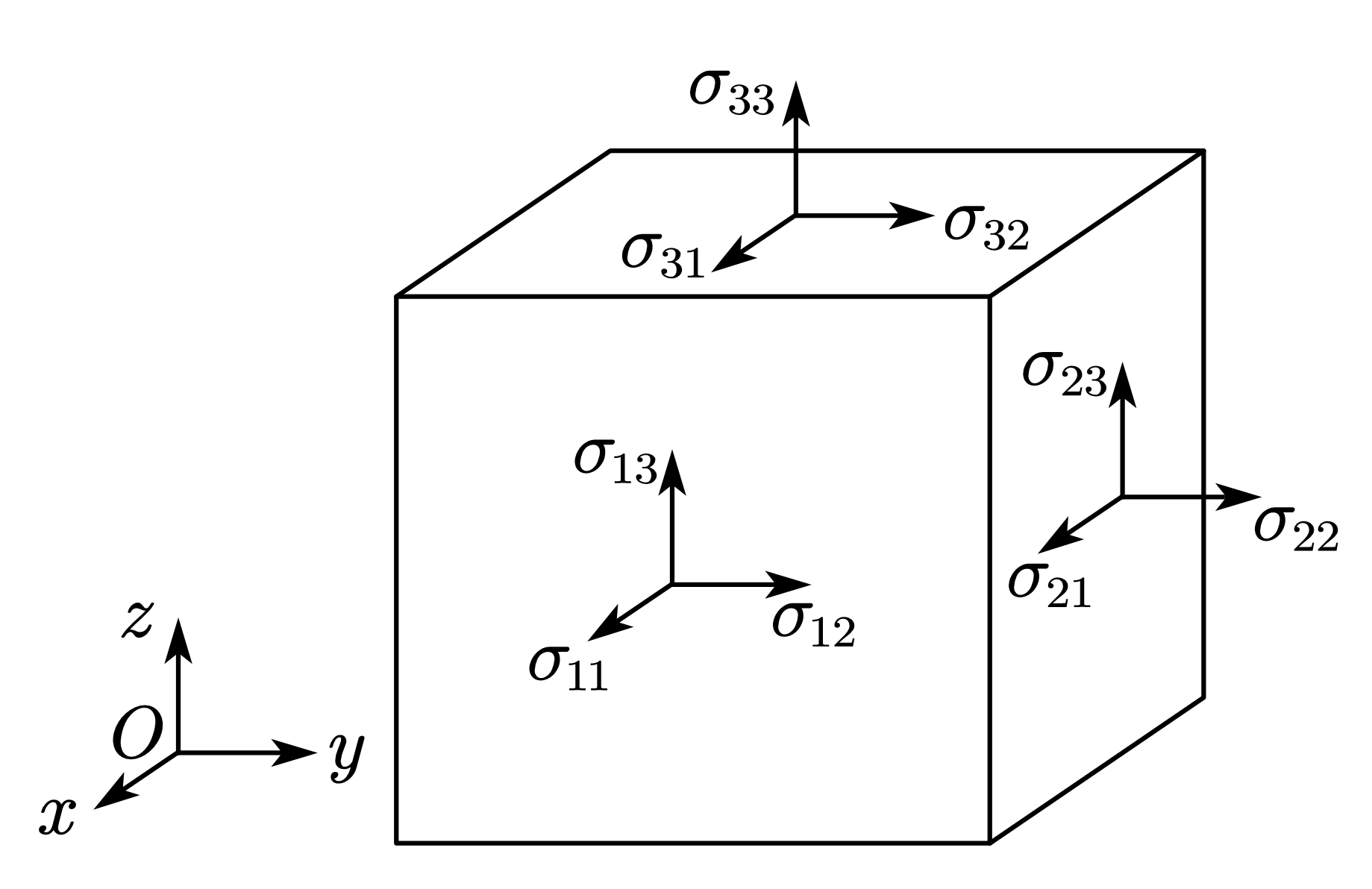}
\par\end{centering}
\caption{The 9 surface forces acting on the 3 positive sides of an infinitesimal
volume}
\label{fig:surface_force}
\end{figure}

In this study, the physical models of piano strings and soundboard
are based on the full or reduced versions of the linear theory of
elasticity \cite{ConstitutiveEquations}. In a 3D Cartesian coordinate
system $(x,y,z)$, consider a material defined over a space $\Omega\subset\mathbb{R}^{3}$
with a boundary $\Gamma\subset\mathbb{R}^{3}$, with homogenous or
heterogenous density $\rho(x,y,z)$, and dynamic displacements in
the 3 directions as 
\begin{equation}
\boldsymbol{u}(x,y,z,t)=\left[u(x,y,z,t),v(x,y,z,t),w(x,y,z,t)\right]^{\top}.\label{eq:displacement}
\end{equation}
Based on the displacement field, we shall perform a force analysis
of the material, considering two kinds of forces: surface force and
body force.

Stress is a major source of surface force for elastic material, and
can be derived from the stress-strain relation. The strain matrix
(second-order symmetric tensor) is expressed by the gradients of displacements
as 
\begin{equation}
\boldsymbol{\epsilon}=\left[\begin{array}{ccc}
\epsilon_{11} & \epsilon_{12} & \epsilon_{13}\\
\epsilon_{12} & \epsilon_{22} & \epsilon_{23}\\
\epsilon_{13} & \epsilon_{23} & \epsilon_{33}
\end{array}\right]=\left[\begin{array}{ccc}
\partial_{x}u & \frac{1}{2}\left(\partial_{y}u+\partial_{x}v\right) & \frac{1}{2}\left(\partial_{z}u+\partial_{x}w\right)\\
\frac{1}{2}\left(\partial_{y}u+\partial_{x}v\right) & \partial_{y}v & \frac{1}{2}\left(\partial_{z}v+\partial_{y}w\right)\\
\frac{1}{2}\left(\partial_{z}u+\partial_{x}w\right) & \frac{1}{2}\left(\partial_{z}v+\partial_{y}w\right) & \partial_{z}w
\end{array}\right],\label{eq:strain_mat}
\end{equation}
which is the linearized Green-Lagrange strain. It can also be written
in vector form as 
\begin{equation}
\overrightarrow{\boldsymbol{\epsilon}}=\left[\begin{array}{c}
\epsilon_{11}\\
\epsilon_{22}\\
\epsilon_{33}\\
2\epsilon_{12}\\
2\epsilon_{13}\\
2\epsilon_{23}
\end{array}\right]=\left[\begin{array}{c}
\partial_{x}u\\
\partial_{y}v\\
\partial_{z}w\\
\partial_{y}u+\partial_{x}v\\
\partial_{z}u+\partial_{x}w\\
\partial_{z}v+\partial_{y}w
\end{array}\right]=\sum_{i=1}^{3}\boldsymbol{H}_{i}\nabla u_{i},\label{eq:strain_vec}
\end{equation}
where 
\begin{equation}
\boldsymbol{H}_{1}=\left[\begin{array}{ccc}
1 & 0 & 0\\
0 & 0 & 0\\
0 & 0 & 0\\
0 & 1 & 0\\
0 & 0 & 1\\
0 & 0 & 0
\end{array}\right],\;\boldsymbol{H}_{2}=\left[\begin{array}{ccc}
0 & 0 & 0\\
0 & 1 & 0\\
0 & 0 & 0\\
1 & 0 & 0\\
0 & 0 & 0\\
0 & 0 & 1
\end{array}\right],\;\boldsymbol{H}_{3}=\left[\begin{array}{ccc}
0 & 0 & 0\\
0 & 0 & 0\\
0 & 0 & 1\\
0 & 0 & 0\\
1 & 0 & 0\\
0 & 1 & 0
\end{array}\right].\label{eq:strain_where}
\end{equation}
From the generalized Hooke's law, the stress-strain relationship is
(here the stress is Cauchy stress) 
\begin{equation}
\overrightarrow{\boldsymbol{\sigma}}=\left[\begin{array}{c}
\sigma_{11}\\
\sigma_{22}\\
\sigma_{33}\\
\sigma_{12}\\
\sigma_{13}\\
\sigma_{23}
\end{array}\right]=\boldsymbol{D}\overrightarrow{\boldsymbol{\epsilon}},\;\boldsymbol{D}=\left[\begin{array}{cccccc}
D_{11} & D_{12} & D_{13} & D_{14} & D_{15} & D_{16}\\
 & D_{22} & D_{23} & D_{24} & D_{25} & D_{26}\\
 &  & D_{33} & D_{34} & D_{35} & D_{36}\\
 &  &  & D_{44} & D_{45} & D_{46}\\
 &  &  &  & D_{55} & D_{56}\\
\mathrm{Sym.} &  &  &  &  & D_{66}
\end{array}\right],\label{eq:stress_strain}
\end{equation}
and the stress matrix (second-order symmetric tensor) is defined as
\begin{equation}
\boldsymbol{\sigma}=\left[\begin{array}{ccc}
\sigma_{11} & \sigma_{12} & \sigma_{13}\\
\sigma_{12} & \sigma_{22} & \sigma_{23}\\
\sigma_{13} & \sigma_{23} & \sigma_{33}
\end{array}\right].\label{eq:stress_mat}
\end{equation}
As shown in figure \ref{fig:surface_force}, there are 9 stress forces
acting on the 3 positive surfaces (as well as 9 stress forces on the
3 negative surfaces not shown) of an infinitesimal volume of material.
$\boldsymbol{D}$ is called the constitutive matrix and its inverse
$\boldsymbol{D}^{-1}$ is called the compliance matrix. The $i$ th
row or column of the stress matrix can be expressed as 
\begin{equation}
\boldsymbol{\sigma}_{i}=\sum_{j=1}^{3}\boldsymbol{A}_{ij}\nabla u_{j},\;\boldsymbol{A}_{ij}=\boldsymbol{H}_{i}^{\top}\boldsymbol{D}\boldsymbol{H}_{j},\;\boldsymbol{A}=\left[\begin{array}{c}
\boldsymbol{A}_{1}\\
\boldsymbol{A}_{2}\\
\boldsymbol{A}_{3}
\end{array}\right]=\left[\begin{array}{ccc}
\boldsymbol{A}_{11} & \boldsymbol{A}_{12} & \boldsymbol{A}_{13}\\
\boldsymbol{A}_{21} & \boldsymbol{A}_{22} & \boldsymbol{A}_{23}\\
\boldsymbol{A}_{31} & \boldsymbol{A}_{32} & \boldsymbol{A}_{33}
\end{array}\right],\label{eq:stress_gradient}
\end{equation}
where block matrix $\boldsymbol{A}$ is symmetric. As $\boldsymbol{\sigma}_{i}$
represents surface forces, converting it to body force would result
in $\nabla\cdot\boldsymbol{\sigma}_{i}$. 

Prestress is considered as the initial stress of material in static
equilibrium, which is not included in the above analysis of stress.
Incorporating prestress into our analysis would require some additional
work as presented in appendix \ref{sec:Prestrain-and-prestress},
which we shall refer to. Define the static tension field matrix (second-order
symmetric tensor) and its vector form as 
\begin{equation}
\boldsymbol{T}(x,y,z)=\left[\begin{array}{ccc}
T_{11} & T_{12} & T_{13}\\
T_{12} & T_{22} & T_{23}\\
T_{13} & T_{23} & T_{33}
\end{array}\right],\;\overrightarrow{\boldsymbol{T}}(x,y,z)=\left[\begin{array}{c}
T_{11}\\
T_{22}\\
T_{23}\\
T_{12}\\
T_{13}\\
T_{23}
\end{array}\right],\label{eq:tension_matrix}
\end{equation}
which can be visualized by figure \ref{fig:surface_force} similar
to stress. Since $\boldsymbol{T}$ is the prestress in static equilibrium,
$\nabla\cdot\boldsymbol{T}=\boldsymbol{0}$ should hold (ignoring
gravity). As per (\ref{eq:total_stress_vec}), the contribution of
prestress to the total stress consists of the static prestress $\boldsymbol{T}$,
and the dynamic prestress $\boldsymbol{\tau}$ which is also a symmetric
tensor. Denote $\boldsymbol{T}_{i}$ and $\boldsymbol{\tau}_{i}$
as the $i$ th row or column of $\boldsymbol{T}$ and $\boldsymbol{\tau}$
respectively, then we find 
\begin{equation}
\boldsymbol{\tau}_{i}=\sum_{j=1}^{3}\boldsymbol{B}_{ij}\nabla u_{j},\;\boldsymbol{B}_{ij}=\Psi_{0}\boldsymbol{H}_{i}^{\top}\boldsymbol{\Psi}_{1}^{\top}\boldsymbol{\Psi}_{2}^{\top}\boldsymbol{D}\boldsymbol{\Psi}_{2}\boldsymbol{\Psi}_{1}\boldsymbol{H}_{j},\;\boldsymbol{B}=\left[\begin{array}{c}
\boldsymbol{B}_{1}\\
\boldsymbol{B}_{2}\\
\boldsymbol{B}_{3}
\end{array}\right]=\left[\begin{array}{ccc}
\boldsymbol{B}_{11} & \boldsymbol{B}_{12} & \boldsymbol{B}_{13}\\
\boldsymbol{B}_{21} & \boldsymbol{B}_{22} & \boldsymbol{B}_{23}\\
\boldsymbol{B}_{31} & \boldsymbol{B}_{32} & \boldsymbol{B}_{33}
\end{array}\right],\label{eq:prestress_gradient}
\end{equation}
where block matrix $\boldsymbol{B}$ is symmetric; the scalar $\Psi_{0}$
and symmetric matrices $\boldsymbol{\Psi}_{1}$ and $\boldsymbol{\Psi}_{2}$,
all constant w.r.t. unknowns, are defined in appendix \ref{sec:Prestrain-and-prestress}.
Similar to $\boldsymbol{\sigma}_{i}$ previously discussed, $\boldsymbol{T}_{i}$
and $\boldsymbol{\tau}_{i}$ contain the 3 sources of prestress in
the same $i$ th axis. For static equilibrium, $\nabla\cdot\boldsymbol{T}_{i}=0$
should hold; then in dynamic states, $\nabla\cdot\boldsymbol{\tau}_{i}$
is the body force of prestress in the $i$ th axis.

Besides stress and prestress which are conservative, non-conservative
forces like damping force often exist. For elastic non-metallic solid,
viscoelastic damping is often the predominant damping \cite{chaigne2016acoustics}.
We hereby adopt a simple viscoelastic model: the viscous damping force
is positively proportional to the first-order time derivative of stress
and prestress\footnote{An explanation for this is that stress and prestress, along with viscous
damping force, are resistance to deformation, and so their relations
should be positive. }. Occurring on surfaces of small volumes, the damping forces can be
visualized by figure \ref{fig:surface_force} similar to stress. The
vector of damping forces parallel to the $x_{i}$ axis is defined
as 
\begin{equation}
\boldsymbol{\varsigma}_{i}=2\mu\partial_{t}(\boldsymbol{\sigma}_{i}+\boldsymbol{\tau}_{i}),
\end{equation}
where $2\mu>0$ is the damping coefficient. Other damping models can
also be incoporated into our model, e.g. structural damping that adds
imaginary parts to elasticity coefficients \cite{bader2022impact}\footnote{The original real consitutive matrix is symmetric. If we add imaginary
values to it, it may be beneficial to make it an Hermitan matrix.
This way energy may be dissipated to the imaginary part of the system
but not outside the system, i.e. energy is still conserved within
the system.}. As for non-conservative forces besides damping, we represent them
as a single force $\boldsymbol{F}=[F_{1},F_{2},F_{3}]^{\top}$ and
will investigate them for different physical systems later.

Having derived all the relevant forces acting on an infinitesimal
volume of solid material, we can invoke Newton's second law to derive
3 PDEs as 
\begin{equation}
\rho\partial_{tt}u_{i}=\nabla\cdot\boldsymbol{G}_{i}+F_{i},\;i=1,2,3.\label{eq:pde}
\end{equation}
where vector $\boldsymbol{G}_{i}=\boldsymbol{\sigma}_{i}+\boldsymbol{\tau}_{i}+\boldsymbol{\varsigma}_{i}$
is defined as the $i$ th row or column of symmetric tensor $\boldsymbol{G}$
representing all surface forces except the static prestress. The above
equation can also be derived via a Lagrange formulation of the virtual
work and the variations of kinetic and potential energy. 

\subsection{Weak form (ODEs)}

We then seek for a weak form of (\ref{eq:pde}) via variational formulation.
Multiplying an arbitrary test function vector $\boldsymbol{\psi}(x,y,z)$
on both sides of it and integrating over $\Omega$ yields 
\begin{align}
\int_{\Omega}\boldsymbol{\psi}\rho\partial_{tt}u_{i}\mathrm{d}V-\int_{\Omega}\boldsymbol{\psi}\nabla\cdot\boldsymbol{G}_{i}\mathrm{d}V & =\int_{\Omega}\boldsymbol{\psi}F_{i}\mathrm{d}V\label{eq:weak_form_1}\\
\int_{\Omega}\boldsymbol{\psi}\rho\partial_{tt}u_{i}\mathrm{d}V+\int_{\Omega}\left(\nabla\boldsymbol{\psi}\right)\boldsymbol{G}_{i}\mathrm{d}V & =\int_{\Omega}\boldsymbol{\psi}F_{i}\mathrm{d}V\label{eq:weak_form_2}
\end{align}
where $\mathrm{d}V=\mathrm{d}x\mathrm{d}y\mathrm{d}z$ and applying
the gradient operator $\nabla$ to a vector results in its Jacobian
matrix. In the above formulation, integration by parts is utilized
with Neumann boundary condition imposed as 
\begin{equation}
\boldsymbol{G}_{i}\cdot\mathrm{d}\boldsymbol{\Gamma}=0,\label{eq:Neumann_boundary}
\end{equation}
where $\mathrm{d}\boldsymbol{\Gamma}$ is the outward normal vector
of the tangent plane of any point on $\Gamma$. Note that this condition
applies only when the test function is non-zero on the boundary, viz.
Neumann boundary conditions need not be satisfied at points where
Dirichlet boundary conditions are present. 

We now use space discretization and the Galerkin method to transform
(\ref{eq:weak_form_2}) into second-order ODE. Based on (\ref{eq:displacement}),
define the displacement field of the soundboard as $\boldsymbol{u}^{*}\approx\boldsymbol{u}=[u,v,w]^{\top}$
(here superscript $^{*}$ means the exact solution) as 
\begin{align}
u(x,y,z,t) & =\boldsymbol{\varphi}_{1}(x,y,z)\cdot\left[\boldsymbol{S}_{1}\boldsymbol{\xi}(t)+\boldsymbol{S}_{0,1}\boldsymbol{\xi}_{0}(t)\right],\nonumber \\
v(x,y,z,t) & =\boldsymbol{\varphi}_{2}(x,y,z)\cdot\left[\boldsymbol{S}_{2}\boldsymbol{\xi}(t)+\boldsymbol{S}_{0,2}\boldsymbol{\xi}_{0}(t)\right],\nonumber \\
w(x,y,z,t) & =\boldsymbol{\varphi}_{3}(x,y,z)\cdot\left[\boldsymbol{S}_{3}\boldsymbol{\xi}(t)+\boldsymbol{S}_{0,3}\boldsymbol{\xi}_{0}(t)\right],\label{eq:displacement_space_discretization}
\end{align}
where the sizes of vectors and matrices are 
\begin{align}
 & \boldsymbol{\xi}:N\times1,\;\boldsymbol{\xi}_{0}:N_{0}\times1,\nonumber \\
 & \boldsymbol{\varphi}_{1}:N_{1}\times1,\;\boldsymbol{S}_{1}:N_{1}\times N,\;\boldsymbol{S}_{0,1}:N_{0,1}\times N_{0},\nonumber \\
 & \boldsymbol{\varphi}_{2}:N_{2}\times1,\;\boldsymbol{S}_{2}:N_{2}\times N,\;\boldsymbol{S}_{0,2}:N_{0,2}\times N_{0},\nonumber \\
 & \boldsymbol{\varphi}_{3}:N_{3}\times1,\;\boldsymbol{S}_{3}:N_{3}\times N,\;\boldsymbol{S}_{0,3}:N_{0,3}\times N_{0}.
\end{align}
In (\ref{eq:displacement_space_discretization}), each scalar displacement
unknown is expressed as the dot product of a space function vector
and time function vector. In FEM, $\boldsymbol{\varphi}_{i}$ is the
vector of shape functions (also known as interpolation functions)
for variable $u_{i}$, and $\boldsymbol{\vartheta}_{i}=\boldsymbol{S}_{i}\boldsymbol{\xi}+\boldsymbol{S}_{0,i}\boldsymbol{\xi}_{0}$
is the vector of coefficients of shape functions, often interpreted
as nodal displacements. The only unknown here is $\boldsymbol{\xi}(t)$,
the vector of unknown DOFs, whereas other vectors and matrices are
known. $\boldsymbol{\xi}(t)$ is defined so that through some linear
transformation by $\boldsymbol{S}_{1}$ and $\boldsymbol{S}_{0,1}\boldsymbol{\xi}_{0}(t)$,
the nodal displacements $\boldsymbol{\vartheta}_{i}$ can be obtained.
The reason we do not simply define $u_{i}=\boldsymbol{\varphi}_{i}\cdot\boldsymbol{\xi}_{i}$
but consider a linear transformation is that it provides additional
flexibility when imposing Dirichlet boundary conditions. For instance,
when $u+v$ rather than $u$ or $v$ is known to be some nonzero functions
on the space boundary, $\boldsymbol{\xi}$ only needs to contain $u$
and $u+v$ is incorporated in $\boldsymbol{\xi}_{0}$. We can thus
see that total DOF, viz. the number of time functions we need to solve,
is $N$ that does not necessarily equal to $N_{1}+N_{2}+N_{3}$. Often
in simple cases, matrix $\boldsymbol{S}=[\boldsymbol{S}_{1}^{\top},\boldsymbol{S}_{2}^{\top},\boldsymbol{S}_{3}^{\top}]^{\top}$
is diagonal with many ones and some zeros on the diagonal.

To express the displacement and its gradient in the DOF vector, we
find the below relations:
\begin{align}
 & u_{i}=\boldsymbol{P}_{i}\boldsymbol{\xi}+\boldsymbol{P}_{0,i}\boldsymbol{\xi}_{0},\;\boldsymbol{u}=\boldsymbol{P}\boldsymbol{\xi}+\boldsymbol{P}_{0}\boldsymbol{\xi}_{0},\;\nabla u_{i}=\boldsymbol{Q}_{i}\boldsymbol{\xi}+\boldsymbol{Q}_{0,i}\boldsymbol{\xi}_{0},\nonumber \\
 & \boldsymbol{P}=\left[\begin{array}{c}
\boldsymbol{P}_{1}\\
\boldsymbol{P}_{2}\\
\boldsymbol{P}_{3}
\end{array}\right]=\left[\begin{array}{c}
\boldsymbol{\varphi}_{1}^{\top}\boldsymbol{S}_{1}\\
\boldsymbol{\varphi}_{2}^{\top}\boldsymbol{S}_{2}\\
\boldsymbol{\varphi}_{3}^{\top}\boldsymbol{S}_{3}
\end{array}\right],\;\boldsymbol{P}_{0}=\left[\begin{array}{c}
\boldsymbol{P}_{0,1}\\
\boldsymbol{P}_{0,2}\\
\boldsymbol{P}_{0,3}
\end{array}\right]=\left[\begin{array}{c}
\boldsymbol{\varphi}_{1}^{\top}\boldsymbol{S}_{0,1}\\
\boldsymbol{\varphi}_{2}^{\top}\boldsymbol{S}_{0,2}\\
\boldsymbol{\varphi}_{3}^{\top}\boldsymbol{S}_{0,3}
\end{array}\right],\nonumber \\
 & \boldsymbol{Q}=\left[\begin{array}{c}
\boldsymbol{Q}_{1}\\
\boldsymbol{Q}_{2}\\
\boldsymbol{Q}_{3}
\end{array}\right]=\left[\begin{array}{c}
\left(\nabla\boldsymbol{\varphi}_{1}\right)^{\top}\boldsymbol{S}_{1}\\
\left(\nabla\boldsymbol{\varphi}_{2}\right)^{\top}\boldsymbol{S}_{2}\\
\left(\nabla\boldsymbol{\varphi}_{3}\right)^{\top}\boldsymbol{S}_{3}
\end{array}\right],\;\boldsymbol{Q}_{0}=\left[\begin{array}{c}
\boldsymbol{Q}_{0,1}\\
\boldsymbol{Q}_{0,2}\\
\boldsymbol{Q}_{0,3}
\end{array}\right]=\left[\begin{array}{c}
\left(\nabla\boldsymbol{\varphi}_{1}\right)^{\top}\boldsymbol{S}_{0,1}\\
\left(\nabla\boldsymbol{\varphi}_{2}\right)^{\top}\boldsymbol{S}_{0,2}\\
\left(\nabla\boldsymbol{\varphi}_{3}\right)^{\top}\boldsymbol{S}_{0,3}
\end{array}\right].\label{eq:space_discretization_map}
\end{align}
Combining (\ref{eq:space_discretization_map}) with (\ref{eq:stress_gradient})
(\ref{eq:prestress_gradient}) , the following relations are found:
\begin{align}
\boldsymbol{\sigma}_{i} & =\boldsymbol{A}_{i}\boldsymbol{Q}\boldsymbol{\xi}+\boldsymbol{A}_{i}\boldsymbol{Q}_{0}\boldsymbol{\xi}_{0},\nonumber \\
\boldsymbol{\tau}_{i} & =\boldsymbol{B}_{i}\boldsymbol{Q}\boldsymbol{\xi}+\boldsymbol{B}_{i}\boldsymbol{Q}_{0}\boldsymbol{\xi}_{0},\label{eq:space_discretization_stress_prestress}
\end{align}
Now we subsitute (\ref{eq:space_discretization_map}) (\ref{eq:space_discretization_stress_prestress})
into (\ref{eq:pde}) (\ref{eq:weak_form_1}) (\ref{eq:weak_form_2}),
and use $\boldsymbol{P}_{i}^{\top}$ as a vector of test functions
for the $i$ th PDE. These would yield 3 groups of weak-form equations,
each group having $N$ equations. Summing the 3 groups into 1 group,
we derive a system of $N$ coupled second-order ODEs as 
\begin{equation}
\boldsymbol{M}\ddot{\boldsymbol{\xi}}(t)+\boldsymbol{C}\dot{\boldsymbol{\xi}}(t)+\boldsymbol{K}\boldsymbol{\xi}(t)=\boldsymbol{f}(t),\label{eq:ODEs}
\end{equation}
where 
\begin{align}
 & \boldsymbol{M}=\int_{\Omega}\rho\boldsymbol{P}^{\top}\boldsymbol{P}\mathrm{d}V,\;\boldsymbol{M}_{0}=\int_{\Omega}\rho\boldsymbol{P}^{\top}\boldsymbol{P}_{0}\mathrm{d}V,\nonumber \\
 & \boldsymbol{K}=\int_{\Omega}\boldsymbol{Q}^{\top}\left(\boldsymbol{A}+\boldsymbol{B}\right)\boldsymbol{Q}\mathrm{d}V,\;\boldsymbol{K}_{0}=\int_{\Omega}\boldsymbol{Q}^{\top}\left(\boldsymbol{A}+\boldsymbol{B}\right)\boldsymbol{Q}_{0}\mathrm{d}V,\nonumber \\
 & \boldsymbol{C}=2\mu\boldsymbol{K},\;\boldsymbol{C}_{0}=2\mu\boldsymbol{K}_{0},\nonumber \\
 & \boldsymbol{f}(t)=\boldsymbol{f}_{0}(t)+\boldsymbol{f}_{1}(t),\nonumber \\
 & \boldsymbol{f}_{0}(t)=-\boldsymbol{M}_{0}\ddot{\boldsymbol{\xi}}_{0}-\boldsymbol{C}_{0}\dot{\boldsymbol{\xi}_{0}}-\boldsymbol{K}_{0}\boldsymbol{\xi}_{0},\;\boldsymbol{f}_{1}(t)=\int_{\Omega}\boldsymbol{P}^{\top}\boldsymbol{F}\mathrm{d}V\label{eq:ODEs_where}
\end{align}
are called the mass matrix, stiffness matrix, damping matrix and force
vector respectively. In the next following sections, we shall apply
the above 3D elastic solid material model to various parts of the
piano physical system, including soundboard, string, room material
and hammer felt. 

\section{Model for piano soundboard}\label{sec:Model-for-piano-soundboard}

\begin{figure}
\begin{centering}
\includegraphics{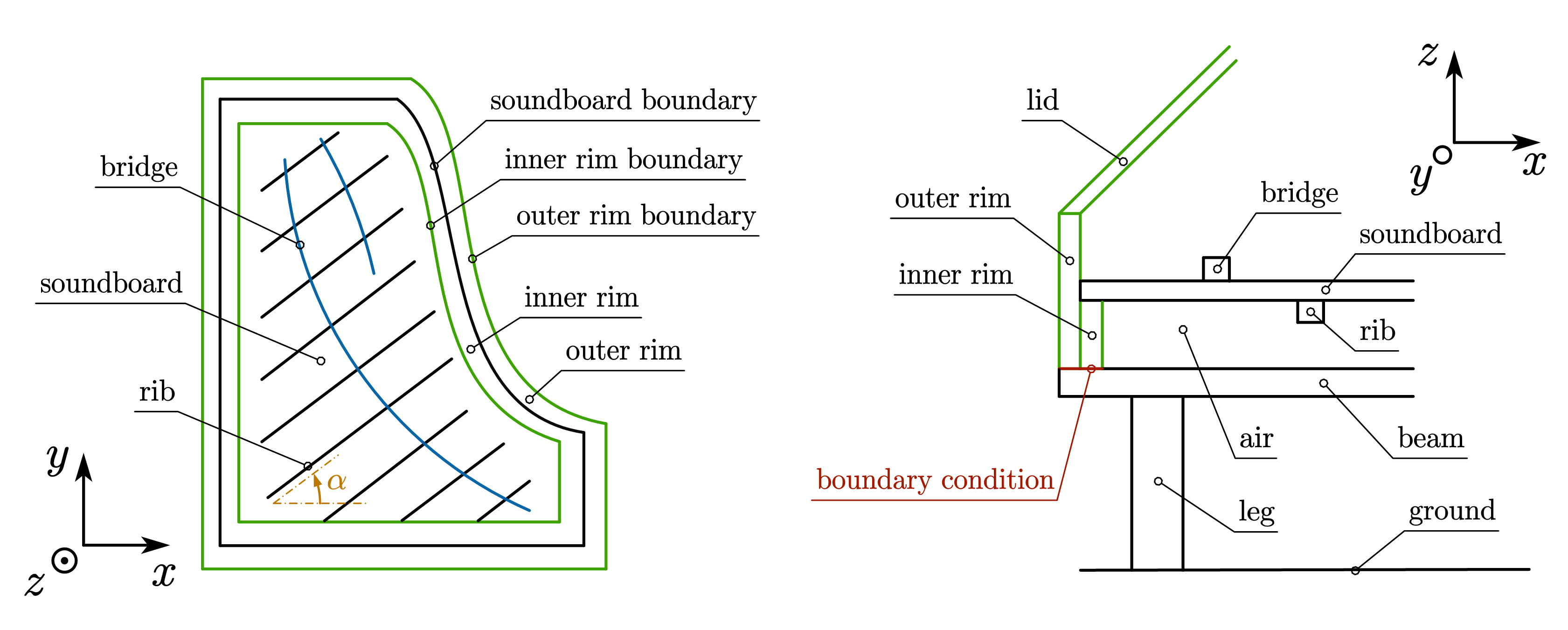}
\par\end{centering}
\caption{The piano soundboard in top view (left) and side view (right)}
\label{fig:soundboard}
\end{figure}

The study treats the grand piano soundboard as a multi-layer plate
with irregular geometry, and physically models it as a 3D structure.
In \cite{chabassier2013modeling,chabassier2014time,chabassier2016time},
the soundboard was modeled as a Reissner-Mindlin plate, accounting
for the ribs and bridges by making the thickness, density, and elastic
coefficients position-dependent; however, it is unclear how the different
orthotropic angles of each layer were handled. In \cite{trevisan2017modal},
the soundboard was modeled as a Kirchhoff-Love plate, considering
a 90 degrees orthotropic rotation of the ribs; however, as the Kirchhoff-Love
plate specifies only 1 unknown for the 3D displacement field, it may
not be accurate enough, especially regarding the ribs and bridges
which are much thicker than the board. In \cite{valiente2022modeling,miranda2024influence},
the soundboard was modeled as a 3D structure using tetrahedral elements,
then reduced it into a modal model; this is an advancement from 2D
to 3D soundboard, but the authors have not presented much details
regarding the 3D equations. Moreover, some features of the soundboard
seem still under-evaluated: the prestress exerted by the string on
the soundboard; the role of rim and lid in the soundboard's vibration;
the more exact 2D Dirichlet boundary conditions for the 3D soundboard,
different from the case of 1D boundary conditions for 2D soundboard. 

To better account for the soundboard's multi-layer feature, the current
study implements a fully 3D soundboard model. Figure \ref{fig:soundboard}
shows a simplified soundboard sketch based on \cite{igrec2013pianos}.
In our physical model, the soundboard is an entity composed of 5 parts:
board, ribs, bridges, rim, lid. The board is parallel to the $xOy$
plane, with the ribs attached to its under side and the bridges attached
to its upper side. The horizontal fibers of the board and bridges
are approximately parallel, while the horizontal fibers of ribs are
almost orthogonal to that of the board. The rim consists of 2 parts:
the inner rim connects the board's boundary from below using bolts
and dowels, and the outer rim encases the board and the lower rim.
The lid, as large as the board and with an angle to it, is an extension
of the outer rim through a wooden stick (considered part of the lid)
and hinges. All parts of the soundboard are treated as a whole in
computation, meaning that the displacements of part intersections
are uniform.

The soundboard's governing PDEs follow (\ref{eq:pde}) and ODEs follow
(\ref{eq:ODEs}), so we do not need to present them again. We only
need to discuss some parameters, initial conditions and boundary conditions
here. Define the occupied space of soundboard as $\Omega\subset\mathbb{R}^{3}$,
and the boundaries of soundboard as $\Gamma_{1},\Gamma_{2}\subset\Gamma\subset\mathbb{R}^{3}$.
Here $\Gamma$ contains all the 2D surfaces of the 3D volume of soundboard;
$\Gamma_{1}$ contains only the underside surfaces of the inner and
outer rims, as marked red at the right of figure 2; $\Gamma_{2}$
is $\Gamma$ excluding $\Gamma_{1}$, the vibrating parts. We shall
see in the following that $\Gamma_{1}$ and $\Gamma_{2}$ are where
the Dirichlet boundary conditions apply to the soundboard and the
air respectively.

Given that all parts of the soundboard are mainly made of wood, an
orthotropic material whose orthotropic directions are determined by
fibers \cite{bucur2017acoustics}, the compliance matrix writes 
\begin{equation}
\boldsymbol{D}_{\mathrm{orth}}^{-1}=\left[\begin{array}{cccccc}
\frac{1}{E_{x}} & -\frac{\nu_{xy}}{E_{x}} & -\frac{\nu_{xz}}{E_{x}} & 0 & 0 & 0\\
 & \frac{1}{E_{y}} & -\frac{\nu_{yz}}{E_{y}} & 0 & 0 & 0\\
 &  & \frac{1}{E_{z}} & 0 & 0 & 0\\
 &  &  & G_{xy} & 0 & 0\\
 &  &  &  & G_{xz} & 0\\
\mathrm{Sym.} &  &  &  &  & G_{yz}
\end{array}\right],\label{eq:soundboard_elasticity_coef}
\end{equation}
where $E$, $G$, $\nu$ are the young's modulus, shear modulus and
Poisson's ratio. Note that different layers would have different elastic
coefficients. It is then necessary to consider that for a certain
layer like ribs, the global axes $x,y$ may need to be rotated by
an angle $\alpha$, as denoted in figure \ref{fig:soundboard}, to
become the material orthotropic axes $x',y'$ and for a certain layer.
The actual constitutive matrix is then 
\begin{equation}
\boldsymbol{D}=\boldsymbol{Z}^{\top}\boldsymbol{D}_{\mathrm{orth}}\boldsymbol{Z},\;\boldsymbol{Z}=\left[\begin{array}{cccccc}
C^{2} & S^{2} & 0 & SC & 0 & 0\\
S^{2} & C^{2} & 0 & -SC & 0 & 0\\
0 & 0 & 1 & 0 & 0 & 0\\
-2SC & 2SC & 0 & C^{2}-S^{2} & 0 & 0\\
0 & 0 & 0 & 0 & C & S\\
0 & 0 & 0 & 0 & -S & C
\end{array}\right],
\end{equation}
where $S=\sin\alpha,C=\cos\alpha$. Readers may refer to \cite{nettles1994basic}
for a detailed deduction. After rotation, the constitutive matrix
would have 13 non-zero entries for its upper triangular part. 

As for initial and boundary conditions, equation (\ref{eq:pde}) is
constrained by: 
\begin{equation}
\boldsymbol{u}|_{\boldsymbol{x}\in\Gamma_{1}}=\boldsymbol{0},\label{eq:soundboard_boundary}
\end{equation}
\begin{equation}
\boldsymbol{u}|_{t=0}=\partial_{t}\boldsymbol{u}|_{t=0}=\boldsymbol{0}.\label{eq:soundboard_initial}
\end{equation}
The choice of this boundary condition stems from that the whole soundboard
is hard supported by beams and legs on the under side of rim (see
figure 2). Thanks to the flexiblility of our 3D model, the rim can
be treated as a natural extension of the main board, thereby capturing
more nuanced boundary conditions compared to the clamped or simply
supported cases in 2D models. Given approriate boundary conditions,
the mass and stiffness matrices are symmetric positive definite.

\section{Model for piano strings}

\begin{figure}
\begin{centering}
\includegraphics{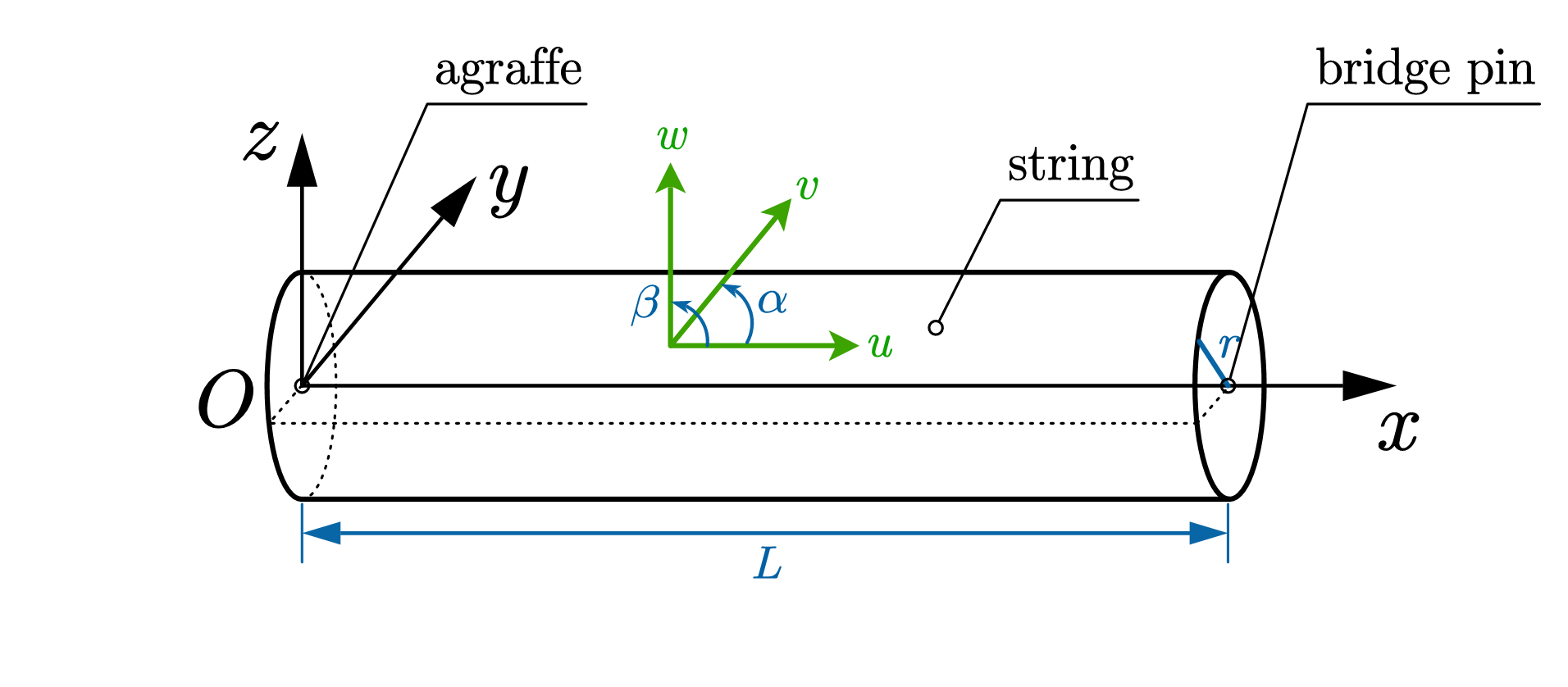}
\par\end{centering}
\caption{The piano string in 3D view}
\label{fig:string}
\end{figure}

The symbols in this section follow sections 2 and 3 if not explicitly
defined. Definitions for scalars, vectors and matrices with subscript
$_{0}$, if not explicited stated, are automatically inferred from
their counterparts without subscript $_{0}$. 

The study treats a grand piano string as a cylinder material, and
physically models it as a 1D beam prestressed longitudinally. As shown
in figure \ref{fig:string}, the string is fixed at the agraffe end
(immobile) and coupled to the bridge end (mobile) \cite{castera2023piano}
through pins. Define that the central line of string range start from
the tunning pin $[-L_{0},0,0]^{\top}$, then go through the agraffe
$\boldsymbol{x}_{0}=[0,0,0]^{\top}$, the front bridge pin $\boldsymbol{x}_{1}=[L_{1},0,0]^{\top}$,
the rear bridge pin $\boldsymbol{x}_{2}=[L_{2},0,0]^{\top}$, and
finally the hitch pin $\boldsymbol{x}_{3}=[L_{3},0,0]^{\top}$, where
$L_{1}$ is the so-called ``speaking length''; let $r$ be the radius
of cross-section and $\rho$ be the homogenous density per unit volume.
It is known that the vibration of piano strings is primarily vertical
($z$ direction) \cite{bank2018model}. Nevertheless, logitudinal
vibration ($x$ direction) \cite{bank2005generation} and horizontal
vibration ($y$ direction) \cite{tan2017piano} may also be essential,
as they contribute respectively to the sound precursor \cite{castera2023piano}
and double-decay \cite{tan2015double} phenomena. Two more kinds of
vibration that require attention are the rotations of the string's
cross-section towards the $y$ and $z$ axes, so as to account for
the stiffness of piano strings that may result in slightly inharmonic
sounds \cite{chabassier2013modeling}. 

To incorporate all these essential kinds of vibrations, define the
displacement field of a piano string as $\boldsymbol{u}^{*}\approx\boldsymbol{u}=[u,v,w]^{\top}$
in a similar notion to (\ref{eq:displacement_space_discretization}).
In addition, we specify that 
\begin{align}
 & \boldsymbol{\varphi}_{1}(x,y,z)=\left[\begin{array}{c}
\boldsymbol{\varphi}_{11}(x)\\
-y\boldsymbol{\varphi}_{12}(x)\\
-z\boldsymbol{\varphi}_{13}(x)
\end{array}\right],\;\boldsymbol{S}_{1}=\left[\begin{array}{c}
\boldsymbol{S}_{11}\\
\boldsymbol{S}_{12}\\
\boldsymbol{S}_{13}
\end{array}\right],\nonumber \\
 & \boldsymbol{\varphi}_{2}(x,y,z)=\boldsymbol{\varphi}_{2}(x),\;\boldsymbol{\varphi}_{3}(x,y,z)=\boldsymbol{\varphi}_{3}(x),\label{eq:string_displacement}
\end{align}
where subscripts $_{11},_{2},_{3},_{12},_{13}$ refer to vectors or
matrices defined for longitudinal, horizontal, vertical, horizontal
rotational and vertical rotational vibrations respectively, as marked
$u,v,w,\alpha,\beta$ in figure \ref{fig:string}. Definition for
$\boldsymbol{\varphi}_{0,1}$ and $\boldsymbol{S}_{0,1}$ are similar
to the above. From a 3D perspective, it can be seen that with regard
to the $y$ and $z$ axes, the $x$ displacement is first-order modeled,
while the $y$ and $z$ displacements are zeroth-order modeled. As
the steel used to manufacture piano strings can be considered as isotropic
material, the elasticity coeffieicnts in (\ref{eq:soundboard_elasticity_coef})
simplifies to 
\begin{align}
E & =E_{x}=E_{y}=E_{z},\nonumber \\
\nu & =\nu_{xy}=\nu_{xz}=\nu_{yz},\nonumber \\
G & =\frac{E}{2(1+\nu)}=G_{xy}=G_{xz}=G_{yz}.
\end{align}
It now suffices to derive a system of $N$ coupled second-order ODEs
similar to (\ref{eq:ODEs}) (\ref{eq:ODEs_where}). In order that
reasonable solutions can be found, equation (\ref{eq:pde}) is at
least constrained by the following Dirichlet boundary conditions:
\begin{equation}
\boldsymbol{u}|_{x=0}=\boldsymbol{u}|_{x=L_{3}}=\boldsymbol{0},\label{eq:string_boundary_1}
\end{equation}
which means the motion of string vanishes at the agraffe point and
the hitch point. Another Dirichlet boundary condition at the coupling
point $\boldsymbol{x}_{1}$ will be discussed in section \ref{sec:Model-for-coupling}.
The initial conditions are
\begin{equation}
\boldsymbol{u}|_{t=0}=\partial_{t}\boldsymbol{u}|_{t=0}=\boldsymbol{0}.\label{eq:string_initial}
\end{equation}

\section{Model for sound radiation in the air}\label{sec:Model-for-sound-radiation-in-the-air}

\subsection{Model for the air}

\begin{figure}
\begin{centering}
\includegraphics{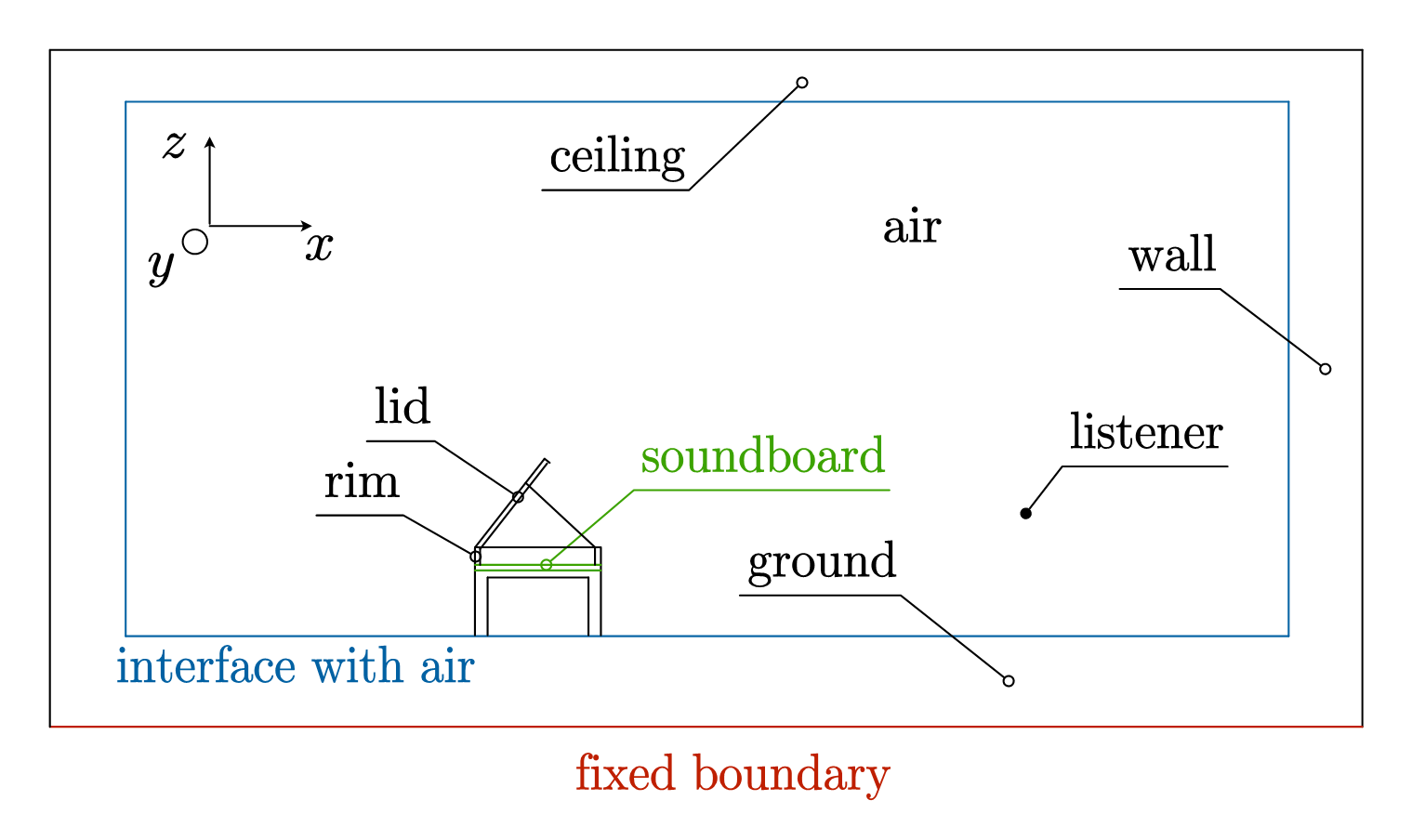}
\par\end{centering}
\caption{The piano in a room}
\label{fig:room}
\end{figure}

The study adopts a 3D acoustic wave model to simulate piano sound
radiation in the air. For simulating wave propagation from the piano
soundboard to a listener situated at a particular location, \cite{elie2022physically}
used Rayleigh integral to compute the acoustic pressure field, which
is able to simulate the different delays and decays of sound at different
positions, but unable to account for boundary conditions. \cite{chabassier2013modeling,chabassier2014time,chabassier2016time}
used 2 first-order linear acoustic equations for 4 coupled unknowns:
acoustic pressure (1 unknown) and acoustic velocity (3 unknowns).
This approach excels at capturing reflections at spatial boundaries
like walls and soundboard-air coupling, but seems unable to produce
decaying room impulse response due to the absence of damping terms\footnote{It seems the acoustic model in \cite{chabassier2014time} relies on
soundboard-air coupling to produce the damping. When the acoustic
equations are fully coupled to the soundboard equations, the soundboard's
damping mechanisms may apply to the air as well. This means even if
the acoustic equations contains no damping, the overall energy may
be stll stable and waves with infinite amplitudes would not occur.
However, we choose to add viscous damping terms to the acoustic equations
for two reasons. Firstly, it can be observed in audio recording that
a very short impulse in a room gets a decaying response, which attributes
to the reverberation effect. Secondly, the coupling computation approach
we shall introduce later actually relies on decoupling and iterative
strategies, meaning that without the room itself's damping mechanisms
the computed pressure may exhibit infinite amplitudes.}. 

Fluid dynamics exhibit both linear and nonlinear behaviours as decribed
by the Navier-Stokes equation, but considering that sound pressure
fluctuations in air are often small, linearized models for sound radiation
in the air can achieve a satisfactory level of accuracy for us. The
acoustic radiation equations we utilize are based on the linearized
fluid dynamics equations \cite{dunn2015springer}, including conservation
of mass and momentum. It works for isotropic, compressible, viscous,
adiabatic flow with homogenous ambient states, in Eulerian description.
Damping due to heat conduction is simplified away here and readers
may refer to \cite{hamilton2020fdtd,hamilton2021time} for incorporating
it. The governing PDEs write 
\begin{align}
 & \frac{1}{\rho c^{2}}\dot{p}+\nabla\cdot\boldsymbol{u}=0,\label{eq:air_pde_coupled_1}\\
 & \rho\dot{\boldsymbol{u}}=\nabla\cdot\boldsymbol{G}+\boldsymbol{F},\label{eq:air_pde_coupled_2}
\end{align}
where $\boldsymbol{u}(x,y,z,t)$ is the acoustic velocity field; $p(x,y,z,t)$
is the perturbation of acoustic pressure field, also should be the
final digital audio signal; $\rho$ is the air density; $c$ is the
sound propagation speed in the air; $\boldsymbol{G}$ is the symmetric
surface force tensor; the gradient operator $\nabla$ applied to a
vector returns its Jacobian matrix, and the divergence operator $\nabla\cdot$
applied to a matrix returns a vector of each row's divergence; $\boldsymbol{F}=[F_{1},F_{2},F_{3}]^{\top}$
is the external force (viz. body force) treated as zero in this study.
According to the Navier-Stokes equation, the surface force tensor
$\boldsymbol{G}$ is defined as 
\begin{equation}
\boldsymbol{G}=-p\boldsymbol{I}+\mu_{1}\left(\nabla\boldsymbol{u}+(\nabla\boldsymbol{u})^{\top}\right)+\mu_{2}(\nabla\cdot\boldsymbol{u})\boldsymbol{I},\label{eq:air_pde_surface_force}
\end{equation}
where $\mu_{1}$ and $\mu_{B}$ are dynamic viscosity and bulk viscosity
coefficients accounting for energy dissipation, and $\mu_{2}=\mu_{B}-\frac{2}{3}\mu_{1}$.
To express the divergence of surface force using acoustic velocity
gradients, we find 
\begin{align}
 & \nabla\cdot\boldsymbol{G}_{i}=-\nabla p+\nabla\cdot\boldsymbol{\varsigma}_{i},\;\boldsymbol{\varsigma}_{i}=\sum_{j=1}^{3}\left(\mu_{1}\boldsymbol{A}_{ij}+\mu_{2}\boldsymbol{B}_{ij}\right)\nabla u_{j},\nonumber \\
 & \boldsymbol{A}=\left[\begin{array}{c}
\boldsymbol{A}_{1}\\
\boldsymbol{A}_{2}\\
\boldsymbol{A}_{3}
\end{array}\right]=\left[\begin{array}{ccc}
\boldsymbol{A}_{11} & \boldsymbol{A}_{12} & \boldsymbol{A}_{13}\\
\boldsymbol{A}_{21} & \boldsymbol{A}_{22} & \boldsymbol{A}_{23}\\
\boldsymbol{A}_{31} & \boldsymbol{A}_{32} & \boldsymbol{A}_{33}
\end{array}\right],\;\boldsymbol{B}=\left[\begin{array}{c}
\boldsymbol{B}_{1}\\
\boldsymbol{B}_{2}\\
\boldsymbol{B}_{3}
\end{array}\right]=\left[\begin{array}{ccc}
\boldsymbol{B}_{11} & \boldsymbol{B}_{12} & \boldsymbol{B}_{13}\\
\boldsymbol{B}_{21} & \boldsymbol{B}_{22} & \boldsymbol{B}_{23}\\
\boldsymbol{B}_{31} & \boldsymbol{B}_{32} & \boldsymbol{B}_{33}
\end{array}\right],\nonumber \\
 & \boldsymbol{A}=\left[\begin{array}{ccc|ccc|ccc}
2 & 0 & 0 & 0 & 0 & 0 & 0 & 0 & 0\\
0 & 1 & 0 & 1 & 0 & 0 & 0 & 0 & 0\\
0 & 0 & 1 & 0 & 0 & 0 & 1 & 0 & 0\\
\hline 0 & 1 & 0 & 1 & 0 & 0 & 0 & 0 & 0\\
0 & 0 & 0 & 0 & 2 & 0 & 0 & 0 & 0\\
0 & 0 & 0 & 0 & 0 & 1 & 0 & 1 & 0\\
\hline 0 & 0 & 1 & 0 & 0 & 0 & 1 & 0 & 0\\
0 & 0 & 0 & 0 & 0 & 1 & 0 & 1 & 0\\
0 & 0 & 0 & 0 & 0 & 0 & 0 & 0 & 2
\end{array}\right],\;\boldsymbol{B}=\left[\begin{array}{ccc|ccc|ccc}
1 & 0 & 0 & 0 & 1 & 0 & 0 & 0 & 1\\
0 & 0 & 0 & 0 & 0 & 0 & 0 & 0 & 0\\
0 & 0 & 0 & 0 & 0 & 0 & 0 & 0 & 0\\
\hline 0 & 0 & 0 & 0 & 0 & 0 & 0 & 0 & 0\\
1 & 0 & 0 & 0 & 1 & 0 & 0 & 0 & 1\\
0 & 0 & 0 & 0 & 0 & 0 & 0 & 0 & 0\\
\hline 0 & 0 & 0 & 0 & 0 & 0 & 0 & 0 & 0\\
0 & 0 & 0 & 0 & 0 & 0 & 0 & 0 & 0\\
1 & 0 & 0 & 0 & 1 & 0 & 0 & 0 & 1
\end{array}\right],
\end{align}
where $\boldsymbol{G}_{i}$, $\boldsymbol{\varsigma}_{i}$ is the
$i$ th column or row of $\boldsymbol{G}$, $\boldsymbol{\varsigma}$;
here $\boldsymbol{\varsigma}$ is the damping force tensor. The 2
PDEs (\ref{eq:air_pde_coupled_1}) (\ref{eq:air_initial}) then become\footnote{It is also viable to decouple the acoustic equations into a single
second-order equation containing only $\boldsymbol{u}$ as unknown.
We choose to not do so because it would result in a damping matrix
not diagonalizable by the mass or stiffness matrices. To fully decouple
the system ODEs would then require doubling the DOFs (equivalent to
6 variables) to transform into a first-order system. This appears
suboptimal to us as the original first-order system of acoustic equations
has only 4 varaibles.} 
\begin{align}
 & \rho\dot{u}_{i}-\nabla\cdot\boldsymbol{\varsigma}_{i}+\partial_{x_{i}}p=F_{i}\;(i=1,2,3),\label{eq:air_pde_decoupled_1}\\
 & \dot{p}+\rho c^{2}\nabla\cdot\boldsymbol{u}=0,\label{eq:air_pde_decoupled_2}
\end{align}
where the divergence of velocity can be expressed in velocity gradients
as 
\begin{align}
 & \nabla\cdot\boldsymbol{u}=\sum_{j=1}^{3}\boldsymbol{H}_{j}\nabla u_{j},\nonumber \\
 & \boldsymbol{H}=\left[\begin{array}{ccc}
\boldsymbol{H}_{1} & \boldsymbol{H}_{2} & \boldsymbol{H}_{3}\end{array}\right]=\left[\begin{array}{ccc|ccc|ccc}
1 & 0 & 0 & 0 & 0 & 0 & 0 & 0 & 0\\
0 & 0 & 0 & 0 & 1 & 0 & 0 & 0 & 0\\
0 & 0 & 0 & 0 & 0 & 0 & 0 & 0 & 1
\end{array}\right].
\end{align}
For acoustic modes the vorticity $\nabla\times\boldsymbol{u}=\boldsymbol{0}$
may be assumed zero so that a decoupled equation containing only $p$
as unknown may be obtained, but we choose to not do so because the
velocity field of any solid to couple with may not satisfy zero vorticity.

Now we seek for the weak forms of (\ref{eq:air_pde_decoupled_1})
(\ref{eq:air_pde_decoupled_2}) via the Galerkin method. Multiplying
an arbitrary test function vector $\boldsymbol{\psi}$ on both sides
of (\ref{eq:air_pde_decoupled_1}) yields 
\begin{align}
\int_{\Omega}\boldsymbol{\psi}\rho\dot{u}_{i}\mathrm{d}V-\int_{\Omega}\boldsymbol{\psi}\left(\nabla\cdot\boldsymbol{\varsigma}_{i}\right)\mathrm{d}V & +\int_{\Omega}\boldsymbol{\psi}\partial_{x_{i}}p\mathrm{d}V=\int_{\Omega}\boldsymbol{\psi}F_{i}\mathrm{d}V\nonumber \\
\int_{\Omega}\boldsymbol{\psi}\rho\dot{u}_{i}\mathrm{d}V+\int_{\Omega}\left(\nabla\boldsymbol{\psi}\right)\boldsymbol{\varsigma}_{i}\mathrm{d}V & +\int_{\Omega}\boldsymbol{\psi}\partial_{x_{i}}p\mathrm{d}V=\int_{\Omega}\boldsymbol{\psi}F_{i}\mathrm{d}V.\label{eq:air_pde_weak_form_1}
\end{align}
When using integration by parts in the above, Neumann boundary condition
is imposed as 
\begin{equation}
\boldsymbol{\varsigma}_{i}\cdot\mathrm{d}\boldsymbol{\Gamma}=0,\label{eq:air_Neumann}
\end{equation}
where $\mathrm{d}\boldsymbol{\Gamma}$ is the outward normal vector
of the tangent plane of any point on the boundary of acoustic space.
This condition means the damping force should vanish on the non-Dirichlet
boundaries. Similarly for (\ref{eq:air_pde_decoupled_2}), multiplying
a test function vector results in 
\begin{equation}
\int_{\Omega}\boldsymbol{\psi}\dot{p}\mathrm{d}V+\int_{\Omega}\boldsymbol{\psi}\rho c^{2}\left(\nabla\cdot\boldsymbol{u}\right)\mathrm{d}V=0.\label{eq:air_pde_weak_form_2}
\end{equation}
To derive the ODEs, we define space discretization similar to (\ref{eq:displacement_space_discretization}),
but with one more variable $p$, as 
\begin{align}
u(x,y,z,t) & =\boldsymbol{\varphi}_{1}(x,y,z)\cdot\left[\boldsymbol{S}_{1}\boldsymbol{\xi}(t)+\boldsymbol{S}_{0,1}\boldsymbol{\xi}_{0}(t)\right],\nonumber \\
v(x,y,z,t) & =\boldsymbol{\varphi}_{2}(x,y,z)\cdot\left[\boldsymbol{S}_{2}\boldsymbol{\xi}(t)+\boldsymbol{S}_{0,2}\boldsymbol{\xi}_{0}(t)\right],\nonumber \\
w(x,y,z,t) & =\boldsymbol{\varphi}_{3}(x,y,z)\cdot\left[\boldsymbol{S}_{3}\boldsymbol{\xi}(t)+\boldsymbol{S}_{0,3}\boldsymbol{\xi}_{0}(t)\right],\nonumber \\
p(x,y,z,t) & =\boldsymbol{\varphi}_{4}(x,y,z)\cdot\left[\boldsymbol{S}_{4}\boldsymbol{\xi}(t)+\boldsymbol{S}_{0,4}\boldsymbol{\xi}_{0}(t)\right].
\end{align}
And similar to (\ref{eq:space_discretization_map}), we define convenience
matrices to express that 
\begin{align}
 & \boldsymbol{u}=\boldsymbol{P}\boldsymbol{\xi}+\boldsymbol{P}_{0}\boldsymbol{\xi}_{0},\;u_{i}=\boldsymbol{P}_{i}\boldsymbol{\xi}+\boldsymbol{P}_{0,i}\boldsymbol{\xi}_{0},\;\nabla u_{i}=\boldsymbol{Q}_{i}\boldsymbol{\xi}+\boldsymbol{Q}_{0,i}\boldsymbol{\xi}_{0},\;\nonumber \\
 & \boldsymbol{P}=\left[\begin{array}{c}
\boldsymbol{P}_{1}\\
\boldsymbol{P}_{2}\\
\boldsymbol{P}_{3}
\end{array}\right]=\left[\begin{array}{c}
\boldsymbol{\varphi}_{1}^{\top}\boldsymbol{S}_{1}\\
\boldsymbol{\varphi}_{2}^{\top}\boldsymbol{S}_{2}\\
\boldsymbol{\varphi}_{3}^{\top}\boldsymbol{S}_{3}
\end{array}\right],\;\boldsymbol{P}_{0}=\left[\begin{array}{c}
\boldsymbol{P}_{0,1}\\
\boldsymbol{P}_{0,2}\\
\boldsymbol{P}_{0,3}
\end{array}\right]=\left[\begin{array}{c}
\boldsymbol{\varphi}_{1}^{\top}\boldsymbol{S}_{0,1}\\
\boldsymbol{\varphi}_{2}^{\top}\boldsymbol{S}_{0,2}\\
\boldsymbol{\varphi}_{3}^{\top}\boldsymbol{S}_{0,3}
\end{array}\right],\nonumber \\
 & \boldsymbol{Q}=\left[\begin{array}{c}
\boldsymbol{Q}_{1}\\
\boldsymbol{Q}_{2}\\
\boldsymbol{Q}_{3}
\end{array}\right]=\left[\begin{array}{c}
\left(\nabla\boldsymbol{\varphi}_{1}\right)^{\top}\boldsymbol{S}_{1}\\
\left(\nabla\boldsymbol{\varphi}_{2}\right)^{\top}\boldsymbol{S}_{2}\\
\left(\nabla\boldsymbol{\varphi}_{3}\right)^{\top}\boldsymbol{S}_{3}
\end{array}\right],\;\boldsymbol{Q}_{0}=\left[\begin{array}{c}
\boldsymbol{Q}_{0,1}\\
\boldsymbol{Q}_{0,2}\\
\boldsymbol{Q}_{0,3}
\end{array}\right]=\left[\begin{array}{c}
\left(\nabla\boldsymbol{\varphi}_{1}\right)^{\top}\boldsymbol{S}_{0,1}\\
\left(\nabla\boldsymbol{\varphi}_{2}\right)^{\top}\boldsymbol{S}_{0,2}\\
\left(\nabla\boldsymbol{\varphi}_{3}\right)^{\top}\boldsymbol{S}_{0,3}
\end{array}\right],\nonumber \\
 & \boldsymbol{P}_{4}=\boldsymbol{\varphi}_{4}^{\top}\boldsymbol{S}_{4},\;\boldsymbol{P}_{0,4}=\boldsymbol{\varphi}_{4}^{\top}\boldsymbol{S}_{0,4},\;\boldsymbol{Q}_{4}=\left(\nabla\boldsymbol{\varphi}_{4}\right)^{\top}\boldsymbol{S}_{4}\;,\boldsymbol{Q}_{0,4}=\left(\nabla\boldsymbol{\varphi}_{4}\right)^{\top}\boldsymbol{S}_{0,4}.
\end{align}
Then similar to (\ref{eq:space_discretization_stress_prestress}),
we find 
\begin{align}
 & \sum_{j=1}^{3}\boldsymbol{A}_{ij}\nabla u_{j}=\boldsymbol{A}_{i}\boldsymbol{Q}\boldsymbol{\xi}+\boldsymbol{A}_{i}\boldsymbol{Q}_{0}\boldsymbol{\xi}_{0},\nonumber \\
 & \sum_{j=1}^{3}\boldsymbol{B}_{ij}\nabla u_{j}=\boldsymbol{B}_{i}\boldsymbol{Q}\boldsymbol{\xi}+\boldsymbol{B}_{i}\boldsymbol{Q}_{0}\boldsymbol{\xi}_{0},\nonumber \\
 & \partial_{x_{i}}p=\partial_{x_{i}}\boldsymbol{\varphi}_{4}\cdot\boldsymbol{S}_{4}\boldsymbol{\xi}+\partial_{x_{i}}\boldsymbol{\varphi}_{4}\cdot\boldsymbol{S}_{0,4}\boldsymbol{\xi}_{0},\label{eq:air_stress_damping_force}
\end{align}
so that space partial derivatives can be expressed as linear transformations
of the DOF vector $\boldsymbol{\xi}(t)$ or its time derivatives.
It now sufficies to derive the system of second-order ODEs similar
to (\ref{eq:ODEs}) (\ref{eq:ODEs_where}). We use $\boldsymbol{P}_{i}^{\top}$
as a vector of test functions for the $i$ th weak form equation in
(\ref{eq:air_pde_weak_form_1}) and use $\boldsymbol{P}_{4}^{\top}$
to test the weak form (\ref{eq:air_pde_weak_form_2}), which would
yield 4 groups of equations, each group having $N$ equations. Summing
the 4 groups into 1 group, we derive a system of $N$ coupled first-order
ODEs as 
\begin{equation}
\boldsymbol{M}\dot{\boldsymbol{\xi}}(t)+\boldsymbol{K}\boldsymbol{\xi}(t)=\boldsymbol{f}(t),\label{eq:air_ODEs}
\end{equation}
where 
\begin{align}
 & \boldsymbol{M}=\int_{\Omega}\left(\rho\boldsymbol{P}^{\top}\boldsymbol{P}+\boldsymbol{P}_{4}^{\top}\boldsymbol{P}_{4}\right)\mathrm{d}V,\nonumber \\
 & \boldsymbol{M}_{0}=\int_{\Omega}\left(\rho\boldsymbol{P}^{\top}\boldsymbol{P}_{0}+\boldsymbol{P}_{4}^{\top}\boldsymbol{P}_{0,4}\right)\mathrm{d}V,\nonumber \\
 & \boldsymbol{K}=\int_{\Omega}\left[\boldsymbol{Q}^{\top}\left(\mu_{1}\boldsymbol{A}+\mu_{2}\boldsymbol{B}\right)\boldsymbol{Q}+\rho c^{2}\boldsymbol{P}_{4}^{\top}\boldsymbol{H}\boldsymbol{Q}+\boldsymbol{P}^{\top}\boldsymbol{Q}_{4}\right]\mathrm{d}V,\nonumber \\
 & \boldsymbol{K}_{0}=\int_{\Omega}\left[\boldsymbol{Q}^{\top}\left(\mu_{1}\boldsymbol{A}+\mu_{2}\boldsymbol{B}\right)\boldsymbol{Q}_{0}+\rho c^{2}\boldsymbol{P}_{4}^{\top}\boldsymbol{H}\boldsymbol{Q}_{0}+\boldsymbol{P}^{\top}\boldsymbol{Q}_{0,4}\right]\mathrm{d}V,\nonumber \\
 & \boldsymbol{f}(t)=\int_{\Omega}\boldsymbol{P}^{\top}\boldsymbol{F}\mathrm{d}V-\boldsymbol{M}_{0}\dot{\boldsymbol{\xi}_{0}}(t)-\boldsymbol{K}_{0}\boldsymbol{\xi}_{0}(t).\label{eq:air_ODEs_where}
\end{align}
Here $\boldsymbol{M}$ and $\boldsymbol{K}$ may not be as meaningful
as the mass and stiffness matrices in the second-order ODEs case.
The initial conditions are 
\begin{equation}
\boldsymbol{u}|_{t=0}=\boldsymbol{0},\;p|_{t=0}=0.\label{eq:air_initial}
\end{equation}

The acoustic space is often finite, which means boundary conditions
are essential to shape the solutions of acoustic equations. Let $\Gamma^{(c)}\in\mathbb{R}^{3}$
be the acoustic space's whole solid boundary, of which $\Gamma_{1}^{(c)},\Gamma_{2}^{(c)}\in\Gamma^{(c)}$
are the room boundary (grounds, walls, ceilings) and piano soundboard
parts respectively, as shown in figure \ref{fig:room}. In \cite{chabassier2014time},
the walls were assumed rigid, which can produce sound reflection effects
but may miss sound absorption effects. This motivates our introduction
of vibratory acoustic boundaries to account for various phenomena
like reflection, scattering, absorption, diffusion, resonance etc.

\subsection{Model for the room}

In this section, we delve into the specification of room boundary
models, which serve as the cornerstone for the subsequent analysis
of solid-air coupling dynamics. For the sake of conciseness, we adopt
a simplified representation of the room environment as a shoebox-shaped
3D space accommodating a piano. It is worth noting, however, that
this room model remains versatile and can be readily adapted to accommodate
irregular room geometries. The room's boundaries---encompassing floors,
walls, and ceilings---play pivotal roles in shaping the acoustic
behavior within, acting as both reflectors and absorbers of sound
waves. In the context of physical modeling, these boundaries are treated
as 3D elastic materials, akin to the piano's soundboard model. This
methodological alignment facilitates the straight application of the
general 3D elasticity model and the soundboard model to the room,
obviating the need for reformulations of the governing equations and
weak forms.

The primary specialization within this framework lies in the stipulation
of boundary conditions of room barriers. As highlighted red in figure
\ref{fig:room}, the inner side of ground, wall and ceiling (blue
color), defined as $\Gamma_{1}^{(f)}$, is the room material's interface
with the air; the underside of ground material (red color), defined
as $\Gamma_{2}^{(f)}$, is where the Dirichlet boundary condition
applies. Firmly rooted in the earth, the ground material should experience
zero displacement along $\Gamma_{2}^{(f)}$. This condition is written
as 
\begin{equation}
\boldsymbol{u}^{(f)}|_{\boldsymbol{x}\in\Gamma_{2}^{(f)}}=\boldsymbol{0}.
\end{equation}
In case the inner concrete layers of walls and ceilings are considered
rigid, displacements on the surfaces of them should be zero too. Another
specialization is that prestress need not be considered for room barriers.
Even if it exists, the dynamic deformation may not be large enough
to induce significant effect of prestress on vibration.

\section{Model for coupling between different parts of the piano}\label{sec:Model-for-coupling}

\subsection{Model for hammer-string coupling}

\begin{figure}
\begin{centering}
\includegraphics{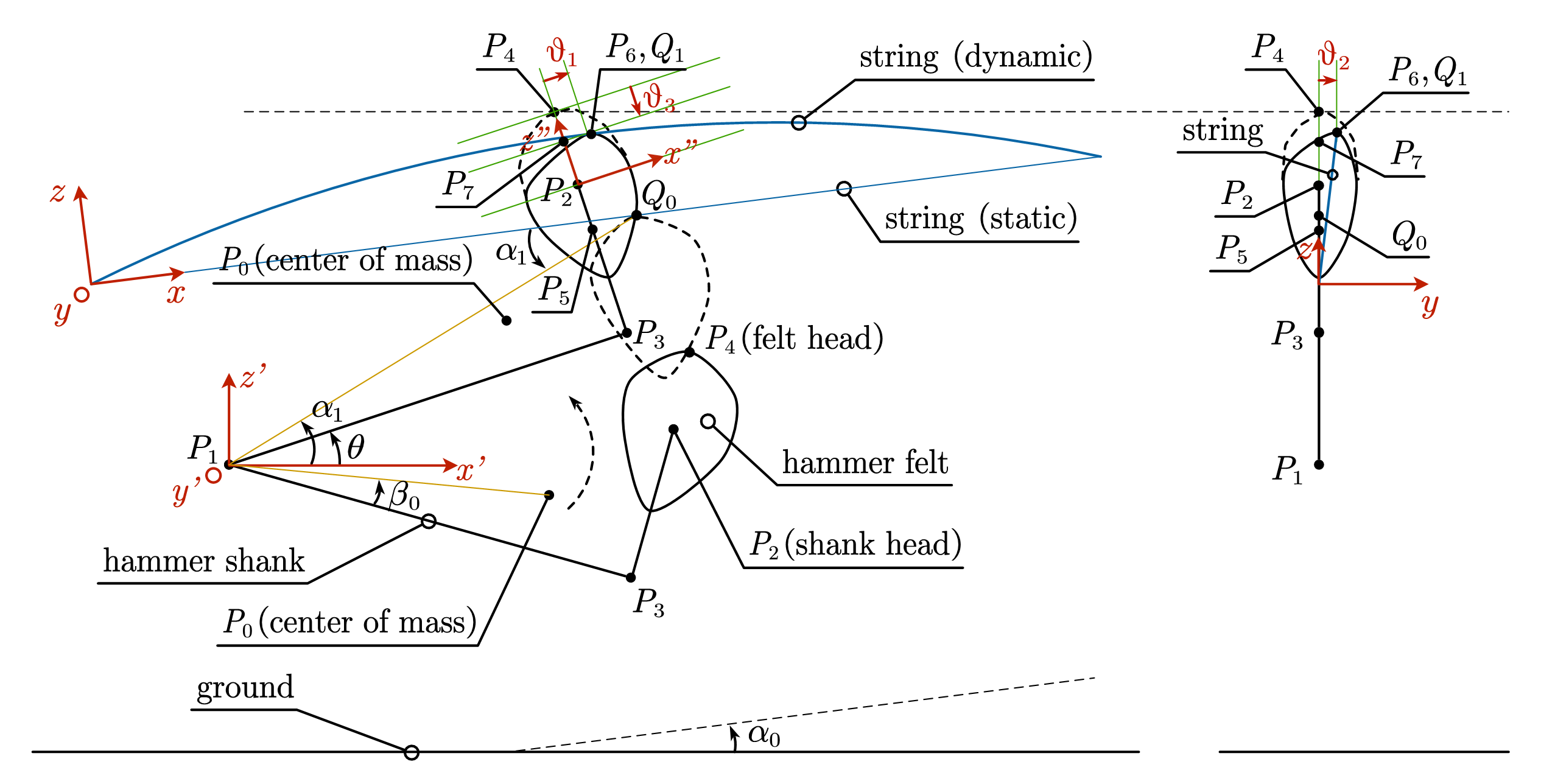}
\par\end{centering}
\caption{The piano hammer view from $y$ (left) and $x$ (right) direction}
\label{fig:hammer}
\end{figure}

The piano hammer positioned below the string functions the excitation
of string vibration, which is known to be a highly nonlinear process
\cite{ducceschi2022real}. We consider two main parts of the hammer
relevant to this excitation process: the wooden hammer shank that
can be approximated as non-deformable; the hammer felt that is deformable,
impacting and exerting force on the string. Normally, the hammer moves
in a circle when triggered by piano player's key action, striking
the string at certain point in during a very short period of time,
and is also pushed back by the vibrating string.

\subsubsection{One hammer striking one string}

For the case of one hammer striking one string, the study follows
the 0D nonlinear hammer-string interaction model in \cite{chabassier2014time},
with some refinements: for the case of one hammer striking one string,
the rotational movement of hammer \cite{chabassier2014energy} and
the horizontal interaction force are taken into account. 

As shown in figure \ref{fig:hammer}, $xyz$ is the coordinate system
of piano string, where the $x$ axis of the string system has an angle
$\alpha_{0}$ to the ground. To better describe the hammer shank motion
which is rotation around point $P_{1}$, we will not often the use
the $xyz$ system. Instead, we define a static coordinate system $x'y'z'$
with origin $P_{1}$, $y'$ axis the same as $y$ axis, $z'$ axis
with negative direction the same as gravity, and $x'$ axis perpendicular
to the $y'O'z'$ plane. The motion of hammer shank as a rigid body
can then be described as rotating around the $y'$ axis and not moving
in the $y'$ direction. We can thus use a single variable $\theta(t)$,
the angle of $\overrightarrow{P_{1}P_{3}}$ to the $x'$ axis, to
fuuly describe the hammer shank motion. The hammer shank drives the
overall motion of the shank head $P_{2}$ and felt head $P_{4}$.
Nevertheless, when the felt head is in contact with the string's $Q_{0}$
point, it undergoes a compression $\boldsymbol{\vartheta}=[\vartheta_{1},\vartheta_{2},\vartheta_{3}]^{\top}$
from $P_{4}$ to $P_{6}$, which also overlaps with the dynamic string
point $Q_{1}$, which also contributes to its motion. To perform a
reasonable orthogonal decomposition of this compression, we define
a dynamic coordinate system $x''y''z''$ with origin $P_{3}$, $x''$
axis with positive direction $\overrightarrow{P_{5}Q_{0}}$, $y''$
axis with positive direction the same as $y$ axis, and $z''$ axis
with positive direction $\overrightarrow{P_{5}P_{2}}$. The hammer
compression can then be decomposed onto the $x'',y'',z''$ axes as
marked as marked $\vartheta_{1},\vartheta_{2},\vartheta_{3}$ respectively
in figure \ref{fig:hammer}. 

Denote the displacement of string point $Q_{0}$ with coordinate $\boldsymbol{x}_{0}^{(a)}$
in the $xyz$ system as $\boldsymbol{u}_{0}^{(a)}=\boldsymbol{u}^{(a)}(\boldsymbol{x}_{0}^{(a)},t)$.
Denote the lengths of $Q_{0}P_{1}$, $P_{0}P_{1}$, $P_{1}P_{3}$,
$P_{2}P_{3}$, $P_{1}P_{2}$, $P_{2}P_{4}$ as $L$, $L_{0}$, $L_{1}$,
$L_{2}$, $L_{3}$, $L_{4}$ respectively, and the angle of $\overrightarrow{P_{1}Q_{0}}$
to the $x'$ axis as $\alpha_{1}$. Our geometric analysis finds the
coordinates of $Q_{1}$ (equivalent to $P_{6}$ in case of compression)
in the $x'y'z'$ system and $x''y''z''$ system, denoted $\boldsymbol{x}_{1}^{(a')}$
and $\boldsymbol{x}_{1}^{(a'')}$, are 
\begin{align}
 & \boldsymbol{x}_{1}^{(a')}=\boldsymbol{r}_{0}+\boldsymbol{R}_{0}\boldsymbol{u}_{0}^{(a)},\;\boldsymbol{x}_{1}^{(a'')}=\boldsymbol{r}+\boldsymbol{R}(\theta)\boldsymbol{x}_{1}^{(a')},\nonumber \\
 & \boldsymbol{r}_{0}=\left[\begin{array}{c}
L\cos\alpha_{1}\\
0\\
L\sin\alpha_{1}
\end{array}\right],\;\boldsymbol{R}_{0}=\left[\begin{array}{ccc}
\cos\alpha_{0} & 0 & -\sin\alpha_{0}\\
0 & 1 & 0\\
\sin\alpha_{0} & 0 & \cos\alpha_{0}
\end{array}\right],\nonumber \\
 & \boldsymbol{r}=\left[\begin{array}{c}
-L_{1}\\
0\\
-L_{2}
\end{array}\right],\;\boldsymbol{R}(\theta)=\left[\begin{array}{ccc}
\cos\theta & 0 & \sin\theta\\
0 & 1 & 0\\
-\sin\theta & 0 & \cos\theta
\end{array}\right].\label{eq:hammer_string_point_coordinate}
\end{align}
And converting $\boldsymbol{x}_{1}^{(a'')}$ to $\boldsymbol{x}_{1}^{(a')}$
results in 
\begin{align}
 & \boldsymbol{x}_{1}^{(a')}=\boldsymbol{s}(\theta)+\boldsymbol{S}(\theta)\boldsymbol{x}_{1}^{(a'')},\nonumber \\
 & \boldsymbol{s}(\theta)=\left[\begin{array}{c}
L_{1}\cos\theta-L_{2}\sin\theta\\
0\\
L_{1}\sin\theta+L_{2}\cos\theta
\end{array}\right],\;\boldsymbol{S}(\theta)=\left[\begin{array}{ccc}
\cos\theta & 0 & -\sin\theta\\
0 & 1 & 0\\
\sin\theta & 0 & \cos\theta
\end{array}\right].\label{eq:hammer_string_point_coordinate_inv}
\end{align}

As previously discussed, we assume $Q_{0}$ and $P_{2}$ is non-deformable
whereas $P_{4}$ is deformable. This means when the hammer is in contact
with string, $P_{4}$ moves to $P_{6}$; when contact is absent, the
position of $P_{4}$ is determined by the shank but not the string.
The condition to satisfy that the hammer is in contact with string
should be $z_{1}^{(a'')}\leq L_{4}$, viz. the $z''$ direction distance
between string point and shank head is less than the static thickness
of hammer felt. The hammer felt compression is then 
\begin{equation}
\boldsymbol{\vartheta}(\theta)=\left\{ \begin{array}{l}
\boldsymbol{x}_{1}^{(a'')}-\boldsymbol{L}_{4},\;z_{1}^{(a'')}\leq L_{4}\\
\boldsymbol{0},\;z_{0}^{(a'')}>L_{4}
\end{array}\right.\label{eq:hammer_compression}
\end{equation}
where $\boldsymbol{L}_{4}=[0,0,L_{4}]^{\top}$; negative value of
$\vartheta_{i}$ means the hammer felt is compressed towards the negative
direction of $x_{i}''$ axis, and vice versa. Due to compression,
the hammer felt exerts an interaction force $\boldsymbol{F}^{(a'')}(\theta)=[F_{1}^{(a'')},F_{2}^{(a'')},F_{3}^{(a'')}]^{\top}$
(in $x''y''z''$ system) on the string. Reciprocally, the hammer felt
suffers $-\boldsymbol{F}^{(a'')}$ from the string according to Newton's
third law\footnote{Note that we ignore here the string's normal and tangent stress and
prestress on the interaction surface exerting on the hammer felt (as
well as the relevant strain on the string side), which should already
be zero per the Neumann boundary condition specified in (\ref{eq:Neumann_boundary})
if DOFs are given to string displacements on this surface. Even if
they should not be zero, it is relatively acceptable to ignore them
as they are probably less contributive than the hammer's compression
force. Treating them as non-zero would require imposing Dirichlet
boundary conditions on some relevant DOFs of the string's side, which
is a computational challenge.}. Based on \cite{chabassier2014time}, the hammer-string interaction
force in the $x''_{i}$ axis is modeled as a nonlinear function of
compression as 
\begin{equation}
F_{i}^{(a'')}=-\mathrm{sgn}(\vartheta_{i})\left[k_{i}|\vartheta_{i}|^{p_{i}}+r_{i}k_{i}\partial_{t}\left(|\vartheta_{i}|^{p_{i}}\right)\right],\;i=1,2,3,\label{eq:hammer_force}
\end{equation}
where $k_{i}$ is the stiffness of hammer felt; $p_{i}$ is a positive
exponent accounting for nonlinearity; $r_{i}$ is the relaxation coefficient
accounting for the hammer's hysteretic and dissipative behaviour;
function $\mathrm{sgn}(\cdot)$ returns the sign of a real number.
Converting the interaction force to $x'y'z'$ and $x'y'z'$ coordinates
yields $\boldsymbol{F}^{(a')}=\boldsymbol{S}(\theta)\boldsymbol{F}^{(a'')}$
and $\boldsymbol{F}^{(a)}=\boldsymbol{R}_{0}^{\top}\boldsymbol{F}^{(a')}$.

To couple the hammer force with string motion, we need equations in
the $x'y'z'$ system governing the motion of hammer shank, a rigid
body. Denote the coordinates of shank mass centre $P_{0}$, shank
rotation centre $P_{1}$, shank head $P_{2}$ in the $x'y'z'$ system
as $\boldsymbol{x}_{0}^{(d')}$, $\boldsymbol{x}_{1}^{(d')}$, $\boldsymbol{x}_{2}^{(d')}$
respectively. It can be found that $\boldsymbol{x}_{0}^{(d')}=[L_{0}\cos(\beta_{0}+\theta),0,L_{0}\sin(\beta_{0}+\theta)]^{\top}$
and $\boldsymbol{x}_{2}^{(d')}=\boldsymbol{s}(\theta)$, where $\beta_{0}$
is the angle of $\overrightarrow{P_{1}P_{0}}$ to the $\overrightarrow{P_{1}P_{3}}$.
Define the total mass and homogenous line density of the hammer shank
as $m$, $\rho$. It can be found that $m=\rho(L_{1}+L_{2})$, and
the axis-free position of center of mass $P_{0}$ projected on $P_{1}P_{3}$
and $P_{2}P_{3}$ are $(\frac{1}{2}L_{1}^{2}+L_{1}L_{2})/(L_{1}+L_{2})$
and $\frac{1}{2}L_{2}^{2}/(L_{1}+L_{2})$ respectively. It is obvious
that the shank as a rigid body suffers these forces: string reaction
force $\boldsymbol{F}^{(d'')}=-\boldsymbol{F}^{(a'')}$ at $P_{2}$;
a rotation constraint force at $P_{1}$, which has zero torque with
respect to the $y'$ axis; gravity $[0,0,-mg]^{\top}$ at $P_{0}$,
where $g$ is the scalar gravitational acceleration. It follows that
the total torque with respect to the $y'$ axis is $mgL_{0}\cos(\beta_{0}+\theta)+\boldsymbol{l}\cdot\boldsymbol{F}^{(d'')}(\theta)$
(only 1 dimension of torque is needed here), where $\boldsymbol{l}=[L_{2},0,-L_{1}]^{\top}$;
the moment of inertia with respect to the $y''$ axis, denoted $I$,
is 
\begin{equation}
I=\int_{0}^{L_{1}}\rho l^{2}\mathrm{d}l+\int_{0}^{L_{2}}\rho\left(l^{2}+L_{1}^{2}\right)\mathrm{d}l=\rho\left(\frac{1}{3}L_{1}^{3}+\frac{1}{3}L_{2}^{3}+L_{1}^{2}L_{2}\right)
\end{equation}
which does not depend on $\theta$. Applying Newton's second law for
rotation, the differential equation of hammer shank motion is 
\begin{equation}
I\ddot{\theta}=-\mu\dot{\theta}+mgL_{0}\cos(\beta_{0}+\theta)+\boldsymbol{l}\cdot\boldsymbol{F}^{(d'')}(\theta),\label{eq:hammer_shank_eq}
\end{equation}
where $-\mu\dot{\theta}$ is the simplified damping force. Here $t=0$
is the moment when the felt head first contacts the string and compression
is still zero, and initial angle and angular velocity at this time
should be known. 

From the aforementioned deductions, we can abstract the hammer-string
interaction force $\boldsymbol{F}^{(e)}$ as a nonlinear operator
$\mathcal{F}(\boldsymbol{u}_{0}^{(a)},\theta):(\mathbb{R}^{3},\mathbb{R})\rightarrow\mathbb{R}^{3}$.
This force should be added to the rhs of (\ref{eq:pde}) as a non-conservative
force. Then, (\ref{eq:pde}) and (\ref{eq:hammer_shank_eq}) forms
2 sets of coupled equations for 2 sets of variables $(\boldsymbol{u}^{(a)},\theta)$,
which provides a foundation for obtaining a solution theoretically.
Due to the highly nonlinear nature of hammer-string coupling, particularly
in that Taylor series approximation may not work well for potentially
non-smooth functions in (\ref{eq:hammer_force}), it would be inappropriate
to use the common perturbation method. In section \ref{subsec:Explicit-time-discretization}
we shall introduce an explicit time discretization method to efficiently
solve the hammer-string coupling.

\subsubsection{One hammer striking two or three strings}

The case of one hammer striking multiple, say three strings can simply
be treated it as if three independent felt heads were striking their
corresponding strings. Three independent interaction forces $\boldsymbol{F}_{i=1,2,3}^{(e')}$
are computed from three independent compressions. The string force
exerted on the shank is the sum $\boldsymbol{F}^{(d')}=-\sum_{i=1}^{3}\boldsymbol{F}_{i}^{(e')}$,
assuming the 3 compression forces apply to the same point of shank
head. However, this approach may lose the dynamic interaction between
different striking points of the hammer felt.

A 3D hammer model may be more accurate in capturing the interplay
of different hammer striking points and the different positions of
compression forces. The hammer felt is now modeled as a 3D elastic
material with space discretization. Two kinds of spacial boundaries
consist in the hammer felt: one contains the felt's 3 potential contact
points with the 3 strings, represented as $P_{4,i}$ with coordinates
$\boldsymbol{x}_{4,i=1,2,3}^{(e'')}$; another is the contact surface
$\Gamma_{1}^{(e'')}\in\mathbb{R}^{2}$ with the hammer shank. The
hammer shank is considered as a rigid 3D object, which means no inner
space discretization is needed for it. Above all, one needs to pay
attention to the choice of coordinate system for the 3D geometry of
felt. A recommended choice here is the dynamic $x''y''z''$ system,
which eliminates the felt's rigid body movement component and is thus
suitable for FEM computation. Also, displacement fields $\boldsymbol{u}^{(e'')}(\boldsymbol{x}^{(e'')},t)$
for hammer felt and $\boldsymbol{u}^{(d'')}(\boldsymbol{x}^{(d'')},t)$
for hammer shank need to be defined.

For more realistic modeling that the contact is dynamically distributed
over a region rather than occurring at only some dimensionless points,
readers can refer to appendix \ref{sec:Elastic-material-contacting}
which describes an iterative algorithm for contact problems directly
integrated into FEM, supporting time-varying contact boundary and
boundary coupling. Another more analytical (but also seems more simplified)
approach for contact problems was introduced in \cite{van2021finite},
where additional dynamic variables including pressure, density, stiffness
and damping of the contact surface along with their governing equations
are formulated.

Derivation of the felt compression dynamics is now based on the 3D
elastic material model and soundboard model previously introduced,
without the need for modeling prestress. We first consider the boundary
conditions needed for the felt model. Denote the coordinate of $Q_{1,i}$
in the $x''y''z''$ system as $\boldsymbol{x}_{1,i}^{(a'')}$, which
can be computed similar to (\ref{eq:hammer_string_point_coordinate}).
To represent hammer felt compression, Dirichlet boundary condition
should be imposed on these 3 contact points as 
\begin{equation}
\boldsymbol{u}^{(e'')}(\boldsymbol{x}_{4,i}^{(e'')},t)=\left\{ \begin{array}{l}
\boldsymbol{x}_{1,i}^{(a'')}-\boldsymbol{x}_{4,i}^{(e'')},\;z_{1,i}^{(a'')}\leq L_{4}\\
\boldsymbol{0},\;z_{1,i}^{(a'')}>L_{4}
\end{array}\right.,\label{eq:hammer_felt_3D_boundary_string}
\end{equation}
where $z_{1,i}^{(a'')}\leq L_{4}$ means contact is present and otherwise
not. This determined displacement would occur on the rhs of system
ODEs like (\ref{eq:ODEs}) as a source term. It is assumed without
contact, not only $P_{4,i}$ but also the entire felt will not experience
deformation; with contact, the position of $P_{4,i}$, i.e. $P_{6,i}$,
is the same as $Q_{1,i}$, triggering motion and compression of the
hammer felt. However, this assumption is a simplification that when
the hammer felt leaves the string, its compression immediately recovers,
and may result in a problem that Neumann boundary condition is not
satisfied at $P_{4,i}$ when there is no contact. For a more accurate
treatment of the contact boundary problem, readers can go to appendix
\ref{sec:Elastic-material-contacting}. There we describe an elastic-rigid
3D contact algorithm able to account for the time-varying contact
surface and the time-varying switch between Dirichlet and Neumann
boundary conditions, but with increased algorithmic complexity and
numeric convergence uncertainty. As for another boundary $\Gamma_{1}^{(e'')}$
which interfaces the shank, Dirichlet boundary condition should be
imposed on as 
\begin{equation}
\boldsymbol{u}^{(e'')}(\boldsymbol{x}^{(e'')},t)=\boldsymbol{0},\;\boldsymbol{x}^{(e'')}\in\Gamma_{1}^{(e'')},\label{eq:hammer_felt_3D_boundary_shank}
\end{equation}
because the rigid shank should have no deformation.

We then consider the dynamics of string and shank impacted by the
felt. At the contact point with string, the felt's nonzero surface
forces $\boldsymbol{G}^{(e'')}(\boldsymbol{x}_{4,i}^{(e'')},t)$ should
be transmitted to the string as 
\begin{align}
 & \boldsymbol{F}_{i}^{(a)}=-\left[\begin{array}{ccc}
0 & Z_{12} & Z_{13}\\
Z_{12} & Z_{22} & Z_{23}\\
Z_{13} & Z_{23} & Z_{33}
\end{array}\right]\left[\begin{array}{c}
1\\
1\\
1
\end{array}\right],\nonumber \\
 & \left\{ Z_{ij}\right\} =\boldsymbol{R}_{0}^{\top}\boldsymbol{R}(\theta)^{\top}\boldsymbol{G}^{(e'')}(\boldsymbol{x}_{4,i}^{(e'')},t)\boldsymbol{R}(\theta)\boldsymbol{R}_{0},\label{eq:hammer_felt_3D_force_string}
\end{align}
where $Z_{11}$ is not transmitted considering that the hammer felt
seems to have no direct contact with the string's surface whose normal
is in the $x$ axis (longitudinal direction). As for the felt's surface
force exerting on the shank along the boundary $\Gamma_{1}^{(e'')}=\Gamma_{1}^{(d'')}$,
previous practice of directly transmitting the felt-string interaction
force to the shank is not applicable here due to the 3D nature of
felt and shank. For a point $\boldsymbol{x}^{(e'')}=\boldsymbol{x}^{(d'')}$
in $\Gamma_{1}^{(e'')}$, with $\boldsymbol{n}^{(e'')}(\boldsymbol{x}^{(e'')})$
as the outward normal of its tangent plane, the felt would exert a
surface force 
\begin{equation}
\boldsymbol{F}^{(d'')}(\boldsymbol{x}^{(d'')},t)=-\boldsymbol{G}^{(e'')}(\boldsymbol{x}^{(e'')},t)\boldsymbol{n}^{(e'')}(\boldsymbol{x}^{(e'')})\label{eq:hammer_felt_3D_force_shank}
\end{equation}
 on the shank. Then the total contribution of felt force to the torque
of shank with respect to the $y''$ axis is 
\begin{align}
T_{1}(\theta) & =\int_{\Gamma_{1}^{(d')}}\left(\boldsymbol{E}_{0}\boldsymbol{F}^{(d')}(\boldsymbol{x}^{(d')},t)\cdot\boldsymbol{x}^{(d')}\right)\mathrm{d}V\nonumber \\
 & =\int_{\Gamma_{1}^{(d'')}}\left[\boldsymbol{E}_{0}\boldsymbol{S}(\theta)\boldsymbol{F}^{(d'')}(\boldsymbol{x}^{(d'')},t)\cdot\left(\boldsymbol{s}(\theta)+\boldsymbol{S}(\theta)\boldsymbol{x}^{(d'')}\right)\right]\mathrm{d}V\nonumber \\
 & =\int_{\Gamma_{1}^{(d'')}}\left[\boldsymbol{F}^{(d'')}(\boldsymbol{x}^{(d'')},t)\cdot\left(\boldsymbol{l}+\boldsymbol{E}\boldsymbol{x}^{(d'')}\right)\right]\mathrm{d}V,
\end{align}
where 
\begin{equation}
\boldsymbol{l}=\left[\begin{array}{c}
L_{2}\\
0\\
-L_{1}
\end{array}\right],\boldsymbol{E}_{0}=\left[\begin{array}{ccc}
0 & 0 & -1\\
0 & 0 & 0\\
1 & 0 & 0
\end{array}\right],\boldsymbol{E}=\left[\begin{array}{ccc}
0 & 0 & 1\\
0 & 0 & 0\\
-1 & 0 & 0
\end{array}\right]
\end{equation}
are constant quantities. Note that the only dependence of $T_{1}$
on $\theta$, though not explicitly written, consists in $\boldsymbol{F}^{(d'')}(\boldsymbol{x}^{(d'')},t)$,
tracing back to the conversion from $xyz$ to $x''y''z''$ for string
displacements. The total contribution of gravity to the torque of
shank with respect to the $y''$ axis is still $mgL_{0}\cos(\beta_{0}+\theta)$
if no relevant modifications are made when switching to the 3D hammer
model. The moment of inertia with respect to the $y''$ axis, constant
through time, is 
\begin{equation}
I=\int_{\Omega^{(d')}}\rho\left(x^{2}+z^{2}\right)\mathrm{d}V,
\end{equation}
where $\Omega^{(d')}\in\mathbb{R}^{3}$ is the shank's volume domain.
Applying Newton's second law for rotation, the differential equation
of hammer shank motion is 
\begin{equation}
I\ddot{\theta}=-\mu\dot{\theta}+mgL_{0}\cos(\beta_{0}+\theta)+T_{1}(\theta).\label{eq:hammer_shank_3D_eq}
\end{equation}

Now the whole model of 3D hammer-string coupling has been established.
The felt-string interaction force $\boldsymbol{F}_{i}^{(e)}$ in (\ref{eq:hammer_felt_3D_force_string})
is computed from the 3D hammer felt model with Dirichlet boundary
conditions (\ref{eq:hammer_felt_3D_boundary_string}) (\ref{eq:hammer_felt_3D_boundary_shank}),
and added to the rhs of (\ref{eq:pde}) as a non-conservative force.
Then, (\ref{eq:pde}) and (\ref{eq:hammer_shank_3D_eq}) forms 2 sets
of coupled equations for 2 sets of variables $(\boldsymbol{u}^{(a)},\theta)$,
which provides a foundation for obtaining a solution theoretically.
Despite the linearity of 3D elasticity model, the felt's force on
the string is still nonlinear because of the interaction judging condition.

\subsection{Model for string-soundboard coupling}

\begin{figure}
\begin{centering}
\includegraphics{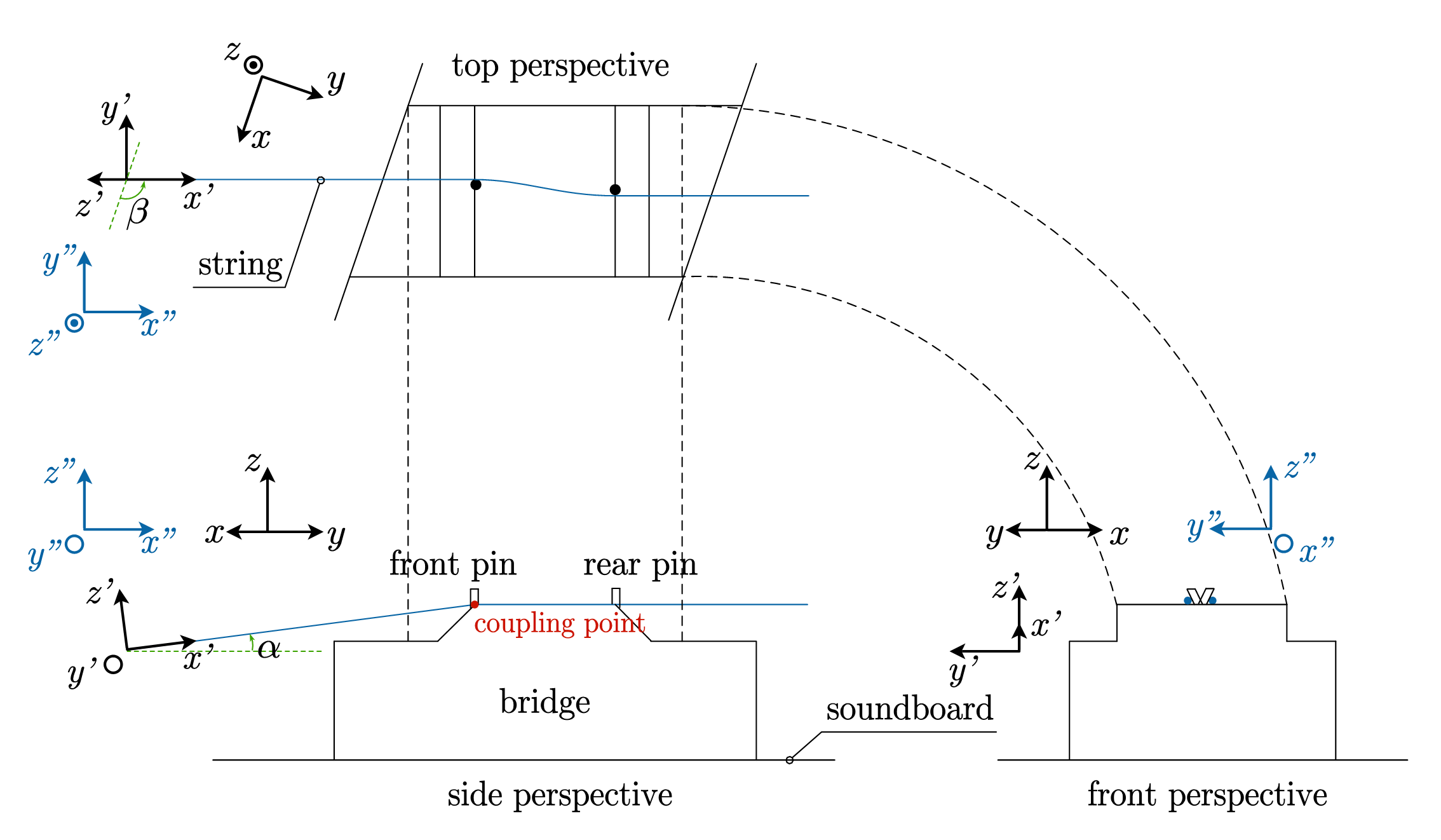}
\par\end{centering}
\caption{The piano bridge view from 3 directions}
\label{fig:bridge}
\end{figure}

Piano strings are coupled to the soundboard's bridge part, which terminates
and transmits string vibration to the soundboard. Figure \ref{fig:bridge}
visualizes this coupling at the bridge from 3 perspectives, where
$xyz$ and $x'y'z'$ are the coordinate systems for soundboard and
string respectively. Converting a vector (not a point) $\boldsymbol{v}$
in the $xyz$ system into $\boldsymbol{v}'$ in the $x'y'z'$ system,
according to the angles $\alpha$ and $\beta$ marked in figure \ref{fig:bridge},
yields $\boldsymbol{v}'=\boldsymbol{R}\boldsymbol{v}$ where $\boldsymbol{R}=\boldsymbol{R}_{1}\boldsymbol{R}_{2}$
and 
\begin{equation}
\boldsymbol{R}_{1}=\left[\begin{array}{ccc}
\cos\alpha & 0 & \sin\alpha\\
0 & 1 & 0\\
-\sin\alpha & 0 & \cos\alpha
\end{array}\right],\;\boldsymbol{R}_{2}=\left[\begin{array}{ccc}
\cos\beta & \sin\beta & 0\\
-\sin\beta & \cos\beta & 0\\
0 & 0 & 1
\end{array}\right]\label{eq:bridge_rotation_matrix}
\end{equation}
are the rotation matrices. The orthogonal property of rotation matrix
yields $\boldsymbol{v}=\boldsymbol{R}^{\top}\boldsymbol{v}'$.

We conjecture from observations that the string-soundboard coupling
is achieved via several mechanisms. The first mechanism is ``bridge
hump coupling'', as illustrated in the side perspective of figure
\ref{fig:bridge}. The middle part of bridge is constructed a bit
higher than the agraffe, the starting point of the speaking string.
As a result, the string experiences some upward pressure from the
bridge and the bridge experiences some downward pressure from the
string, both statically and dynamically. The second mechanism is ``bridge
pin horizontal coupling'', as illustrated in the top perspective
of figure \ref{fig:bridge}. Two bridge pins for one string are drilled
into the bridge at positions that would bend the string a bit horizontally,
bringing mutual horizontal tension between the string and bridge,
and restricting the string's horizontal movement. Made of hard metal
like steel, brass or even titanium, the bridge pins can be treated
as rigid bodies with zero strain and stress. The third mechanism is
``bridge pin vertical coupling'', as illustrated in the front perspective
of figure \ref{fig:bridge}. The two bridge pins lean towards their
respective string sides, blocking the string from moving upwards beyond
the pins. Also, the notches of bridge ensure that bridge hump coupling
and bridge pin coupling occur at almost the same position, viz. the
front bridge pin (the one closer to the agraffe). 

On the string's side, the coupling point is defined as $\boldsymbol{x}_{1}^{(a)}=[L_{1}^{(a)},0,-r^{(a)}]^{\top}$,
the lowest point of string at the front pin. While other relevant
points for coupling may be $[L_{1}^{(a)},-r^{(a)},0]^{\top}$, the
closest point of string to the bridge pin, the string's displacements
and surface forces at these points are identical, assuming $u^{(a)}$
does not vary with $y^{(a)}$ or $z^{(a)}$ at $x^{(a)}=L_{1}^{(a)}$
(no rotation) for simplicity\footnote{This is also true for the string's central point $\boldsymbol{x}_{1}^{(a)}=[L_{1}^{(a)},0,0]^{\top}$
previously defined.}. The coordinate of this coupling point on the soundboard's side is
denoted $\boldsymbol{x}_{1}^{(b)}$.

\subsubsection{Static coupling}

For coupling before motion is initiated, the string and soundboard
should have non-zero prestress fields that attain static balance.
It is sufficient to specify that the string prestress only acts perpendicular
to its cross-sections and is constant through its speaking length.
This means in the string's tension matrix $\boldsymbol{T}^{(a)}$,
$T_{11}$ is a constant and other entries are zero, immediately satisfying
the static balance. For the soundboard, a full space dependent tension
matrix $\boldsymbol{T}^{(b)}$ with 6 unique elements is necessary.
The static balance condition can then be expressed as 3 partial differential
equations 
\begin{equation}
\nabla\cdot\boldsymbol{T}^{(b)}=\boldsymbol{0},
\end{equation}
which can be solved numerically by FEM using nodal surface forces
as time-independent DOFs. Nonetheless, continuity between string and
soundboard tension fields is required. To write the continuity equations,
we define an intermediate coordinate system $x''y''z''$ resulting
from rotating the $x'y'z'$ system by $\boldsymbol{R}_{1}^{\top}$
or rotating the $xyz$ system by $\boldsymbol{R}_{2}$, as marked
blue in figure \ref{fig:bridge}. At the coupling point, the string's
tension on surfaces with outward normals in the $y''$ and $z''$
axes exert on the bridge, whereas that in the $x''$ axis exerts on
the farther hitch pins that seem to not connect the soundboard. This
leads to one Dirichlet boundary condition for each string coupled
to the soundboard as 
\begin{equation}
\left(\boldsymbol{R}_{2}\boldsymbol{T}^{(b)}(\boldsymbol{x}_{1}^{(b)})\boldsymbol{R}_{2}^{\top}\right)_{[22,33,12,13,23]}=\left(\boldsymbol{R}_{1}^{\top}\boldsymbol{T}^{(a)}(\boldsymbol{x}_{1}^{(a)})\boldsymbol{R}_{1}\right)_{[22,33,12,13,23]},\label{eq:string_soundboard_coupling_static}
\end{equation}
where subscript $_{[i_{1},i_{2},...]}$ means taking vector or matrix
elements at indices $i_{1},i_{2},...$ to form a vector. This condition
can be deduced into 5 equations, leaving 1 DOF for the coupling point. 

\subsubsection{Dynamic coupling}

For coupling in motion, the string transmits its surface forces (dynamic
prestress, stress and damping force) to the soundboard. At coupling
point, the string's total surface force, excluding the static prestress
which is already accounted for on the soundboard's side, is $\boldsymbol{G}^{(a)}(\boldsymbol{x}_{1},t)$.
Similar to static coupling, only forces on surfaces with outward normals
in the $y''$ and $z''$ axes are transmitted to the soundboard at
the coupling point. This leads to 
\begin{align}
 & \boldsymbol{F}^{(b)}(\boldsymbol{x}^{(b)},t)=\delta\left(\boldsymbol{x}^{(b)}-\boldsymbol{x}_{1}^{(b)}\right)\left(\boldsymbol{R}_{2}^{\top}\left[\begin{array}{ccc}
0 & Z_{12} & Z_{13}\\
Z_{12} & Z_{22} & Z_{23}\\
Z_{13} & Z_{23} & Z_{33}
\end{array}\right]\boldsymbol{R}_{2}\right)\left[\begin{array}{c}
1\\
1\\
1
\end{array}\right],\nonumber \\
 & \left\{ Z_{ij}\right\} =\boldsymbol{R}_{1}^{\top}\boldsymbol{G}^{(a)}(\boldsymbol{x}_{1}^{(a)},t)\boldsymbol{R}_{1},\label{eq:string_soundboard_coupling_force_dynamic}
\end{align}
where $\delta(\boldsymbol{x})$ is a 3D Dirac delta function indicating
a point load. The above equations treats the string's surface force
as a non-conservative force input to the soundboard system.

Another coupling condition arising from observation is that the string
and soundboard should have the same $y''$ and $z''$ direction displacements\footnote{Here displacement continuity should be equivalent to velocity continuity,
since both the string's and soundboard's displacements are based on
a $[0,0,0]^{\top}$ point.} at the coupling point, written as 
\begin{equation}
\left(\boldsymbol{R}_{1}^{\top}\partial_{t}\boldsymbol{u}^{(a)}(\boldsymbol{x}_{1}^{(a)},t)\right)_{[2,3]}=\left(\boldsymbol{R}_{2}\partial_{t}\boldsymbol{u}^{(b)}(\boldsymbol{x}_{1}^{(b)},t)\right)_{[2,3]}.\label{eq:string_soundboard_coupling_displacement_dynamic}
\end{equation}
The reason for this is that the the aforementioned three coupling
mechanisms constitute support for the string's surfaces with outward
normals in the $y''$ and $z''$ axes at the coupling point. This
may not hold for the $x''$ axis, but if it is needed\footnote{Maybe in case of high enough viscosity or friction, the string can
not slip more than the soundboard in the $x''$ direction. Restricting
this movement may also make the solution of string's PDE more stable.
This needs to be tested.}, the subscript $_{[2,3]}$ in above equations can be removed to impose
a stronger displacement continuity condition. Anyway, restriction
of the string's $x''$ axis (longitudinal) motion is always present
at the farther hitch point which may not connect the soundboard, see
(\ref{eq:string_boundary_1}). The same-displacement condition also
provides a basis for establishing the string's displacement at the
coupling point (Dirichlet boundary condition), so that the string's
PDE (\ref{eq:pde}) has an appropriate solution.

\subsubsection{More discussions on coupling}

String-soundboard coupling exhibits complex mechanisms that we find
challenging to discover and describe. Some of these mechanisms that
we observe but not covered in the aforementioned coupling model are
discussed in this subsection. These discussions may not cover much
details of computation due to their complexities.

Firstly, we only specified the coupling point at the bridge's front
pin but ignored the rear pin. As coupling at the rear pin position
potentially exists, the string in our model may need to be extented
in length to cover the segment between the front pin position $\boldsymbol{x}_{1}=[L_{1}^{(a)},0,-r^{(a)}]^{\top}$
and the rear pin position $\boldsymbol{x}_{2}=[L_{2}^{(a)},0,-r^{(a)}]^{\top}$
(we call it ``this segment'' is this paragraph). At this segment,
the horizontal bending of string may be accouted for in the prestress,
but not necessarily represented in the geometrical volume for simplicity.
In static state, the prestress field coupling should be computed for
this segment paying attention that only at the two pin points are
$y''$ axis surface forces coupled. In dynamic state, transmission
of the string's surface forces should be computed for this segment,
paying attention that only at the two pin points are $y''$ axis surface
forces transmitted. A maybe more accurate computation of surface force
transmission is, that at the two pins only when the string's $y''$
axis surface forces are towards the front or rear pins should they
be transmitted, and that at between (excluding) the two pins only
when the string's $z''$ axis surface forces are upwards should they
be transmitted. This may lead to a highly nonlinear function akin
to the case of hammer-string interaction (\ref{eq:hammer_compression}),
and may make imposing boundary conditions tougher due to the inequality
rather than equality nature. For displacement coupling at this segment,
at the two pins when the string's $y''$ direction displacements should
not exceed the pins, whereas the $z''$ direction displacements are
fully coupled to the pins; at between (excluding) the string's $z''$
direction displacements should not be under the soundboard plane,
whereas the $y''$ direction displacements are unrestricted.

Secondly, we considered coupling as occurring at a point, but it may
actually occur in a small contact surface (we call it ``this surface''
is this paragraph). In such a case, the Dirichlet boundary condition
of prestress field coupling should be specified over this surface;
the dynamic surface force transmission should be treated as surface
load rather than point load; the Dirichlet boundary condition of displacement
coupling should also be specified over this surface.

Thirdly, we still lack geometric details regarding the string's notable
vertical and horizontal bending at the bridge. This bending changes
the string's longitudinal direction, and thereby the 3 directions
of vibration. Though string vibration should be terminated at the
bridge pin, it seems only at the hitch pin that longitudinal vibration
is fully restricted. Therefore, the string's segment from front bridge
pin to hitch pin may require investigation, which may concern the
duplex scale phenomena \cite{miranda2024influence}. We can design
a multi-segment geometric model for the string, each segment having
differently rotated constitutive matrices. Also, the position-dependent
static tension can be specified parallel to the central line of each
string segment.

Finally, we ignored the soundboard's normal and tangent surface forces
exerting on the string at coupling point. This is similar to the case
of hammer-string coupling where the string's surface forces exerting
on the hammer are ignored. This is mainly for practical considerations,
as we would want to solve the soundboard's equations for only once.
Since we did not impose Dirichlet boundary condition for the soundboard
at each coupling point with the around 200 strings, Neumann boundary
conditions apply here restricting the soundboard's conservative surface
forces to be zero at coupling points. It may be acceptable given the
intuitive feeling that ``string → soundboard'' transmission should
dominate ``soundboard → string'' transmission. If the latter is
essential, some DOFs may be need to be removed from the coupling point
on the soundboard's side, which is a challenge for imposing Dirichlet
boundary condition reasonably for both the string and the soundboard.

\subsection{Model for soundboard-air and room-air coupling}

Modeling soundboard-air and room-air coupling is crucial for arriving
at the final digital audio sound to the listeners. Generally in the
context of solid-fluid coupling dynamics, the interaction surface
should have continuous normal and tangential velocities, as well as
continuous normal and tangential surface forces on both sides \cite{dunn2015springer}. 

As per section \ref{sec:Model-for-sound-radiation-in-the-air}, room-air
coupling occurs on surface $\Gamma_{1}^{(f)}$ (room side) and $\Gamma_{1}^{(c)}$
(air side). Define coordinate transformation $\boldsymbol{x}^{(f)}=\boldsymbol{r}_{1}+\boldsymbol{R}_{1}\boldsymbol{x}^{(c)}$
from air coordinates to soundboard coordinates, where $\boldsymbol{r}_{1}$
and $\boldsymbol{R}_{1}$ are the shift vector and rotation matrix.
We impose Dirichlet boundary condition of velocity continuity that
\begin{equation}
\boldsymbol{u}^{(c)}(\boldsymbol{x}^{(c)},t)=\boldsymbol{R}_{1}^{\top}\dot{\boldsymbol{u}}^{(f)}(\boldsymbol{x}^{(f)},t),\;\boldsymbol{x}^{(c)}\in\Gamma_{1}^{(c)},\;\boldsymbol{x}^{(f)}=\boldsymbol{r}_{1}+\boldsymbol{R}_{1}\boldsymbol{x}^{(c)}.\label{eq:room_air_coupling_boundary}
\end{equation}
We also specify the contribution of air surface force to the room
equation's source term on the rhs of (\ref{eq:pde}) as 
\begin{align}
 & \boldsymbol{F}_{1}^{(f)}(\boldsymbol{x}^{(f)},t)=-\boldsymbol{G}^{(c)}(\boldsymbol{x}^{(c)},t)\boldsymbol{n}^{(c)}(\boldsymbol{x}^{(c)}),\;\boldsymbol{x}^{(f)}\in\Gamma_{1}^{(f)},\;\boldsymbol{x}^{(c)}=-\boldsymbol{R}_{1}^{\top}\boldsymbol{r}_{1}+\boldsymbol{R}_{1}^{\top}\boldsymbol{x}^{(f)},\label{eq:room_air_coupling_force}
\end{align}
where $\boldsymbol{n}^{(c)}(\boldsymbol{x}^{(c)})$ is the outward
normal of the tangent plane of $\boldsymbol{x}^{(c)}$.

As per sections \ref{sec:Model-for-piano-soundboard} and \ref{sec:Model-for-sound-radiation-in-the-air},
soundboard-air coupling occurs on surface $\Gamma_{2}^{(b)}$ (soundboard
side) and $\Gamma_{2}^{(c)}$ (air side). Define coordinate transformation
$\boldsymbol{x}^{(b)}=\boldsymbol{r}_{2}+\boldsymbol{R}_{2}\boldsymbol{x}^{(c)}$
from air coordinates to soundboard coordinates, where $\boldsymbol{r}_{2}$
and $\boldsymbol{R}_{2}$ are the shift vector and rotation matrix.
We impose Dirichlet boundary condition of velocity continuity that
\begin{equation}
\boldsymbol{u}^{(c)}(\boldsymbol{x}^{(c)},t)=\boldsymbol{R}_{2}^{\top}\dot{\boldsymbol{u}}^{(b)}(\boldsymbol{x}^{(b)},t),\;\boldsymbol{x}^{(c)}\in\Gamma_{2}^{(c)},\;\boldsymbol{x}^{(b)}=\boldsymbol{r}_{2}+\boldsymbol{R}_{2}\boldsymbol{x}^{(c)}.\label{eq:soundboard_air_coupling_boundary}
\end{equation}
We also specify the contribution of air surface force to the soundboard
equation's source term on the rhs of (\ref{eq:pde}) as 
\begin{align}
 & \boldsymbol{F}_{1}^{(b)}(\boldsymbol{x}^{(b)},t)=-\boldsymbol{G}^{(c)}(\boldsymbol{x}^{(c)},t)\boldsymbol{n}^{(c)}(\boldsymbol{x}^{(c)}),\;\boldsymbol{x}^{(b)}\in\Gamma_{2}^{(b)},\;\boldsymbol{x}^{(c)}=-\boldsymbol{R}_{2}^{\top}\boldsymbol{r}_{2}+\boldsymbol{R}_{2}^{\top}\boldsymbol{x}^{(b)},\label{eq:soundboard_air_coupling_force}
\end{align}
where $\boldsymbol{n}^{(c)}(\boldsymbol{x}^{(c)})$ is the outward
normal of the tangent plane of $\boldsymbol{x}^{(c)}$.

\section{Numeric schemes}

\subsection{Modal superposition method for solving coupled ODEs}\label{subsec:Modal-superposition-method}

\subsubsection{Modal transformation of second-order ODEs}

For solving coupled second-order ODEs like (\ref{eq:ODEs}), time
discretization using finite-difference is a direct approach. However,
given that damping matrix is diagonalizable by the mass and stiffness
matrices, decoupling and dimension reduction of the system by means
of modal superposition is preferred. To do so, we first define the
following eigenvalue problem 
\begin{equation}
\boldsymbol{K}\boldsymbol{\phi}_{i}=\lambda_{i}\boldsymbol{M}\boldsymbol{\phi}_{i}\Longleftrightarrow\boldsymbol{M}^{-1}\boldsymbol{K}\boldsymbol{\phi}_{i}=\lambda_{i}\boldsymbol{\phi}_{i}\;(i=1,...,N),\label{eq:second_ODEs_eigenvalue_problem}
\end{equation}
and the eigen decomposition 
\begin{equation}
\boldsymbol{K}=\boldsymbol{M}\boldsymbol{\Phi}\boldsymbol{\Lambda}\boldsymbol{\Phi}^{-1}\Longleftrightarrow\boldsymbol{M}^{-1}\boldsymbol{K}=\boldsymbol{\Phi}\boldsymbol{\Lambda}\boldsymbol{\Phi}^{-1},\label{eq:second_ODEs_eigen_decomposition}
\end{equation}
where $\lambda_{i}$ is a (maybe complex) eigenvalue, $\boldsymbol{\phi}_{i}$
is a $N\times1$ (maybe complex) eigenvector, $\boldsymbol{\Lambda}=\mathrm{diag}(\lambda_{1},...,\lambda_{N})$
is a diagonal matrix of eigenvalues, $\boldsymbol{\Phi}=[\boldsymbol{\phi}_{1},...,\boldsymbol{\phi}_{N}]$
is a $N\times N$ matrix of eigenvectors. Note that since both $\boldsymbol{M}$
and $\boldsymbol{K}$ are sparse matrices with large dimensions, it
is preferrable to solve the generalized eigenvalue problem on the
left side of (\ref{eq:second_ODEs_eigenvalue_problem}), avoiding
explicit inversion of $\boldsymbol{M}$ that would otherwise result
in a large dense matrix. We can leverage existing softwares of sparse
eigensolvers like Arpack and FEAST to solve the generalized eigenvalue
problem. Nevertheless, for solving ODEs of the acoustic system with
large DOFs, it may be unwise to store all rows of eigenvectors at
the same time which may otherwise lead to memory overload. A re-implementation
of existing sparse eigensolver algorithms may be desired so as to
store partial rows or columns of eigenvectors in a ``rolling'' way.

Since eigenvectors are linearly independent, that is to say $\boldsymbol{\Phi}$
is invertible, we can write the solution in the form $\boldsymbol{\xi}(t)=\boldsymbol{\Phi}\boldsymbol{q}(t)$
where $\boldsymbol{q}(t)$ is an $N\times1$ vector we call as modal
DOFs. Note here $\boldsymbol{\xi}(t)$ should be real (in the complex
domain), but $\boldsymbol{q}(t)$ may be complex because $\boldsymbol{\Phi}^{-1}$
may be complex. Then (\ref{eq:ODEs}) can be decoupled as 
\begin{subequations}
\begin{align}
\boldsymbol{M}\boldsymbol{\Phi}\ddot{\boldsymbol{q}}(t)+2\mu\boldsymbol{K}\boldsymbol{\Phi}\dot{\boldsymbol{q}}(t)+\boldsymbol{K}\boldsymbol{\Phi}\boldsymbol{q}(t) & =\boldsymbol{f}(t)\\
\boldsymbol{\Phi}^{\mathrm{H}}\boldsymbol{M}\boldsymbol{\Phi}\ddot{\boldsymbol{q}}(t)+2\mu\boldsymbol{\Phi}^{\mathrm{H}}\boldsymbol{K}\boldsymbol{\Phi}\dot{\boldsymbol{q}}(t)+\boldsymbol{\Phi}^{\mathrm{H}}\boldsymbol{K}\boldsymbol{\Phi}\boldsymbol{q}(t) & =\boldsymbol{\Phi}^{\mathrm{H}}\boldsymbol{f}(t)\\
\bar{\boldsymbol{M}}\ddot{\boldsymbol{q}}(t)+2\mu\bar{\boldsymbol{M}}\boldsymbol{\Lambda}\dot{\boldsymbol{q}}(t)+\bar{\boldsymbol{M}}\boldsymbol{\Lambda}\boldsymbol{q}(t) & =\boldsymbol{\Phi}^{\mathrm{H}}\boldsymbol{f}(t)\label{eq:second_ODEs_decoupled_1}\\
\ddot{\boldsymbol{q}}(t)+2\mu\boldsymbol{\Lambda}\dot{\boldsymbol{q}}(t)+\boldsymbol{\Lambda}\boldsymbol{q}(t) & =\boldsymbol{p}(t)\label{eq:second_ODEs_decoupled_2}
\end{align}
\end{subequations}
 where 
\begin{equation}
\bar{\boldsymbol{M}}=\boldsymbol{\Phi}^{\mathrm{H}}\boldsymbol{M}\boldsymbol{\Phi},\;\boldsymbol{p}(t)=\bar{\boldsymbol{M}}^{-1}\boldsymbol{\Phi}^{\mathrm{H}}\boldsymbol{f}(t);\label{eq:second_ODEs_modal_mass_force_1}
\end{equation}
are called the modal mass matrix and the modal force vector respectively.
Through eigen decomposition, the $N$ coupled ODEs have now been transformed
into $N$ uncoupled ODEs. Note that in case the damping matrix can
not be diagonalized by the mass and stiffness matrices, we should
rewrite the second-order ODEs as first-order ODEs with doubled number
of DOFs, so that it can be decoupled as will be shown in section \ref{subsec:Modal-transformation-of-first}.

It is well-understood that the analytical solution of the $i$ th
uncoupled second-order ODE without the rhs nonhomogenous term is a
sinusoidal signal with a single eigenfrequency positively correlated
to the magnitude of $i$ th eigenvalue. Assuming eigenvalues in $\boldsymbol{\Lambda}$
are ascendingly sorted by their real magitudes, we can select only
the lowest $M$ eigenvalues and discard the rest, because most human
ears are insensitive to eigenfrequencies above a certain threshold
(often 10 kHz). This is also for practical considerations that the
number of DOFs is often too large for numeric computation, making
dimension reduction desirable. Consequently, we obtain a $M\times1$
vector $\boldsymbol{q}'(t)$ with only the first $M$ entries of $\boldsymbol{q}(t)$,
and the $M+1$ to $N$ entries are all approximated as zeros and discarded.
Subsituting $\boldsymbol{q}(t)$ by $\boldsymbol{q}'(t)$ (more exactly,
$\boldsymbol{q}'(t)$ should be padded $N-M$ zeros at the end) in
(\ref{eq:second_ODEs_decoupled_1}), and omitting the last $N-M$
equations and unknowns, yields 
\begin{subequations}
\begin{align}
\boldsymbol{S}'\ddot{\boldsymbol{q}}'(t)+2\mu\boldsymbol{S}'\boldsymbol{\Lambda}'\dot{\boldsymbol{q}}'(t)+\boldsymbol{S}'\boldsymbol{\Lambda}'\boldsymbol{q}'(t) & =\boldsymbol{\Phi}'{}^{\top}\boldsymbol{f}(t)\label{eq:second_ODEs_decoupled_3}\\
\ddot{\boldsymbol{q}}'(t)+2\mu\boldsymbol{\Lambda}'\dot{\boldsymbol{q}}'(t)+\boldsymbol{\Lambda}'\boldsymbol{q}'(t) & =\boldsymbol{p}'(t)\label{eq:second_ODEs_decoupled_4}
\end{align}
\end{subequations}
 where $\boldsymbol{\Lambda}'$ has dimension $M\times M$ containing
the first $M$ eigenvalues and $\boldsymbol{\Phi}'$ has dimension
$N\times M$ containing the first $M$ eigenvectors. The large numbe
of DOFs is now approximated as the linear combination of a smaller
number of modal DOFs as $\boldsymbol{\xi}(t)\approx\boldsymbol{\Phi}'\boldsymbol{q}'(t)$.
The reduced modal mass and modal force are 
\begin{equation}
\bar{\boldsymbol{M}}'=\boldsymbol{\Phi}'{}^{\mathrm{H}}\boldsymbol{M}\boldsymbol{\Phi}',\;\boldsymbol{p}(t)=\bar{\boldsymbol{M}}'{}^{-1}\boldsymbol{\Phi}'{}^{\mathrm{H}}\boldsymbol{f}(t).\label{eq:second_ODEs_modal_mass_force_2}
\end{equation}
For around 2500 modes of the soundboard system as estimated in \cite{chabassier2013modeling},
the fully dense modal mass matrix (128 bit complex type) would require
about 95 MB memory which is normally acceptable in both storage and
computation aspects. But this dense matrix can be avoided if the mass
and stiffness matrices are both symmetric because in this case the
eigenvectors can be normalized so that the modal mass matrix equals
to an identity matrix, making its inversion much easier. This benefit
is achievable in our 3D elastic solid model because the constitutive
matrix is symmetric even in the presence of prestres. A symmetric
stiffness matrix often means that energy is conserved within the system
if there is no other non-conservative forces like damping force.

Now we consider the analytical solution of the $i$ th uncoupled nonhomogenous
ODE 
\begin{equation}
\ddot{q}(t)+2\mu\lambda\dot{q}(t)+\lambda q(t)=p(t),\label{eq:second_ODE}
\end{equation}
where the subscripts $_{i}$ and superscripts $'$ are dropped for
convenience. Applying Laplace transform to this equation yields 
\begin{align}
s^{2}Q(s)+2\mu\lambda sQ(s)+\lambda Q(s) & =P(s)+Q_{0}(s),\nonumber \\
s^{2}G(s)+2\mu\lambda sG(s)+\lambda G(s)-G_{0}(s) & =1\nonumber \\
\ddot{g}(t)+2\mu\lambda\dot{g}(t)+\lambda g(t) & =\delta(t)\label{eq:second_ODE_Green_function}
\end{align}
where $s$ is a complex variable, $\delta(t)$ is the Dirac delta
function; the initial parts of Laplace transforming derivatives are
defined as 
\begin{align}
Q_{0}(s) & =sq(0)+\dot{q}(0)+2\mu\lambda q(0)\nonumber \\
G_{0}(s) & =sg(0)+\dot{g}(0)+2\mu\lambda g(0);\label{eq:second_ODE_Laplace_initial_term}
\end{align}
the well-known Green's function (frequency domain), the response to
a unit impulse, is defined as 
\begin{equation}
G(s)=\frac{Q(s)\left[1+G_{0}(s)\right]}{P(s)+Q_{0}(s)}\Rightarrow Q(s)=\frac{G(s)\left[P(s)+Q_{0}(s)\right]}{1+G_{0}(s)}.\label{eq:second_ODE_Laplace_map}
\end{equation}
To solve $g(t)$, we first specify that response should not exist
during zero and negative time, viz. $g(t)=\dot{g}(t)=0$ for $t\leq0$;
then notice when $t>0$, (\ref{eq:second_ODE_Green_function}) becomes
homogenous with solution 
\[
g(t)=C_{1}\exp(z_{1}t)+C_{2}\exp(z_{2}t),\;z_{1},z_{2}=-\mu\lambda\pm\sqrt{(\mu^{2}\lambda-1)\lambda}.
\]
To determine complex constants $C_{1}$, $C_{2}$, we first notice
that $g(t)$ should be continuous at $t=0$ in the presence of second
derivative in (\ref{eq:second_ODE_Green_function}), thus 
\begin{equation}
g(0^{+})=g(0)\Rightarrow C_{1}+C_{2}=0.\label{eq:second_ODE_Green_initial_1}
\end{equation}
Then, integrating (\ref{eq:second_ODE_Green_function}) over $(-\infty,+\infty)$
yields 
\begin{align}
 & \int_{-\infty}^{+\infty}\left[\ddot{g}(t)+2\mu\lambda\dot{g}(t)+\lambda g(t)\right]\mathrm{d}t=\int_{-\infty}^{+\infty}\delta(t)\mathrm{d}t\nonumber \\
\Rightarrow & \int_{0^{-}}^{0^{+}}\ddot{g}(t)\mathrm{d}t=\dot{g}(0^{+})-\dot{g}(0^{-})=\dot{g}(0^{+})=1\nonumber \\
\Rightarrow & C_{1}z_{1}+C_{2}z_{2}=1,\label{eq:second_ODE_Green_initial_2}
\end{align}
which is the so-called jump discontinuity condition for first derivative;
the continuity condition for the zero-order has been applied again
here. From (\ref{eq:second_ODE_Green_initial_1}) (\ref{eq:second_ODE_Green_initial_2})
the complex constants are found to be 
\begin{equation}
C_{1}=\frac{1}{z_{1}-z_{2}},\;C_{2}=\frac{1}{z_{2}-z_{1}}.
\end{equation}
Subsituting $g(0)=0$ and $\dot{g}(0)=0$ into (\ref{eq:second_ODE_Laplace_initial_term})
yields $G_{0}(s)=0$. Then from (\ref{eq:second_ODE_Laplace_map})
we have 
\begin{align}
Q(s) & =P(s)G(s)+\left[\dot{q}(0)+2\mu\lambda q(0)\right]G(s)+sq(0)G(s)\nonumber \\
q(t) & =p(t)*g(t)+\left[\dot{q}(0)+2\mu\lambda q(0)\right]g(t)+q(0)\dot{g}(t),
\end{align}
where $*$ is the convolution operator over $[0,+\infty)$. For the
general case of initial conditions $q(0)=\dot{q}(0)=0$, the solution
reduces to $q(t)=p(t)*g(t)$. It is now clear that the nonhomogenous
ODEs can be solved by convolving the modal force $p(t)$ (source signal)
with the Green's function $g(t)$ (response signal), which can be
efficiently computed using fast Fourier transform (FFT) convolution
in the frequency domain. However, the analytical solution may not
be actually useful or efficient in the presence of coupling between
systems and the explicit time discretization scheme introduced in
section \ref{subsec:Explicit-time-discretization} will be an alternative.
Nevertheless, it provides an understanding of the characteristics
of vibration modes.

\subsubsection{Modal transformation of first-order ODEs}\label{subsec:Modal-transformation-of-first}

Solving coupled first-order ODEs like (\ref{eq:air_ODEs}) by means
of modal superposition is similar to the second-order case, and we
only discuss some particularities here. Firstly, the eigenvalue problem
is defined in the same way for $\boldsymbol{M}$ (now for first-order)
and $\boldsymbol{K}$. The decoupled ODEs write 
\begin{subequations}
\begin{align}
\bar{\boldsymbol{M}}\dot{\boldsymbol{q}}(t)+\bar{\boldsymbol{M}}\boldsymbol{\Lambda}\boldsymbol{q}(t) & =\boldsymbol{\Phi}^{\mathrm{H}}\boldsymbol{f}(t)\label{eq:first_ODEs_decoupled_1}\\
\dot{\boldsymbol{q}}(t)+\boldsymbol{\Lambda}\boldsymbol{q}(t) & =\boldsymbol{p}(t)\label{eq:first_ODEs_decoupled_2}
\end{align}
\end{subequations}
 where 
\begin{equation}
\bar{\boldsymbol{M}}=\boldsymbol{\Phi}^{\mathrm{H}}\boldsymbol{M}\boldsymbol{\Phi},\;\boldsymbol{p}(t)=\bar{\boldsymbol{M}}^{-1}\boldsymbol{\Phi}^{\mathrm{H}}\boldsymbol{f}(t)\label{eq:first_ODEs_modal_mass_force_1}
\end{equation}
are the modal mass and the modal force. Specific attension should
be paid to the analytical solution of the $i$ th uncoupled nonhomogenous
ODE 
\begin{equation}
\dot{q}(t)+\lambda q(t)=p(t).\label{eq:first_ODE}
\end{equation}
Laplace transform yields 
\begin{align}
sQ(s)+\lambda Q(s) & =P(s)+q(0),\nonumber \\
sG(s)+\lambda G(s)-g(0) & =1\nonumber \\
\dot{g}(t)+\lambda g(t) & =\delta(t)\label{eq:first_ODE_Green_function}
\end{align}
where the Green's function is defined as 
\begin{equation}
G(s)=\frac{Q(s)\left[1+g(0)\right]}{P(s)+q(0)}\Rightarrow Q(s)=\frac{G(s)\left[P(s)+q(0)\right]}{1+g(0)}.
\end{equation}
Note that $g(t)=\dot{g}(t)=0$ for $t\leq0$ is still required, but
continuity $g(0)=g(0^{+})$ is unnecessary because the highest order
of derivative in (\ref{eq:first_ODE_Green_function}) is only one.
Nevertheless, continuity of $\int g(t)\mathrm{d}t$ at $t=0$ can
be easily satisfied because the primitive function can have an arbitrary
constant added. Then integrating (\ref{eq:first_ODE_Green_function}),
we can find $g(0^{+})=1$ and the solution of Green's function 
\begin{equation}
g(t)=-\frac{1}{\lambda}\exp(-\lambda t).
\end{equation}
It follows that 
\begin{align}
Q(s) & =P(s)G(s)+q(0)G(s)\nonumber \\
q(t) & =p(t)*g(t)+q(0)g(t),
\end{align}
is the solution of the uncoupled nonhomogenous first-order ODE.

\subsection{Explicit time discretization for coupling between systems}\label{subsec:Explicit-time-discretization}

Having discussed the weak form ODEs for each subsystem of the physical
piano, our question is then how a numerical treatment of coupling
between system may be achieved that attains a sensible tradeoff between
feasibility and accuracy. For most subsystems presented in the previous
sections, there exists an external force term on the rhs of PDEs.
This, along with the predetermined displacements on the boundary,
form the main contributions to the rhs source term of system ODEs. 

Nevertheless, to say ``ODEs'' here is actually indefensible, because
in many coupling cases discussed before the rhs source term $\boldsymbol{f}(t)$
depends linearly or nonlinearly on the lhs unknown DOFs $\boldsymbol{\xi}(t)$
too. In such cases, eigen decomposition of ``pseudo ODEs'' can only
obtain decoupled lhs operators on $\boldsymbol{q}(t)$ on the lhs,
but rhs operators on $\boldsymbol{q}(t)$ often remain coupled, not
only because of the nonlinearity of operators, but also because eigen
decomposition applies to the local subsystem but not the global system
where rhs sources come from. Hammer-string coupling is a typical example:
the string displacement depends on the felt's force, the felt's force
depends on its compression, but this compression depends on string
displacement. Another example is in soundboard-string coupling, the
soundboard displacement at the coupling point depends on the string's
surface force, the string's surface force depends on its displacement,
computing its displacement requires knowing the coupling point displacement
(Dirichlet boundary), but this displacement depends on the soundboard
displacement at the coupling point to satisfy the continuity condition. 

An explanation for these seemingly odd relations is that a theoretically
sound way would seem to do computations as if all subsystems were
a single system, rather than to separate systems and reintroduce coupling
between them. This means all DOFs are integrated into a single vector
of DOFs, all mass (or stiffness) matrices are assembled into a bigger
one, displacements at coupling positions are expressed by the same
unique DOFs behind, in order for full coupling solutions. However,
this is often unacceptable not only because of the high costs of storage
and computation, but also because the inherent heterogeneities of
different subsystems, particularly the nonlinearity of hammer felt
compression, the first-order characteristic of acoustic system, the
different damping mechanisms, may actually not cohere well and may
even be tough to control in a single system. Therefore, we shall still
adopt the framework of separate systems with mutual coupling, and
seek for numeric schemes capable of achieving a level of accuracy
as close to full coupling schemes as possible.

The time domain scheme we shall introduce here is inspired by the
idea of velocity Verlet algorithms, but customized for our case of
modal-transformed first and second order ODEs, with second-order accuracy\footnote{Alternatively, one may consider transforming time domain into frequency
domain to explore coupling from the perspective of mobility \cite{valiente2022modeling}.}. Despite being lower-order compared to the complex time schemes in
\cite{chabassier2013introduction,castera2023numerical}, our scheme
may be more efficient particularly in eliminating the need to invert
or solve big or dense matrices at each step, even when the mass matrix
is non-diagonal. Due to the existence of coupling, the analytical
solutions of first and second order ODEs discussed in section \ref{subsec:Modal-superposition-method}
may be not applicable here. Nevertheless, system decoupling and dimension
reduction via modal decomposition will be shown useful in improving
the efficiency of time-stepping algorithms. 

\subsubsection{Time stepping of second-order ODEs}

In the previously derived second-order decoupled ODEs (\ref{eq:second_ODEs_decoupled_2}),
the rhs source term can rewritten as 
\begin{equation}
\ddot{\boldsymbol{q}}(t)+2\mu\boldsymbol{\Lambda}\dot{\boldsymbol{q}}(t)+\boldsymbol{\Lambda}\boldsymbol{q}(t)=\boldsymbol{p}(t),\;\boldsymbol{p}(t)=\boldsymbol{r}\left(t,\boldsymbol{q}(t),\dot{\boldsymbol{q}}(t)\right),\label{eq:second_ODEs_rhs}
\end{equation}
where $\boldsymbol{r}$ incoporate all DOF-independent and DOF-dependent
contributions to the source term; DOF-dependent contributions consist
of two major sources: non-zero Dirichlet boundary conditions; non-conservative
force (but excluding damping force already on the lhs). 

To perform time discretization, we define the discrete time interval
as $h=\Delta t$, which can be chosen as 1/44100 seconds for producing
44.1 kHz digital audio; define the $n$ th discrete point in time
as $t_{n}=nh$. The notion of discrete points in time form the basis
of time stepping algorithms, where integrations are performed between
time steps. An integral over a small interval $[x_{0},x_{0}+\Delta x]$
can be approximated as area of trapeziud 
\begin{equation}
\int_{x_{0}}^{x_{0}+h}f(x)\mathrm{d}x\approx\frac{h}{2}\left[f(x_{0})+f(x_{0}+h)\right],\label{eq:integral_trapezoid}
\end{equation}
which can achieve a relatively high accuracy as per the mean value
theorem for definite integrals, though $f(x_{0}+h)$ may not be known
beforehand in some contexts. If we seek for a first-order approximation
of $f(x_{0}+h)$, then 
\begin{equation}
\int_{x_{0}}^{x_{0}+h}f(x)\mathrm{d}x\approx hf(x_{0})+\frac{h^{2}}{2}f'(x_{0})\label{eq:integral_taylor}
\end{equation}
is a less accurate approximation. Utilizing these integration strategies,
we integrate (\ref{eq:second_ODEs_rhs}) over $[t_{n},t_{n+1}]$ to
get 
\begin{align}
\dot{\boldsymbol{q}}(t_{n+1})-\dot{\boldsymbol{q}}(t_{n}) & =-2\mu\boldsymbol{\Lambda}\left[\boldsymbol{q}(t_{n+1})-\boldsymbol{q}(t_{n})\right]-s\boldsymbol{\Lambda}\int_{t_{n}}^{t_{n+1}}\boldsymbol{q}(t)\mathrm{d}t+\int_{t_{n}}^{t_{n+1}}\boldsymbol{p}(t)\mathrm{d}t\nonumber \\
\dot{\boldsymbol{q}}(t_{n+1})-\dot{\boldsymbol{q}}(t_{n}) & \approx-2\mu\boldsymbol{\Lambda}\left[\boldsymbol{q}(t_{n+1})-\boldsymbol{q}(t_{n})\right]-\frac{h}{2}\boldsymbol{\Lambda}\left[\boldsymbol{q}(t_{n})+\boldsymbol{q}(t_{n+1})\right]+h\boldsymbol{p}(t_{n})+\frac{h^{2}}{2}\dot{\boldsymbol{p}}(t_{n}).\label{eq:second_ODEs_integral_eq_1}
\end{align}
Also it is obvious that 
\begin{equation}
\boldsymbol{q}(t_{n+1})-\boldsymbol{q}(t_{n})=\int_{t_{n}}^{t_{n+1}}\dot{\boldsymbol{q}}(t)\mathrm{d}t\approx\frac{h}{2}\left[\dot{\boldsymbol{q}}(t_{n})+\dot{\boldsymbol{q}}(t_{n+1})\right].\label{eq:second_ODEs_integral_eq_2}
\end{equation}
The solution of (\ref{eq:second_ODEs_integral_eq_1}) (\ref{eq:second_ODEs_integral_eq_2})
is 
\begin{align}
\boldsymbol{q}(t_{n+1}) & =\boldsymbol{Z}_{1}^{-1}\left[\boldsymbol{Z}_{0}\boldsymbol{q}(t_{n})+2\dot{\boldsymbol{q}}(t_{n})+h\boldsymbol{p}(t_{n})+\frac{h^{2}}{2}\dot{\boldsymbol{p}}(t_{n})\right],\nonumber \\
\dot{\boldsymbol{q}}(t_{n+1}) & =\frac{2}{h}\boldsymbol{q}(t_{n+1})-\frac{2}{h}\boldsymbol{q}(t_{n})-\dot{\boldsymbol{q}}(t_{n}),\label{eq:second_ODEs_integral_next}
\end{align}
where 
\begin{align}
\boldsymbol{Z}_{0} & =\frac{2}{h}\boldsymbol{I}+\left(2\mu-\frac{h}{2}\right)\boldsymbol{\Lambda},\nonumber \\
\boldsymbol{Z}_{1} & =\frac{2}{h}\boldsymbol{I}+\left(2\mu+\frac{h}{2}\right)\boldsymbol{\Lambda}.
\end{align}
are diagonal matrices easy to invert. (\ref{eq:second_ODEs_integral_next})
reveal that $\boldsymbol{q}(t_{n+1})$, $\dot{\boldsymbol{q}}(t_{n+1})$
can be computed with second-order accuracy using $\boldsymbol{q}(t_{n})$,
$\dot{\boldsymbol{q}}(t_{n})$, $\boldsymbol{p}(t_{n})$, $\dot{\boldsymbol{p}}(t_{n})$,
but without using unknowns of the $t_{n+1}$ step. This means our
time discretization scheme is explicit. As for its energy property,
readers can go to appendix \ref{sec:Nonlinear-strain-energy} for
a detailed deduction incorporating nonlinear strain energy. From the
deduction there, our preliminary judgement is that this explicit scheme
is energy stable at least for the kinetic energy, potential energy
and damping, but uncertain for other non-conservative forces on the
rhs. The most significant accuracy loss of this scheme seems to be
the use of (\ref{eq:integral_taylor}) rather than (\ref{eq:integral_trapezoid})
in approximating the integral of $\boldsymbol{p}(t)$, which may also
lead to energy unstability. We would expect the first-order approximation
of $\boldsymbol{p}(t_{n+1})$ to be acceptable with small $h$. This
actually implies that we rely on some historical ($t_{n}$) displacement
and velocity values to predict the current ($t_{n+1}$) input sources,
then the current input sources are updated by the computed current
displacement and velocity values. This process can be repeated several
times for one time step if higher accuracy and better energy stability
is desired, able to use (\ref{eq:integral_trapezoid}) rather than
(\ref{eq:integral_taylor}) in approximating the integral of $\boldsymbol{p}(t)$.

\subsubsection{Time stepping of first-order ODEs}

Recall the previously derived first-order decoupled ODEs (\ref{eq:first_ODEs_decoupled_2}),
the rhs source term can rewritten as 
\begin{equation}
\dot{\boldsymbol{q}}(t)+\boldsymbol{\Lambda}\boldsymbol{q}(t)=\boldsymbol{p}(t),\;\boldsymbol{p}(t)=\boldsymbol{r}\left(t,\boldsymbol{q}(t)\right),\label{eq:first_ODEs_rhs}
\end{equation}
where $\boldsymbol{r}$ incoporate all DOF-independent and DOF-dependent
contributions to the source term; DOF-dependent contributions consist
of two major sources: non-zero Dirichlet boundary conditions; non-conservative
force (but excluding damping force already on the lhs). Integrating
(\ref{eq:first_ODEs_rhs}) over $[t_{n},t_{n+1}]$, the approximate
solution of $\boldsymbol{q}(t_{n+1})$ can be found as 
\begin{align}
\boldsymbol{q}(t_{n+1})-\boldsymbol{q}(t_{n}) & =-\boldsymbol{\Lambda}\int_{t_{n}}^{t_{n+1}}\boldsymbol{q}(t)\mathrm{d}t+\int_{t_{n}}^{t_{n+1}}\boldsymbol{p}(t)\mathrm{d}t\nonumber \\
\boldsymbol{q}(t_{n+1})-\boldsymbol{q}(t_{n}) & \approx-\frac{h}{2}\boldsymbol{\Lambda}\left[\boldsymbol{q}(t_{n})+\boldsymbol{q}(t_{n+1})\right]+h\boldsymbol{p}(t_{n})+\frac{h^{2}}{2}\dot{\boldsymbol{p}}(t_{n})\nonumber \\
\boldsymbol{q}(t_{n+1}) & =\left(\boldsymbol{I}+\frac{h}{2}\boldsymbol{\Lambda}\right)^{-1}\left[\left(\boldsymbol{I}-\frac{h}{2}\boldsymbol{\Lambda}\right)\boldsymbol{q}(t_{n})+h\boldsymbol{p}(t_{n})+\frac{h^{2}}{2}\dot{\boldsymbol{p}}(t_{n})\right].\label{eq:first_ODEs_next}
\end{align}
Therefore, $\boldsymbol{q}(t_{n+1})$ can be computed using $\boldsymbol{q}(t_{n})$,
$\boldsymbol{p}(t_{n})$, $\dot{\boldsymbol{p}}(t_{n})$, but without
using unknowns of the $t_{n+1}$ step.

\subsubsection{Time stepping of hammer shank rotation ODE}

The equation of hammer shank motion (\ref{eq:hammer_shank_3D_eq})
can be written in a more abstract way as 
\begin{equation}
I\ddot{\theta}=-\mu\dot{\theta}+T(\theta),\label{eq:hammer_ODE_rhs}
\end{equation}
where $T(\theta)$ incoporate the contributions of gravity and felt
force to the total torque of shank, dependent on the zero-order value
of $\theta$. Similar to second-order system ODEs, the time discretization
of (\ref{eq:hammer_ODE_rhs}) is found to be 
\begin{align}
 & I\left[\dot{\theta}(t_{n+1})-\dot{\theta}(t_{n})\right]=-\mu\left[\theta(t_{n+1})-\theta(t_{n})\right]+\int_{t_{n}}^{t_{n+1}}T(\theta)\mathrm{d}t\nonumber \\
 & I\left[\frac{2}{h}\theta(t_{n+1})-\frac{2}{h}\theta(t_{n})-2\dot{\theta}(t_{n})\right]\approx-\mu\left[\theta(t_{n+1})-\theta(t_{n})\right]+hT(t_{n})+\frac{h^{2}}{2}\dot{T}(t_{n})\nonumber \\
 & \theta(t_{n+1})=\left(\frac{2}{h}I+\mu\right)^{-1}\left[\left(\frac{2}{h}I+\mu\right)\theta(t_{n})+2\dot{\theta}(t_{n})+hT(t_{n})+\frac{h^{2}}{2}\dot{T}(t_{n})\right],\label{eq:hammer_ODE_next_1}
\end{align}
and 
\begin{equation}
\dot{\theta}(t_{n+1})=\frac{2}{h}\theta(t_{n+1})-\frac{2}{h}\theta(t_{n})-\dot{\theta}(t_{n}).\label{eq:hammer_ODE_next_2}
\end{equation}
Therefore, $\theta(t_{n+1})$ and $\dot{\theta}(t_{n+1})$ can be
computed using $\theta(t_{n})$, $\dot{\theta}(t_{n})$, $T(t_{n})$,
$\dot{T}(t_{n})$, but without using unknowns of the $t_{n+1}$ step.

\subsection{Concatenating the whole model}

With modal transformation and time discretization of the ODEs established,
it now suffices to combine all parts of the piano model together.
The whole computation process consists of two major separate parts:
first do space discretization, then do time discretization. Rooted
in the time-space separation idea of FEM, these two numeric works
do not interfere so computation costs should be acceptable. 

\subsubsection{Space discretization}

\paragraph{Derive ODEs for each subsystem.
\begin{align*}
 & \boldsymbol{M}\ddot{\boldsymbol{\xi}}(t)+\boldsymbol{C}\dot{\boldsymbol{\xi}}(t)+\boldsymbol{K}\boldsymbol{\xi}(t)=\boldsymbol{f}(t),\;\mathrm{superscripts}:a,b,e,f\\
 & \boldsymbol{M}\dot{\boldsymbol{\xi}}(t)+\boldsymbol{K}\boldsymbol{\xi}(t)=\boldsymbol{f}(t),\;\mathrm{superscripts}:c\\
 & \ddot{\theta}=-\mu\dot{\theta}+T(\theta),\;\mathrm{superscripts}:d
\end{align*}
}

\paragraph{Solve generalized eigenvalue problems.}

Only find the lowest $M$ eigenvalues and corresponding eigenvectors,
and integrate them into $\boldsymbol{\Lambda}$, $\boldsymbol{\Phi}$.
\[
\boldsymbol{K}\boldsymbol{\phi}_{i}=\lambda_{i}\boldsymbol{M}\boldsymbol{\phi}_{i},\;i=1,...,M,\;\mathrm{superscripts}:a,b,c,e,f
\]

\paragraph{Compute modal force transformation matrix (dense).}

This is useful for transformation $\boldsymbol{p}(t)=\boldsymbol{S}\boldsymbol{f}(t)$.
\[
\boldsymbol{S}=\left(\boldsymbol{\Phi}^{\mathrm{H}}\boldsymbol{M}\boldsymbol{\Phi}\right)^{-1}\boldsymbol{\Phi}^{\mathrm{H}},\;\mathrm{superscripts}:a,b,c,e,f
\]

\subsubsection{Time discretization}

\paragraph{Update schemes for modal DOFs}

Below lists the previously introduced time discretization shemes to
be later referred to. For each time step $n$, schemes 1a/2a/3a are
preferred over 1b/2b/3b whenever possible, because they utilize values
of the rhs source term for next time step $n+1$.
\begin{itemize}
\item Update scheme 1a:
\[
\boldsymbol{q}(t_{n+1})=\left(\boldsymbol{I}+\frac{h}{2}\boldsymbol{\Lambda}\right)^{-1}\left[\left(\boldsymbol{I}-\frac{h}{2}\boldsymbol{\Lambda}\right)\boldsymbol{q}(t_{n})+\frac{h}{2}\boldsymbol{p}(t_{n})+\frac{h}{2}\boldsymbol{p}(t_{n+1})\right]
\]
\item Update scheme 1b:
\[
\boldsymbol{q}(t_{n+1})=\left(\boldsymbol{I}+\frac{h}{2}\boldsymbol{\Lambda}\right)^{-1}\left[\left(\boldsymbol{I}-\frac{h}{2}\boldsymbol{\Lambda}\right)\boldsymbol{q}(t_{n})+h\boldsymbol{p}(t_{n})+\frac{h^{2}}{2}\dot{\boldsymbol{p}}(t_{n})\right]
\]
\item Update scheme 2a:
\begin{align*}
\boldsymbol{q}(t_{n+1}) & =\boldsymbol{Z}_{1}^{-1}\left[\boldsymbol{Z}_{0}\boldsymbol{q}(t_{n})+2\dot{\boldsymbol{q}}(t_{n})+\frac{h}{2}\boldsymbol{p}(t_{n})+\frac{h}{2}\boldsymbol{p}(t_{n+1})\right]\\
\dot{\boldsymbol{q}}(t_{n+1}) & =\frac{2}{h}\boldsymbol{q}(t_{n+1})-\frac{2}{h}\boldsymbol{q}(t_{n})-\dot{\boldsymbol{q}}(t_{n})
\end{align*}
\item Update scheme 2b:
\begin{align*}
\boldsymbol{q}(t_{n+1}) & =\boldsymbol{Z}_{1}^{-1}\left[\boldsymbol{Z}_{0}\boldsymbol{q}(t_{n})+2\dot{\boldsymbol{q}}(t_{n})+h\boldsymbol{p}(t_{n})+\frac{h^{2}}{2}\dot{\boldsymbol{p}}(t_{n})\right]\\
\dot{\boldsymbol{q}}(t_{n+1}) & =\frac{2}{h}\boldsymbol{q}(t_{n+1})-\frac{2}{h}\boldsymbol{q}(t_{n})-\dot{\boldsymbol{q}}(t_{n})
\end{align*}
\item Update scheme 3a: 
\begin{align*}
\theta(t_{n+1}) & =\left(\frac{2}{h}I+\mu\right)^{-1}\left[\left(\frac{2}{h}I+\mu\right)\theta(t_{n})+2\dot{\theta}(t_{n})+\frac{h}{2}T(t_{n})+\frac{h}{2}T(t_{n+1})\right]\\
\dot{\theta}(t_{n+1}) & =\frac{2}{h}\theta(t_{n+1})-\frac{2}{h}\theta(t_{n})-\dot{\theta}(t_{n})
\end{align*}
\item Update scheme 3b: 
\begin{align*}
\theta(t_{n+1}) & =\left(\frac{2}{h}I+\mu\right)^{-1}\left[\left(\frac{2}{h}I+\mu\right)\theta(t_{n})+2\dot{\theta}(t_{n})+hT(t_{n})+\frac{h^{2}}{2}\dot{T}(t_{n})\right]\\
\dot{\theta}(t_{n+1}) & =\frac{2}{h}\theta(t_{n+1})-\frac{2}{h}\theta(t_{n})-\dot{\theta}(t_{n})
\end{align*}
\end{itemize}

\paragraph{Initialization}
\begin{enumerate}
\item Discrete time interval $h$, number of time steps $n_{1}$.
\item Hammer shank: initial angle and angular velocity $\theta^{(d')}(t_{0})$,
$\dot{\theta}^{(d')}(t_{0})$, from which the initial torque $T^{(d')}(t_{0})$
and its derivative $\dot{T}^{(d')}(t_{0})$ (gravity contribution
only) can be computed.
\item Hammer felt \& string \& soundboard \& air \& room barraiers: initial
modal DOFs and modal forces 
\[
\boldsymbol{q}^{(*)}(t_{0})=\dot{\boldsymbol{q}}^{(*)}(t_{0})=\boldsymbol{p}^{(*)}(t_{0})=\dot{\boldsymbol{p}}^{(*)}(t_{0})=\boldsymbol{0},\;*=a,b,c,e'',f
\]
\end{enumerate}

\paragraph{At each time step $n=0,1,2,...,n_{1}$}
\begin{enumerate}
\item Compute the next rotation of hammer shank using scheme 3b:
\[
\theta^{(d')}(t_{n}),\dot{\theta}^{(d')}(t_{n}),T^{(d')}(t_{n}),\dot{T}^{(d')}(t_{n})\rightarrow\theta^{(d')}(t_{n+1}),\dot{\theta}^{(d')}(t_{n+1})
\]
\item Compute the current modal forces of hammer felt (non-zero boundary
condition) using (\ref{eq:hammer_felt_3D_boundary_string}):
\[
\theta^{(d')}(t_{n}),\dot{\theta}^{(d')}(t_{n}),\boldsymbol{q}^{(a)}(t_{n}),\dot{\boldsymbol{q}}^{(a)}(t_{n})\rightarrow\boldsymbol{p}^{(e'')}(t_{n}),\dot{\boldsymbol{p}}^{(e'')}(t_{n})
\]
Note: $\boldsymbol{p}^{(e'')}(t_{n})$ relates to the displacement
of felt at the contact points with string. The algorithm in appendix
\ref{sec:Elastic-material-contacting} may be more accurate for the
hammer felt, but for simplicity we choose to not integrate it here.
\item Compute the next modal DOFs of hammer felt using scheme 2b: 
\[
\boldsymbol{q}^{(e'')}(t_{n}),\dot{\boldsymbol{q}}^{(e'')}(t_{n}),\boldsymbol{p}^{(e'')}(t_{n}),\dot{\boldsymbol{p}}^{(e'')}(t_{n})\rightarrow\boldsymbol{q}^{(e'')}(t_{n+1}),\dot{\boldsymbol{q}}^{(e'')}(t_{n+1})
\]
\item Compute the next torque of hammer shank using (\ref{eq:hammer_shank_3D_eq}):
\[
\theta^{(d')}(t_{n+1}),\dot{\theta}^{(d')}(t_{n+1}),\boldsymbol{q}^{(e'')}(t_{n+1}),\dot{\boldsymbol{q}}^{(e'')}(t_{n+1})\rightarrow T^{(d')}(t_{n+1}),\dot{T}^{(d')}(t_{n+1})
\]
\item Compute the next modal forces of string (non-conservative force) using
(\ref{eq:hammer_felt_3D_force_string}):
\[
\boldsymbol{q}^{(e'')}(t_{n+1})\rightarrow\boldsymbol{p}_{1}^{(a)}(t_{n+1})
\]
Note: $\boldsymbol{p}_{1}^{(a)}$ relates to hammer felt force, $\boldsymbol{p}_{2}^{(a)}$
relates to bridge point displacement.
\item Compute the next modal DOFs of string using scheme 2a for $\boldsymbol{p}_{1}^{(a)}$
and scheme 2b for $\boldsymbol{p}_{2}^{(a)}$: 
\[
\boldsymbol{q}^{(a)}(t_{n}),\dot{\boldsymbol{q}}^{(a)}(t_{n}),\boldsymbol{p}_{1}^{(a)}(t_{n}),\boldsymbol{p}_{1}^{(a)}(t_{n+1}),\boldsymbol{p}_{2}^{(a)}(t_{n}),\dot{\boldsymbol{p}}_{2}^{(a)}(t_{n})\rightarrow\boldsymbol{q}^{(a)}(t_{n+1}),\dot{\boldsymbol{q}}^{(a)}(t_{n+1})
\]
\item Compute the next modal forces of soundboard (non-conservative force)
using (\ref{eq:string_soundboard_coupling_force_dynamic}):
\[
\boldsymbol{q}^{(a)}(t_{n+1})\rightarrow\boldsymbol{p}_{1}^{(b)}(t_{n+1})
\]
Note: $\boldsymbol{p}_{1}^{(b)}$ relates to string force at the bridge,
$\boldsymbol{p}_{2}^{(b)}$ relates to air pressure on the soundboard.
\item Compute the next modal DOFs of soundboard using scheme 2a for $\boldsymbol{p}_{1}^{(b)}$
and scheme 2b for $\boldsymbol{p}_{2}^{(b)}$: 
\[
\boldsymbol{q}^{(b)}(t_{n}),\dot{\boldsymbol{q}}^{(b)}(t_{n}),\boldsymbol{p}_{1}^{(b)}(t_{n}),\boldsymbol{p}_{1}^{(b)}(t_{n+1}),\boldsymbol{p}_{2}^{(b)}(t_{n}),\dot{\boldsymbol{p}}_{2}^{(b)}(t_{n})\rightarrow\boldsymbol{q}^{(b)}(t_{n+1}),\dot{\boldsymbol{q}}^{(b)}(t_{n+1})
\]
\item Compute the next modal forces of air (non-zero boundary condition)
using (\ref{eq:soundboard_air_coupling_boundary}):
\[
\boldsymbol{q}^{(b)}(t_{n+1})\rightarrow\boldsymbol{p}_{1}^{(c)}(t_{n+1})
\]
Note: $\boldsymbol{p}_{1}^{(c)}$ relates to displacements at the
interface with soundboard, $\boldsymbol{p}_{2}^{(c)}$ relates to
displacements at interface with room barriers.
\item Compute the next modal DOFs of air using scheme 1a for $\boldsymbol{p}_{1}^{(c)}$
and scheme 1b for $\boldsymbol{p}_{2}^{(c)}$: 
\[
\boldsymbol{q}^{(c)}(t_{n}),\boldsymbol{p}_{1}^{(c)}(t_{n}),\boldsymbol{p}_{1}^{(c)}(t_{n+1}),\boldsymbol{p}_{2}^{(c)}(t_{n}),\dot{\boldsymbol{p}}_{2}^{(c)}(t_{n})\rightarrow\boldsymbol{q}^{(c)}(t_{n+1})
\]
\item Compute the next modal forces of soundboard (non-conservative force)
using (\ref{eq:soundboard_air_coupling_force}):
\[
\boldsymbol{q}^{(c)}(t_{n+1})\rightarrow\boldsymbol{p}_{2}^{(b)}(t_{n+1}),\dot{\boldsymbol{p}}_{2}^{(b)}(t_{n+1})
\]
\item Compute the next modal forces of room barriers (non-conservative force)
using (\ref{eq:room_air_coupling_force}):
\[
\boldsymbol{q}^{(c)}(t_{n+1})\rightarrow\boldsymbol{p}^{(f)}(t_{n+1})
\]
Note: $\boldsymbol{p}^{(e'')}(t_{n})$ relates to air pressure on
the room barriers.
\item Compute the next modal DOFs of room barriers using scheme 2a: 
\[
\boldsymbol{q}^{(f)}(t_{n}),\dot{\boldsymbol{q}}^{(f)}(t_{n}),\boldsymbol{p}^{(f)}(t_{n}),\boldsymbol{p}^{(f)}(t_{n+1})\rightarrow\boldsymbol{q}^{(f)}(t_{n+1}),\dot{\boldsymbol{q}}^{(f)}(t_{n+1})
\]
\item Compute the next modal forces of air (non-zero boundary condition)
using (\ref{eq:room_air_coupling_boundary}):
\[
\boldsymbol{q}^{(f)}(t_{n+1}),\dot{\boldsymbol{q}}^{(f)}(t_{n+1})\rightarrow\boldsymbol{p}_{2}^{(c)}(t_{n+1}),\dot{\boldsymbol{p}}_{2}^{(c)}(t_{n+1})
\]
\item If higher accuracy and better energy stability is desired, repeat
the above steps 1 to 14 several times. Note that different from the
first iteration, the subsequent iterations always use schemes 1a,
2a, 3a and do not need to use schemes 1b, 2b, 3b. If no more repetition
is needed, end the current time step and move on to the next time
step. 
\end{enumerate}

\paragraph{Audio output}

Compute the acoustic pressure signals at certaining listening positions
using the stored modal DOFs $\boldsymbol{q}^{(c)}(t_{n})\;(n=0,...,n_{1})$
and modal superposition. These signals are the final output digital
audio of piano simulation model.

\section{Conclusion}

This paper presented a detailed physical model for simulating acoustic
piano sounds. For solid parts of the piano system, viz. strings, soundboard,
room barriers, hammer felt, a 3D prestressed elasticity model is generally
applied. For fluid parts of the piano system, viz. sound radiation
in the air, conservation of mass and Navier-Stokes equation is applied.
For coupling between different subsystems of the piano, mechanisms
of surface force transmission and displacement/velocity continuity
are considered. For numeric simulation, modal superposition and explicit
time discretization schemes are utilized. Despite the complexity of
this whole piano model, we have paid efforts to a straightforward
presentation based more on system ODEs transformed from strong PDEs,
as well as a time domain simulation scheme balancing efficiency and
accuracy.

Below discusses the current study's limitations and our plans or recommendations
for future research.
\begin{itemize}
\item Waiting for numeric simulation results. Due to the complication of
our piano model, we choose to write down theoretic models first as
a guiding framework. Our next step involves implementing the computation
procedures using high-performance, expressive and well-structured
programming languages like Rust, as well as performing result analysis
using convenient and ecologically rich programming languages like
Python. Facing some unknown uncertainties in practice, our model needs
to be further tested and improved.
\item In hammer felt-string coupling and string-soundboard coupling, the
contact was treated as occuring at a point rather than an area. This
simplification would probably lead to a loss in realism. The 3D contact
framework described in appendix \ref{sec:Elastic-material-contacting}
may apply here, but algorithmic complexity and numeric stability needs
to be better addressed.
\item The string-soundboard coupling mechanism was assumed of fixation but
not collision nature. A coupling model similar to the nonlinear hammer-string
interaction may be more suitable, considering that a string seems
to actually be supported between two distanced bridge pins. The 3D
contact framework described in appendix \ref{sec:Elastic-material-contacting}
may apply here, but algorithmic complexity and numeric stability needs
to be better addressed. It also remains to discover how the relative
positions of the two pins affect the vibrations of the string and
soundboard.
\item Treatment of nonlinear coupling between the distinct subsystems of
the piano in time discretization schemes remains challenging. As far
as we understand, only nonlinear implicit schemes can achieve unconditional
energy stability if the two coupling mechanisms we consider, namely
the surface force transmission and the displacement/velocity continuity,
exhibit nonlinearity. A viable tradeoff between accuracy and efficiency
may be the explicit iteration approach we proposed, but we hope to
see better alternatives.
\item The damping in our model is relatively simplified, as only one or
two viscousity coefficients are used. It may be beneficial to introduce
additional unknowns like temperature and additional governing equations
like thermal equations to reflect the damping phenomena more realistically.
\item Some observed phenomena of acoustic pianos are still missing their
representations in our model. For example, strings not striked by
the hammer may also vibrate as long as the sustain pedal is pressed,
which is often called sympathetic resonance. An explanation for this
is that the striked string transmits its vibration to other strings
through the variation of air pressure, which can actually lead to
a string-air coupling model. Also, the transmission of piano player's
key action into hammer shank movement requires further investigation.
\item It remains to discover the relation between physical parameters of
the piano model and the objective (waveforms, spectrums) and subjective
(listener feel) aspects of the final ouput sound, so that these parameters
can be tuned to approach realism or even obtain new sounds without
a real world couterpart. 
\end{itemize}
Finally, we hope this study could contribute to the understanding
of the vibroacoustics of a piano, and to the innovation of digital
musical instruments with desired characteristics of sound.

\listoffigures

\bibliographystyle{plain}
\bibliography{paper_01}

\appendix

\section{Prestrain and prestress}\label{sec:Prestrain-and-prestress}

Consider 3 configurations of a material: natural → initial → current
\cite{pau2015nonlinear}. In the natural configuration, no prestrain
and prestress is present. In the initial configuration, prestrain
and prestress exist. In the current configuration, strain and stress
emerge in addition to prestrain and prestress. Under the same $xyz$
coordinate system, define $\mathbb{R}^{3}$ vectors $\bar{\boldsymbol{x}}$,
$\boldsymbol{x}$, $\hat{\boldsymbol{x}}$ as the coordinates of the
same material particle in the natural, initial and current configurations
respectively. The following relation holds: 
\begin{align}
\boldsymbol{x} & =\bar{\boldsymbol{x}}+\bar{\boldsymbol{u}}(\bar{\boldsymbol{x}}),\nonumber \\
\hat{\boldsymbol{x}} & =\boldsymbol{x}+\boldsymbol{u}(\boldsymbol{x}),\label{eq:deform_map}
\end{align}
where $\bar{\boldsymbol{u}}$ and $\boldsymbol{u}$ are the displacement
vectors from natural to initial and from initial to current configurations
respectively. Denote deformation gradients $\hat{\boldsymbol{Z}}=\frac{\partial\hat{\boldsymbol{x}}}{\partial\bar{\boldsymbol{x}}}$,
$\boldsymbol{Z}=\frac{\partial\hat{\boldsymbol{x}}}{\partial\boldsymbol{x}}$
and $\bar{\boldsymbol{Z}}=\frac{\partial\boldsymbol{x}}{\partial\bar{\boldsymbol{x}}}$;
denote Jacobian matrices $\boldsymbol{J}=\nabla\boldsymbol{u}$, $\bar{\boldsymbol{J}}=\nabla\bar{\boldsymbol{u}}$,.
We then have 
\begin{align}
 & \boldsymbol{Z}=\boldsymbol{I}+\boldsymbol{J},\;\bar{\boldsymbol{Z}}=\boldsymbol{I}+\bar{\boldsymbol{J}},\nonumber \\
 & \hat{\boldsymbol{Z}}=\boldsymbol{Z}\bar{\boldsymbol{Z}}=\left(\boldsymbol{I}+\boldsymbol{J}\right)\left(\boldsymbol{I}+\bar{\boldsymbol{J}}\right).
\end{align}
The Green-Lagrange strain tensor of current configuration with respect
to natural configuration is 
\begin{align}
\hat{\boldsymbol{E}} & =\frac{1}{2}\left(\hat{\boldsymbol{Z}}^{\top}\hat{\boldsymbol{Z}}-\boldsymbol{I}\right)=\left(\boldsymbol{I}+\bar{\boldsymbol{J}}\right)^{\top}\left(\boldsymbol{I}+\boldsymbol{J}\right)^{\top}\left(\boldsymbol{I}+\boldsymbol{J}\right)\left(\boldsymbol{I}+\bar{\boldsymbol{J}}\right)\nonumber \\
 & =\frac{1}{2}\left(\bar{\boldsymbol{J}}+\bar{\boldsymbol{J}}^{\top}+\bar{\boldsymbol{J}}^{\top}\bar{\boldsymbol{J}}\right)+\frac{1}{2}\left(\boldsymbol{I}+\bar{\boldsymbol{J}}^{\top}\right)\left(\boldsymbol{J}+\boldsymbol{J}^{\top}+\boldsymbol{J}^{\top}\boldsymbol{J}\right)\left(\boldsymbol{I}+\bar{\boldsymbol{J}}\right)\nonumber \\
 & \approx\widetilde{\boldsymbol{E}}=\boldsymbol{\epsilon}+\bar{\boldsymbol{E}}+\left(\bar{\boldsymbol{J}}^{\top}\boldsymbol{\epsilon}+\boldsymbol{\epsilon}\bar{\boldsymbol{J}}+\bar{\boldsymbol{J}}^{\top}\boldsymbol{\epsilon}\bar{\boldsymbol{J}}\right),\label{eq:total_strain_1}
\end{align}
where $\bar{\boldsymbol{E}}=\frac{1}{2}\left(\bar{\boldsymbol{J}}+\bar{\boldsymbol{J}}^{\top}+\bar{\boldsymbol{J}}^{\top}\bar{\boldsymbol{J}}\right)$
is the prestrain tensor, and $\boldsymbol{\epsilon}=\frac{1}{2}\left(\boldsymbol{J}+\boldsymbol{J}^{\top}\right)$
is the engineering strain discarding the second-order $\frac{1}{2}\boldsymbol{J}^{\top}\boldsymbol{J}$
term for the sake of linearization. Note this linearization should
. Assuming $\bar{\boldsymbol{J}}=\{\bar{J}_{ij}\}$ is symmetric,
viz. no rigid body rotation, then we have eigen decomposition $\bar{\boldsymbol{J}}=\boldsymbol{U}\mathrm{diag}(\lambda_{1},\lambda_{2},\lambda_{3})\boldsymbol{U}^{\top}$
and 
\begin{align}
\bar{\boldsymbol{J}}+\frac{1}{2}\bar{\boldsymbol{J}}^{2} & =\bar{\boldsymbol{E}}\nonumber \\
\boldsymbol{U}\left[\begin{array}{ccc}
\lambda_{1}+\frac{1}{2}\lambda_{1}^{2}\\
 & \lambda_{2}+\frac{1}{2}\lambda_{2}^{2}\\
 &  & \lambda_{3}+\frac{1}{2}\lambda_{3}^{2}
\end{array}\right]\boldsymbol{U}^{\top} & =\boldsymbol{U}\left[\begin{array}{ccc}
\lambda_{1}'\\
 & \lambda_{2}'\\
 &  & \lambda_{3}'
\end{array}\right]\boldsymbol{U}^{\top},\label{eq:prestrain_gradient}
\end{align}
from which $\lambda_{i}=4\sqrt{\lambda_{i}'+4}-8$ is the solution
and $\bar{\boldsymbol{J}}$ can be computed per eigen decomposition.
The uniqueness of this solution stems from that the principal stretches
$\lambda_{i}'\geq-1$ should hold for normal cases, viz. no negative
compression, and $\lambda_{i}$ should be close to $\lambda_{i}'$
to be consistent with the case of ignoring $\frac{1}{2}\bar{\boldsymbol{J}}^{2}$.
If the prestrain is not large enough to induce non-negligible geometric
nonlinearity, $\frac{1}{2}\bar{\boldsymbol{J}}^{\top}\bar{\boldsymbol{J}}$
can be disregarded and $\bar{\boldsymbol{J}}=\bar{\boldsymbol{E}}$
simply holds. But generally for string instruments like piano, prestrain
may be large and even dominate the post-strain, thus we should cover
the geometric nonlinearity of prestrain. This, however, would not
necessarily make (\ref{eq:total_strain_1}) nonlinear with respect
to $\boldsymbol{J}$, because strain deformation is normally small
enough to make $\frac{1}{2}\boldsymbol{J}^{\top}\boldsymbol{J}$ negligible.
Consequently, with $\bar{\boldsymbol{J}}$ and $\bar{\boldsymbol{E}}$
known, the prestrain model (\ref{eq:total_strain_1}) is reasonably
linear with respect to the unknowns. Converting it into vector form,
we find 
\begin{align}
\overrightarrow{\hat{\boldsymbol{E}}} & \approx\overrightarrow{\widetilde{\boldsymbol{E}}}=\overrightarrow{\bar{\boldsymbol{E}}}+\overrightarrow{\widetilde{\boldsymbol{E}}}_{\mathrm{dyn}},\;\overrightarrow{\widetilde{\boldsymbol{E}}}_{\mathrm{dyn}}=\left(\boldsymbol{I}+\boldsymbol{\Psi}_{2}\boldsymbol{\Psi}_{1}\right)\overrightarrow{\boldsymbol{\epsilon}},\;\boldsymbol{\Psi}_{1}=\left[\begin{array}{cccccc}
2 & 0 & 0 & 0 & 0 & 0\\
0 & 2 & 0 & 0 & 0 & 0\\
0 & 0 & 2 & 0 & 0 & 0\\
0 & 0 & 0 & 1 & 0 & 0\\
0 & 0 & 0 & 0 & 1 & 0\\
\mathrm{0} & 0 & 0 & 0 & 0 & 1
\end{array}\right]\nonumber \\
\boldsymbol{\Psi}_{2} & =\left[\begin{smallmatrix}\bar{J}_{11}\left(\bar{J}_{11}+2\right) & \bar{J}_{12}^{2} & \bar{J}_{13}^{2} & \bar{J}_{12}\left(\bar{J}_{11}+1\right) & \bar{J}_{13}\left(\bar{J}_{11}+1\right) & \bar{J}_{12}\bar{J}_{13}\\
 & \bar{J}_{22}\left(\bar{J}_{22}+2\right) & \bar{J}_{23}^{2} & \bar{J}_{12}\left(\bar{J}_{22}+1\right) & \bar{J}_{12}\bar{J}_{23} & \bar{J}_{23}\left(\bar{J}_{22}+1\right)\\
 &  & \bar{J}_{33}\left(\bar{J}_{33}+2\right) & \bar{J}_{13}\bar{J}_{23} & \bar{J}_{13}\left(\bar{J}_{33}+1\right) & \bar{J}_{23}\left(\bar{J}_{33}+1\right)\\
 &  &  & \bar{J}_{11}\bar{J}_{22}+\bar{J}_{11}+\bar{J}_{12}^{2}+\bar{J}_{22} & \bar{J}_{11}\bar{J}_{23}+\bar{J}_{12}\bar{J}_{13}+\bar{J}_{23} & \bar{J}_{12}\bar{J}_{23}+\bar{J}_{13}\bar{J}_{22}+\bar{J}_{13}\\
 &  &  &  & \bar{J}_{11}\bar{J}_{33}+\bar{J}_{11}+\bar{J}_{13}^{2}+\bar{J}_{33} & \bar{J}_{12}\bar{J}_{33}+\bar{J}_{12}+\bar{J}_{13}\bar{J}_{23}\\
\mathrm{Sym} &  &  &  &  & \bar{J}_{22}\bar{J}_{33}+\bar{J}_{22}+\bar{J}_{23}^{2}+\bar{J}_{33}
\end{smallmatrix}\right]\label{eq:total_strain_vec}
\end{align}
where $\overrightarrow{\widetilde{\boldsymbol{E}}}_{\mathrm{dyn}}$
is the dynamic part of strain vector; all the vectors here are defined
similar to (\ref{eq:strain_vec}). It is now straight forward that
the prestrain functions a linear transformation (addition and scaling)
of the strain at a first approximation. However, as in (\ref{eq:total_strain_vec})
$\bar{E}_{ij}(\bar{\boldsymbol{x}})$ is expressed in natural coordinates,
converting it into initial coordinate representations to align with
$\overrightarrow{\boldsymbol{\epsilon}}$ would be preferrable. In
order for this, define prestress vector $\overrightarrow{\bar{\boldsymbol{S}}}(\bar{\boldsymbol{x}})$,
then $\overrightarrow{\bar{\boldsymbol{E}}}=\boldsymbol{D}^{-1}\overrightarrow{\bar{\boldsymbol{S}}}$.
From (\ref{eq:deform_map}), an inverse mapping $\bar{\boldsymbol{x}}=f(\boldsymbol{x})$
should exist. If we let $\overrightarrow{\boldsymbol{T}}(\boldsymbol{x})=\overrightarrow{\bar{\boldsymbol{S}}}(f(\boldsymbol{x}))$,
the vectorized tension field from (\ref{eq:tension_matrix}), and
subsititute $\overrightarrow{\bar{\boldsymbol{E}}}=\boldsymbol{D}^{-1}\overrightarrow{\boldsymbol{T}}(\boldsymbol{x})$
into (\ref{eq:prestrain_gradient})(\ref{eq:total_strain_vec}), then
the total strain $\hat{\boldsymbol{E}}$ can be expressed in only
the initial configurations. The advantage of this is not only avoiding
finding the intricate natural coordinates, but also making it straightfoward
to satisfy static equilibrium by condition $\nabla\cdot\boldsymbol{T}(\boldsymbol{x})=\boldsymbol{0}$.

Given the strain vector $\overrightarrow{\hat{\boldsymbol{E}}}$,
the dynamic strain energy is 
\begin{align}
U & =\int_{\Omega}\int_{\boldsymbol{0}}^{\overrightarrow{\widetilde{\boldsymbol{E}}}_{\mathrm{dyn}}}\Psi_{0}\overrightarrow{\widetilde{\boldsymbol{S}}}_{\mathrm{dyn}}\cdot\mathrm{d}\overrightarrow{\widetilde{\boldsymbol{E}}}_{\mathrm{dyn}}\mathrm{d}V\nonumber \\
 & =\int_{\Omega}\int_{\boldsymbol{0}}^{\overrightarrow{\widetilde{\boldsymbol{E}}}_{\mathrm{dyn}}}\Psi_{0}\boldsymbol{D}\overrightarrow{\widetilde{\boldsymbol{E}}}_{\mathrm{dyn}}\cdot\mathrm{d}\overrightarrow{\widetilde{\boldsymbol{E}}}_{\mathrm{dyn}}\mathrm{d}V\nonumber \\
 & =\frac{1}{2}\int_{\Omega}\Psi_{0}\overrightarrow{\widetilde{\boldsymbol{E}}}_{\mathrm{dyn}}\cdot\boldsymbol{D}\overrightarrow{\widetilde{\boldsymbol{E}}}_{\mathrm{dyn}}\mathrm{d}V\nonumber \\
 & =\int_{\Omega}\frac{1}{2}\Psi_{0}\overrightarrow{\boldsymbol{\epsilon}}\cdot\left(\boldsymbol{D}+\boldsymbol{\Psi}_{1}^{\top}\boldsymbol{\Psi}_{2}^{\top}\boldsymbol{D}\boldsymbol{\Psi}_{2}\boldsymbol{\Psi}_{1}\right)\overrightarrow{\boldsymbol{\epsilon}}\mathrm{d}V,
\end{align}
where $\Psi_{0}(\boldsymbol{x})=\det(\bar{\boldsymbol{Z}})^{-1}$
is the volume factor linear w.r.t. the unknowns, accounts for volume
change from natural to current configuration, and can be computed
given $\bar{\boldsymbol{J}}$. Note that $\overrightarrow{\widetilde{\boldsymbol{S}}}_{\mathrm{dyn}}$
here is understood as the (dynamic) second Piola-Kirchhoff (PK2) stress
that is energy conjugate to $\overrightarrow{\widetilde{\boldsymbol{E}}}_{\mathrm{dyn}}$.
Though PK2 should be formulated in the natural configuration, we have
transformed it to the initial configuration along with the volume
factor. According to \cite{Stress-and-Equations-of-Motion}, PK2 may
also be changed to Cauchy stress in the initial configuration, but
it is unnecessary here because the strain energy is the same even
without doing so. And anyway, what we seek for here is not PK2, but
the dynamic stress vector $\overrightarrow{\boldsymbol{\sigma}}$
that is energy conjugate to $\overrightarrow{\boldsymbol{\epsilon}}$,
which can be derived from the expression of dynamic strain energy
as 
\begin{equation}
\overrightarrow{\hat{\boldsymbol{\sigma}}}=\left(\boldsymbol{D}\overrightarrow{\boldsymbol{\epsilon}}+\overrightarrow{\boldsymbol{\tau}}\right),\;\overrightarrow{\boldsymbol{\tau}}+\Psi_{0}\boldsymbol{\Psi}_{1}^{\top}\boldsymbol{\Psi}_{2}^{\top}\boldsymbol{D}\boldsymbol{\Psi}_{2}\boldsymbol{\Psi}_{1}\overrightarrow{\boldsymbol{\epsilon}},\label{eq:total_stress_vec}
\end{equation}
where $\overrightarrow{\boldsymbol{\tau}}$ is the dynamic prestress
vector. It is now clear that the effect of prestress can be seen as
adding a symmetrically transformed constitutive matrix $\boldsymbol{\Psi}_{1}^{\top}\boldsymbol{\Psi}_{2}^{\top}\boldsymbol{D}\boldsymbol{\Psi}_{2}\boldsymbol{\Psi}_{1}$. 

Finally, we acknowledge that tn the presence of large deformations
from initial to current configuration, the quadratic term in the Green-Lagrange
strain should be retained, leading to nonlinear operators on the unknowns
for the strain energy, as well as PDEs and ODEs. However, this concern
is irrelevant in case of large deformations from natural to initial
configuration because even if prestrain itself is nonlinear, it is
linear w.r.t. the strain where the unknowns are contained in. Another
more intricate nonlinearity, often known as the geometric nonlinearity,
arises if we want to express stress and prestress in the current configuration.
This would require transforming PK2 to Cauchy stress, yet we have
not encountered the necessity for doing so in the present context.

\section{Lagrangian formulation}

As a complement to the Newtonian formulation, we present a non-conservative
Lagrangian formulation of the 3D elastic material model. To derive
system ODEs (weak form) in a more straightforward way, FEM space discretization
is incorporated into the Lagrangian formulation, skipping the intermediate
PDEs (strong form). As per (\ref{eq:space_discretization_map}), the
kinetic energy and its variation over $[t_{0},t_{1}]$ are 
\begin{align}
E_{k} & =\int_{\Omega}\frac{1}{2}\rho\dot{\boldsymbol{u}}\cdot\dot{\boldsymbol{u}}\mathrm{d}V=\frac{1}{2}\int_{\Omega}\dot{\boldsymbol{\xi}}\cdot\rho\boldsymbol{P}^{\top}\boldsymbol{P}\dot{\boldsymbol{\xi}}\mathrm{d}V\nonumber \\
\int_{t_{0}}^{t_{1}}\delta E_{k}\mathrm{d}t & =\int_{t_{0}}^{t_{1}}\int_{\Omega}\delta\dot{\boldsymbol{\xi}}\cdot\rho\boldsymbol{P}^{\top}\boldsymbol{P}\dot{\boldsymbol{\xi}}\mathrm{d}V\mathrm{d}t=-\int_{t_{0}}^{t_{1}}\int_{\Omega}\delta\boldsymbol{\xi}\cdot\rho\boldsymbol{P}^{\top}\boldsymbol{P}\ddot{\boldsymbol{\xi}}\mathrm{d}V\mathrm{d}t
\end{align}
where integration by parts is utilized assuming 
\begin{equation}
\delta\boldsymbol{\xi}(t_{0})=\delta\boldsymbol{\xi}(t_{1})=\boldsymbol{0}.\label{eq:energy_variation_condition}
\end{equation}
Note that we have excluded $\boldsymbol{\xi}_{0}$ in the derivation
of kinetic energy because nonzero $\boldsymbol{\xi}_{0}$ (boundary
displacements) actually implys the existence some non-conservative
forces (e.g. constraints). Nevertheless, we can compute the ``virtual
kinetic energy'' as 
\begin{equation}
\int_{t_{0}}^{t_{1}}\delta E_{k,0}=\int_{t_{0}}^{t_{1}}\int_{\Omega}\delta\dot{\boldsymbol{\xi}}\cdot\rho\boldsymbol{P}^{\top}\boldsymbol{P}_{0}\dot{\boldsymbol{\xi}}_{0}\mathrm{d}V\mathrm{d}t
\end{equation}
The potential energy and its variation are 
\begin{align}
E_{p} & =-\int_{\Omega}\frac{1}{2}\boldsymbol{u}\cdot\mathrm{Div}\left(\boldsymbol{\sigma}+\boldsymbol{\tau}\right)\mathrm{d}V\nonumber \\
 & =-\int_{\Gamma}\left(\frac{1}{2}\sum_{i=1}^{3}u_{i}\left(\boldsymbol{\sigma}_{i}+\boldsymbol{\tau}_{i}\right)\right)\cdot\mathrm{d}\boldsymbol{\Gamma}+\int_{\Omega}\left(\frac{1}{2}\sum_{i=1}^{3}\nabla u_{i}\cdot\left(\boldsymbol{\sigma}_{i}+\boldsymbol{\tau}_{i}\right)\right)\mathrm{d}V\nonumber \\
 & =\int_{\Omega}\left[\sum_{i=1}^{3}\frac{1}{2}\left(\boldsymbol{\xi}\cdot\boldsymbol{Q}_{i}^{\top}\left(\boldsymbol{A}_{i}+\boldsymbol{B}_{i}\right)\boldsymbol{Q}\boldsymbol{\xi}\right)\right]\mathrm{d}V\nonumber \\
 & =\int_{\Omega}\frac{1}{2}\boldsymbol{\xi}\cdot\boldsymbol{Q}^{\top}\left(\boldsymbol{A}+\boldsymbol{B}\right)\boldsymbol{Q}\boldsymbol{\xi}\mathrm{d}V\nonumber \\
\delta E_{p} & =\int_{\Omega}\delta\boldsymbol{\xi}\cdot\boldsymbol{Q}^{\top}\left(\boldsymbol{A}+\boldsymbol{B}\right)\boldsymbol{Q}\boldsymbol{\xi}\mathrm{d}V
\end{align}
where Dirichlet or Neumann boundary conditions should automatically
apply to ensure the integral on the boundary is zero. Similar to the
kinetic energy, nonzero $\boldsymbol{\xi}_{0}$ does not correspond
to a potential energy but a ``virtual potential energy'' (equivalent
to virtual work but sign flipped here for convenience) as 
\begin{equation}
\delta E_{p,0}=\int_{\Omega}\delta\boldsymbol{\xi}\cdot\boldsymbol{Q}^{\top}\left(\boldsymbol{A}+\boldsymbol{B}\right)\boldsymbol{Q}_{0}\boldsymbol{\xi}_{0}\mathrm{d}V
\end{equation}
As for the damping force which is non-conservative, we can compute
its virtual work as 
\begin{align}
\delta W_{d} & =2\mu\int_{\Omega}\delta\boldsymbol{u}\cdot\mathrm{Div}\left(\dot{\boldsymbol{\sigma}}+\dot{\boldsymbol{\tau}}\right)\mathrm{d}V\nonumber \\
 & =-2\mu\int_{\Omega}\left[\sum_{i=1}^{3}\delta\nabla u_{i}\cdot\left(\dot{\boldsymbol{\sigma}}_{i}+\dot{\boldsymbol{\tau}}_{i}\right)\right]\mathrm{d}V\nonumber \\
 & =-2\mu\int_{\Omega}\delta\boldsymbol{\xi}\cdot\left[\boldsymbol{Q}^{\top}\left(\boldsymbol{A}+\boldsymbol{B}\right)\boldsymbol{Q}\dot{\boldsymbol{\xi}}+\boldsymbol{Q}^{\top}\left(\boldsymbol{A}+\boldsymbol{B}\right)\boldsymbol{Q}_{0}\dot{\boldsymbol{\xi}_{0}}\right]\mathrm{d}V.
\end{align}
It can be verified that the virtual work approach is also applicable
to conservative forces like stress and prestress to derive the identical
potential energy functions. For other non-conservative forces, the
virtual work is 
\begin{equation}
\delta W_{nc}=\int_{\Omega}\delta\boldsymbol{u}\cdot\boldsymbol{F}\mathrm{d}V=\int_{\Omega}\delta\boldsymbol{\xi}\cdot\boldsymbol{P}^{\top}\boldsymbol{F}\mathrm{d}V.
\end{equation}
Applying the principle of virtual work, we derive the energy equation
\begin{equation}
\int_{t_{0}}^{t_{1}}\left(\delta E_{k}+\delta E_{k,0}-\delta E_{p}-\delta E_{p,0}+\delta W_{d}+\delta W_{nc}\right)\mathrm{d}t=-\int_{t_{0}}^{t_{1}}\delta\boldsymbol{\xi}\cdot\left(\boldsymbol{M}\ddot{\boldsymbol{\xi}}+\boldsymbol{C}\dot{\boldsymbol{\xi}}+\boldsymbol{K}\boldsymbol{\xi}-\boldsymbol{f}\right)\mathrm{d}t=0,\label{eq:energy_eq}
\end{equation}
where $\boldsymbol{M}$, $\boldsymbol{C}$, $\boldsymbol{K}$, $\boldsymbol{f}$
are defined in (\ref{eq:ODEs_where}). Since $\delta\boldsymbol{\xi}$
is arbitrary as long as (\ref{eq:energy_variation_condition}) is
satisfied, (\ref{eq:ODEs}) can be derived from (\ref{eq:energy_eq}).

\section{Elastic material contacting rigid surface in 3D setting}\label{sec:Elastic-material-contacting}

We now consider the case of a material A contacting another material
B in a fully 3D setting, where A is elastic with weak form ODEs like
(\ref{eq:ODEs}) and B's contact surface is rigid (hence B may still
be an elastic material). In this case, applying Dirichlet boundary
conditions through $\boldsymbol{\xi}_{0}(t)$ would be infeasible
because not only the contact status (in contact or not) but also the
contact boundary changes with time. As we shall see, the potential
contact boundary is Dirichlet when contact is present meaning that
boundary stress and strain appear, and is Neumann when contact is
absent meaning that boundary stress and strain vanish. This leads
to a time-varying dimension (number of DOFs) of the system ODEs, invalidating
the modal superposition method. To model the dynamic evolution of
contact system, we adopt the master-slave idea and will brief describe
our understanding and application of it. Readers can refer to \cite{wriggers1995finite}\cite{kim2015finite}
for a comprehensive and rigorous FEM formulation of the contact problem. 

Consider a system of second-order ODEs where the rhs source term depends
on the DOFs: 
\begin{equation}
\boldsymbol{M}\ddot{\boldsymbol{\xi}}(t)+\boldsymbol{C}\dot{\boldsymbol{\xi}}(t)+\boldsymbol{K}\boldsymbol{\xi}(t)=\boldsymbol{f}(t),\;\boldsymbol{f}(t)=\boldsymbol{r}\left(t,\boldsymbol{\xi}(t),\dot{\boldsymbol{\xi}}(t)\right).\label{eq:contact_ODEs}
\end{equation}
To derive time stepping algorithms like those in \ref{subsec:Explicit-time-discretization},
we first integrate (\ref{eq:contact_ODEs}) over $[t_{n},t_{n+1}]$
to yield 
\begin{align}
\boldsymbol{M}\left[\dot{\boldsymbol{\xi}}(t_{n+1})-\dot{\boldsymbol{\xi}}(t_{n})\right] & =-\boldsymbol{C}\left[\boldsymbol{\xi}(t_{n+1})-\boldsymbol{\xi}(t_{n})\right]-\boldsymbol{K}\int_{t_{n}}^{t_{n+1}}\boldsymbol{\xi}(t)\mathrm{d}t+\int_{t_{n}}^{t_{n+1}}\boldsymbol{f}(t)\mathrm{d}t\nonumber \\
\boldsymbol{M}\left[\dot{\boldsymbol{\xi}}(t_{n+1})-\dot{\boldsymbol{\xi}}(t_{n})\right] & \approx-\boldsymbol{C}\left[\boldsymbol{\xi}(t_{n+1})-\boldsymbol{\xi}(t_{n})\right]-\frac{h}{2}\boldsymbol{K}\left[\boldsymbol{\xi}(t_{n})+\boldsymbol{\xi}(t_{n+1})\right]+h\boldsymbol{f}(t_{n})+\frac{h^{2}}{2}\dot{\boldsymbol{f}}(t_{n}).
\end{align}
Combining this with 
\begin{equation}
\boldsymbol{\xi}(t_{n+1})-\boldsymbol{\xi}(t_{n})=\int_{t_{n}}^{t_{n+1}}\dot{\boldsymbol{\xi}}(t)\mathrm{d}t\approx\frac{h}{2}\left[\dot{\boldsymbol{\xi}}(t_{n})+\dot{\boldsymbol{\xi}}(t_{n+1})\right],
\end{equation}
the update scheme for DOFs vector is found to be 
\begin{align}
\boldsymbol{\xi}(t_{n+1}) & =\boldsymbol{Z}_{1}^{-1}\left[\boldsymbol{Z}_{0}\boldsymbol{\xi}(t_{n})+2\boldsymbol{M}\dot{\boldsymbol{\xi}}(t_{n})+h\boldsymbol{f}(t_{n})+\frac{h^{2}}{2}\dot{\boldsymbol{f}}(t_{n})\right],\nonumber \\
\dot{\boldsymbol{\xi}}(t_{n+1}) & =\frac{2}{h}\boldsymbol{\xi}(t_{n+1})-\frac{2}{h}\boldsymbol{\xi}(t_{n})-\dot{\boldsymbol{\xi}}(t_{n}),\label{eq:contact_ODEs_integral_next}
\end{align}
where 
\begin{align}
\boldsymbol{Z}_{0} & =\frac{2}{h}\boldsymbol{M}+\boldsymbol{C}-\frac{h}{2}\boldsymbol{K},\nonumber \\
\boldsymbol{Z}_{1} & =\frac{2}{h}\boldsymbol{M}+\boldsymbol{C}+\frac{h}{2}\boldsymbol{K}.
\end{align}
It now suffices to describe the time stepping algorithm for contact
problem as follows:
\begin{enumerate}
\item At time step $t_{n}$, set iteration steps $j=1$ and $j_{\mathrm{max}}>1$;
set $\boldsymbol{\xi}_{j}^{a}(t_{n})=\boldsymbol{\xi}_{j}(t_{n})=\boldsymbol{\xi}(t_{n})$,
$\dot{\boldsymbol{\xi}_{j}^{a}}(t_{n})=\dot{\boldsymbol{\xi}}_{j}(t_{n})=\dot{\boldsymbol{\xi}}(t_{n})$,
$\boldsymbol{f}_{j}^{a}(t_{n})=\dot{\boldsymbol{f}_{j}^{a}}(t_{n})=\boldsymbol{0}$,
$\boldsymbol{M}_{j}^{a}=\boldsymbol{M}$, $\boldsymbol{C}_{j}^{a}=\boldsymbol{C}$,
$\boldsymbol{K}_{j}^{a}=\boldsymbol{K}$. Here we assume the rhs source
only comes from contact between A and B. If any other rhs sources
are present and independent of this contact, add them to $\boldsymbol{f}_{j}^{a}(t_{n})$
and $\dot{\boldsymbol{f}_{j}^{a}}(t_{n})$.
\item Input $\boldsymbol{\xi}_{j}^{a}(t_{n})$, $\dot{\boldsymbol{\xi}_{j}^{a}}(t_{n})$,
$\boldsymbol{f}_{j}^{a}(t_{n})$, $\dot{\boldsymbol{f}_{j}^{a}}(t_{n})$,
$\boldsymbol{M}_{j}^{a}$, $\boldsymbol{C}_{j}^{a}$, $\boldsymbol{K}_{j}^{a}$
into (\ref{eq:contact_ODEs_integral_next}) to compute $\boldsymbol{\xi}_{j}^{a}(t_{n+1})$
and $\dot{\boldsymbol{\xi}_{j}^{a}}(t_{n+1})$.
\item Based on $\boldsymbol{\xi}_{j}^{a}(t_{n+1})$ and $\dot{\boldsymbol{\xi}_{j}^{a}}(t_{n+1})$,
check if A ``penetrates'' B at any potential contact boundary nodes
using a geometry searching algorithm. Here $\boldsymbol{\xi}_{j}^{a}(t_{n+1})$
and $\dot{\boldsymbol{\xi}_{j}^{a}}(t_{n+1})$ are used to compute
A's surface forces acting on B, so as to obtain B's displacements
at $t_{n+1}$. 
\begin{enumerate}
\item If penetration occurs at some nodes:
\begin{enumerate}
\item Set $j\leftarrow j+1$. 
\begin{enumerate}
\item If $j>j_{\mathrm{max}}$, go to step 4.
\item Otherwise, reclassify $\boldsymbol{\xi}_{j}(t_{n})$, $\dot{\boldsymbol{\xi}}_{j}(t_{n})$
(all DOFs, dimension $N$) into $\boldsymbol{\xi}_{j}^{a}(t_{n})$,
$\dot{\boldsymbol{\xi}_{j}^{a}}(t_{n})$ (non-penetrated DOFs, dimension
$N_{j}^{a}$ and $\boldsymbol{\xi}_{j}^{b}(t_{n})$, $\dot{\boldsymbol{\xi}_{j}^{b}}(t_{n})$
(penetrated DOFs, dimension $N_{j}^{b}$). Note here $\boldsymbol{\xi}_{j-1}^{b}(t_{n})$,
$\dot{\boldsymbol{\xi}_{j-1}^{b}}(t_{n})$ (if any) must be classified
as penetrated in $\boldsymbol{\xi}_{j}^{b}(t_{n})$, $\dot{\boldsymbol{\xi}_{j}^{b}}(t_{n})$.
Accordingly, $\boldsymbol{M}$, $\boldsymbol{C}$, $\boldsymbol{K}$
are adjusted to $\boldsymbol{M}_{j}^{a}$, $\boldsymbol{C}_{j}^{a}$,
$\boldsymbol{K}_{j}^{a}$ with dimension $N_{j}^{a}$, in a way as
if $\boldsymbol{\xi}_{j}^{a}$ were $\boldsymbol{\xi}$ (unknown DOFs)
and $\boldsymbol{\xi}_{j}^{b}$ were $\boldsymbol{\xi}_{0}$ (known
DOFs) in (\ref{eq:ODEs_where}). Using $\boldsymbol{\xi}_{j}^{b}(t_{n})$
and $\dot{\boldsymbol{\xi}_{j}^{b}}(t_{n})$ (Dirichlet boundary conditions),
$\boldsymbol{f}_{j}^{a}(t_{n})$ and $\dot{\boldsymbol{f}_{j}^{a}}(t_{n})$
with dimension $N_{j}^{a}$ can be computed. 
\end{enumerate}
\item Input $\boldsymbol{\xi}_{j}^{a}(t_{n})$, $\dot{\boldsymbol{\xi}_{j}^{a}}(t_{n})$,
$\boldsymbol{f}_{j}^{a}(t_{n})$, $\dot{\boldsymbol{f}_{j}^{a}}(t_{n})$,
$\boldsymbol{M}_{j}^{a}$, $\boldsymbol{C}_{j}^{a}$, $\boldsymbol{K}_{j}^{a}$
into (\ref{eq:contact_ODEs_integral_next}) to compute $\boldsymbol{\xi}_{j}^{a}(t_{n+1})$
and $\dot{\boldsymbol{\xi}_{j}^{a}}(t_{n+1})$.
\item Based on $\boldsymbol{\xi}_{j}^{a}(t_{n+1})$ and $\dot{\boldsymbol{\xi}_{j}^{a}}(t_{n+1})$,
check if A ``penetrates'' B at any potential contact boundary nodes
using a geometry searching algorithm. Here $\boldsymbol{\xi}_{j}^{a}(t_{n+1})$
and $\dot{\boldsymbol{\xi}_{j}^{a}}(t_{n+1})$ are used to compute
A's surface forces acting on B, so as to obtain B's displacements
at $t_{n+1}$, as well as $\boldsymbol{\xi}_{j}^{b}(t_{n+1})$ and
$\dot{\boldsymbol{\xi}_{j}^{b}}(t_{n+1})$. Note here if $\dot{\boldsymbol{\xi}_{j}^{b}}(t_{n+1})$
is not well-defined due to non-smoothness, the weak derivative may
be adopted as an alternative. Combine $\boldsymbol{\xi}_{j}^{a}(t_{n+1})$,
$\dot{\boldsymbol{\xi}_{j}^{a}}(t_{n+1})$ and $\boldsymbol{\xi}_{j}^{b}(t_{n+1})$,
$\dot{\boldsymbol{\xi}_{j}^{b}}(t_{n+1})$ to get $\boldsymbol{\xi}_{j}(t_{n+1})$
and $\dot{\boldsymbol{\xi}}_{j}(t_{n+1})$.
\begin{enumerate}
\item If penetration occurs at some nodes, go back to step 3.(a).i.
\item Otherwise, go to step 4.
\end{enumerate}
\end{enumerate}
\item Otherwise, set $\boldsymbol{\xi}_{j}(t_{n+1})=\boldsymbol{\xi}_{j}^{a}(t_{n+1})$,
$\dot{\boldsymbol{\xi}}_{j}(t_{n+1})=\dot{\boldsymbol{\xi}_{j}^{a}}(t_{n+1})$.
\end{enumerate}
\item End the current time step with $\boldsymbol{\xi}(t_{n+1})=\boldsymbol{\xi}_{j}(t_{n+1})$
and $\dot{\boldsymbol{\xi}}(t_{n+1})=\dot{\boldsymbol{\xi}}_{j}(t_{n+1})$,
and move on to the next time step.
\end{enumerate}
The basic idea of this algorithm is to try Neumann first, and if penetration
occurs then change to Dirichlet. This strategy can be justified from
an energy perspective: first seek for displacements minimizing the
Lagrangian and virtual work of the system without any constraint;
if unrealistic penetration occurs, add an equality constraint regarding
Dirichlet boundary displacements, so as to find displacements minimizing
the Lagrangian and virtual work under contact constraint (similar
to the Lagrange multipliers). This check of penetration can lead to
the nonlinearity and non-smoothness of a contact problem. For geometry
searching algorithms needed to check penetration, we refer to the
{[}\href{https://www.comsol.com/support/learning-center/article/Structural-Contact-Modeling-Guidelines-83831}{COMSOL description}{]}.
If the contact boundary is assumed to be only a fixed point, then
only one or two iterations per time step is needed.

Certain contact problems may involve deformations large enough to
induce non-negligible nonlinearity. For nonlinearity of strain, the
Green-Lagrange strain instead of the engineering strain may be needed.
For nonlinearity of stress, the second-Piola--Kirchhoff stress tensor
instead of the Cauchy stress tensor may be needed. This would introduce
another kind of nonlinearity into the contact model. While the nonlinearity
due to contact's one-sided nature still keeps the computation of each
iteration in one time step linear, the nonlinearity due to nonlinear
stress and strain can make the PDEs as well as ODEs nonlinear, meaning
that computation of each iteration is nonlinear if no linearization
is applied. If the nonlinear perturbation to the equations is small
(controlled by some ``small parameters''), the perturbation method
applying linearization several times is acceptable; otherwise, the
perturbation method would suffer significant accuracy loss. 

\section{An energy-stable scheme for nonlinear forces}\label{sec:Nonlinear-strain-energy}

To better solve elastic solid problems subject to deformation nonlinearity,
particularly elastic contact and piano string vibration, we are working
on integrating into our model the invariant energy quadratization
(IEQ) method \cite{zhao2017numerical} and scalar auxiliary variable
(SAV) method \cite{shen2018scalar,shen2019new}. Based on the idea
of IEQ, SAV is a recent advancement in computational physics towards
efficient and energy-stable time discretization scheme for gradient
flows. Both methods have been applied to musical instruments modeling,
achieving discrete energy evolution (conservation or dissipation)
that is consistent with the continuous counterpart, using fully explicit
or linearly implicit update schemes \cite{ducceschi2021linearly,van2024implicit,bhanuprakash2024quadratic}.
These schemes attain the desired discrete energy-stable property while
being more efficient than existing nonlinear solvers many of which
rely on Newton-Raphson like iterations with computationally expensive
Jacobian matrices. 

To introduce nonlinearity into the 3D elastic solid model, we rewrite
(\ref{eq:ODEs}) as 
\begin{equation}
\boldsymbol{M}\ddot{\boldsymbol{\xi}}(t)+\boldsymbol{C}\dot{\boldsymbol{\xi}}(t)+\boldsymbol{K}\boldsymbol{\xi}(t)+\boldsymbol{g}\left[\boldsymbol{\xi}(t)\right]=\boldsymbol{f}(t),\label{eq:nonlinear_ODEs}
\end{equation}
where $\boldsymbol{g}(\boldsymbol{\xi})$ is an operator incorporating
all the ``symmetric'' nonlinear forces of the system, which can
be conservative or non-conservative. The symmetry property we require
here means that $\boldsymbol{g}$ is a self-adjoint operator thus
the variational Jacobian matrix 
\begin{equation}
\boldsymbol{J}\left[\boldsymbol{\xi}(t)\right]=\delta_{\boldsymbol{\xi}}\boldsymbol{g}
\end{equation}
is symmetric. This is equivalent to $\boldsymbol{g}$ being a self-adjoint
operator. For simplicity we have not considered nonlinear damping
forces in the form of $\boldsymbol{g}_{1}(\dot{\boldsymbol{\xi}})$
here, but they can simply be treated in a similar way to $\boldsymbol{g}(\boldsymbol{\xi})$.
An example of nonlinear symmetric forces is the nonlinear stress arising
from the quadratic term in Green-Lagrange strain, as well as the damping
force proportional to its first-order time derivative. Anyway, linear
or nonlinear forces not satisfying this symmetry property are incorporated
in $\boldsymbol{f}(t)$, and unfortunately the coupling forces in
our piano model are probably such examples. The symmetry property
has important implications that one can define a nonlinear energy
functional as 
\begin{equation}
E_{nl}\left[\boldsymbol{\xi}(t)\right]=\int_{0}^{t}\dot{\boldsymbol{\xi}}(t')\cdot\boldsymbol{g}\left[\boldsymbol{\xi}(t')\right]\mathrm{d}t',\label{eq:nonlinear_energy}
\end{equation}
with first-order time derivative 
\begin{equation}
\partial_{t}E_{nl}\left[\boldsymbol{\xi}(t)\right]=\dot{\boldsymbol{\xi}}(t)\cdot\boldsymbol{g}\left[\boldsymbol{\xi}(t)\right],
\end{equation}
and variation 
\begin{align}
\delta E_{nl}\left[\boldsymbol{\xi}(t)\right] & =\int_{0}^{t}\delta\dot{\boldsymbol{\xi}}\cdot\boldsymbol{g}\mathrm{d}t'+\int_{0}^{t}\dot{\boldsymbol{\xi}}\cdot\delta\boldsymbol{g}\mathrm{d}t',\nonumber \\
 & =\delta\boldsymbol{\xi}(t)\cdot\boldsymbol{g}-\delta\boldsymbol{\xi}(0)\cdot\boldsymbol{g}+\int_{0}^{t}\dot{\boldsymbol{\xi}}\cdot\delta\boldsymbol{g}\mathrm{d}t'-\int_{0}^{t}\delta\boldsymbol{\xi}\cdot\dot{\boldsymbol{g}}\mathrm{d}t'\nonumber \\
 & =\delta\boldsymbol{\xi}(t)\cdot\boldsymbol{g}
\end{align}
where $\delta\boldsymbol{\xi}(0)=\boldsymbol{0}$ is imposed and one
can derive $\dot{\boldsymbol{\xi}}\cdot\delta\boldsymbol{g}-\delta\boldsymbol{\xi}\cdot\dot{\boldsymbol{g}}=\boldsymbol{0}$
with the symmetry property of $\boldsymbol{J}$. It is now clear that
under symmetry, the virtual work equates real work. The kinetic energy
functional is 
\begin{align}
E_{k}\left[\boldsymbol{\xi}(t)\right] & =\frac{1}{2}\dot{\boldsymbol{\xi}}(t)\cdot\boldsymbol{M}\dot{\boldsymbol{\xi}}(t),\nonumber \\
\partial_{t}E_{k}\left[\boldsymbol{\xi}(t)\right] & =\dot{\boldsymbol{\xi}}(t)\cdot\boldsymbol{M}\ddot{\boldsymbol{\xi}}(t).
\end{align}
The linear potential energy functional is 
\begin{align}
E_{p}\left[\boldsymbol{\xi}(t)\right] & =\frac{1}{2}\boldsymbol{\xi}(t)\cdot\boldsymbol{K}\boldsymbol{\xi}(t),\nonumber \\
\partial_{t}E_{p}\left[\boldsymbol{\xi}(t)\right] & =\dot{\boldsymbol{\xi}}(t)\cdot\boldsymbol{K}\boldsymbol{\xi}(t).
\end{align}
The overall energy evolution of the system is found to be 
\begin{align}
 & \dot{\boldsymbol{\xi}}(t)\cdot\left(\boldsymbol{M}\ddot{\boldsymbol{\xi}}(t)+\boldsymbol{C}\dot{\boldsymbol{\xi}}(t)+\boldsymbol{K}\boldsymbol{\xi}(t)+\boldsymbol{g}\left[\boldsymbol{\xi}(t)\right]\right)=\dot{\boldsymbol{\xi}}(t)\cdot\boldsymbol{f}(t)\nonumber \\
 & \partial_{t}E=\partial_{t}E_{k}+\partial_{t}E_{p}+\partial_{t}E_{nl}=\dot{\boldsymbol{\xi}}(t)\cdot\boldsymbol{f}(t)-\dot{\boldsymbol{\xi}}(t)\cdot\boldsymbol{C}\dot{\boldsymbol{\xi}}(t)
\end{align}
where $E[\boldsymbol{\xi}(t)]=E_{k}+E_{p}+E_{nl}$. 

To apply SAV \cite{shen2018scalar}, assume the nonlinear potential
energy functional can be rewritten in the square of a scalar variable
$\psi(t)$ as 
\begin{align}
E_{nl}\left[\boldsymbol{\xi}(t)\right] & =\frac{1}{2}\psi(t)^{2},
\end{align}
where 
\begin{align}
\psi(t) & =\sqrt{2}\sqrt{E_{nl}\left[\boldsymbol{\xi}(t)\right]+C_{0}},\label{eq:SAV_zero_order}\\
\dot{\psi}(t) & =\frac{1}{\sqrt{2}}\frac{\dot{\boldsymbol{\xi}}(t)\cdot\boldsymbol{g}\left[a(t),\boldsymbol{\xi}(t),\dot{\boldsymbol{\xi}}(t)\right]}{\sqrt{E_{nl}\left[\boldsymbol{\xi}(t)\right]+C_{0}}},\label{eq:SAV_first_order}
\end{align}
and $E_{nl}\left[\boldsymbol{\xi}(t)\right]\geq-C_{0}$ is assumed
(bounded from below). Utilizing the Crank--Nicolson scheme described
in \cite{shen2019new} we have 
\begin{align}
\int_{t_{n}}^{t_{n+1}}\boldsymbol{g}\mathrm{d}t & =\frac{1}{\sqrt{2}}\int_{t_{n}}^{t_{n+1}}\frac{\psi(t)}{\sqrt{E_{nl}\left[\boldsymbol{\xi}(t)\right]}}\boldsymbol{g}\left[\boldsymbol{\xi}(t)\right]\mathrm{d}t\nonumber \\
 & \approx\frac{h}{2\sqrt{2}}\frac{\psi(t_{n})+\psi(t_{n+1})}{\sqrt{E_{nl}\left[\boldsymbol{\xi}(t_{n+\frac{1}{2}})\right]+C_{0}}}\boldsymbol{g}\left[\boldsymbol{\xi}(t_{n+\frac{1}{2}})\right]\nonumber \\
 & \approx\frac{h}{2\sqrt{2}}\frac{\psi(t_{n})+\psi(t_{n+1})}{\sqrt{E_{nl}\left[\boldsymbol{\xi}(t_{n})\right]+\frac{h}{2}\dot{E}_{nl}\left[\boldsymbol{\xi}(t_{n})\right]+C_{0}}}\boldsymbol{g}\left[\boldsymbol{\xi}(t_{n})+\frac{h}{2}\dot{\boldsymbol{\xi}}(t_{n})\right]\nonumber \\
 & =\frac{h}{2}\left(\psi(t_{n})+\psi(t_{n+1})\right)\boldsymbol{s}\left(\boldsymbol{\xi}(t_{n}),\dot{\boldsymbol{\xi}}(t_{n})\right),\label{eq:nonlinear_integral}
\end{align}
where 
\begin{equation}
\boldsymbol{s}\left(\boldsymbol{\xi}(t_{n}),\dot{\boldsymbol{\xi}}(t_{n})\right):\mathbb{R}^{N}=\frac{\boldsymbol{g}\left[\boldsymbol{\xi}(t_{n})+\frac{h}{2}\dot{\boldsymbol{\xi}}(t_{n})\right]}{\sqrt{2}\sqrt{E_{nl}\left[\boldsymbol{\xi}(t_{n})\right]+\frac{h}{2}\dot{\boldsymbol{\xi}}(t_{n})\cdot\boldsymbol{g}\left[\boldsymbol{\xi}(t_{n})\right]+C_{0}}}.\label{eq:nonlinear_where}
\end{equation}
Here we encountered the problem of how to approximate the values of
$\boldsymbol{g}$ and $E_{nl}$ at mid step $t_{n+\frac{1}{2}}$.
Unlike the several interpolation methods used in \cite{shen2019new},
we chose a more direct first-order approximation approach using $\dot{E}_{nl}[\boldsymbol{\xi}(t_{n})]$
and $\dot{\boldsymbol{\xi}}(t_{n})$. $\dot{\boldsymbol{g}}(\boldsymbol{\xi}(t_{n}))$
was not used because its expression involving $\boldsymbol{J}$ may
be hard to write, and we instead did a slightly wrose approximation
using only $\dot{\boldsymbol{\xi}}(t_{n})$; it is also viable to
approximate $\dot{\boldsymbol{g}}(\boldsymbol{\xi}(t_{n}))$ using
$\boldsymbol{g}(\boldsymbol{\xi}(t_{n-1}))$ and $\boldsymbol{g}(\boldsymbol{\xi}(t_{n}))$.
Notice that in (\ref{eq:nonlinear_integral}) $\psi(t_{n+1})$ is
unknown, so we need to discretize (\ref{eq:SAV_first_order}) for
another update scheme 
\begin{align}
\psi(t_{n+1})-\psi(t_{n}) & \approx\frac{1}{\sqrt{2}}\frac{\left(\boldsymbol{\xi}(t_{n+1})-\boldsymbol{\xi}(t_{n})\right)\cdot\boldsymbol{g}\left(\boldsymbol{\xi}(t_{n+\frac{1}{2}})\right)}{\sqrt{E_{nl}\left[\boldsymbol{\xi}(t_{n+\frac{1}{2}})\right]}}\nonumber \\
 & \approx\left(\boldsymbol{\xi}(t_{n+1})-\boldsymbol{\xi}(t_{n})\right)\cdot\boldsymbol{s}\left(\boldsymbol{\xi}(t_{n}),\dot{\boldsymbol{\xi}}(t_{n})\right).\label{eq:nonlinear_update_3}
\end{align}
It now sufficies to formulate the whole discretization of nonlinear
ODEs as 
\begin{equation}
\boldsymbol{M}\left[\dot{\boldsymbol{\xi}}(t_{n+1})-\dot{\boldsymbol{\xi}}(t_{n})\right]+\boldsymbol{C}\left[\boldsymbol{\xi}(t_{n+1})-\boldsymbol{\xi}(t_{n})\right]+\boldsymbol{K}\int_{t_{n}}^{t_{n+1}}\boldsymbol{\xi}\mathrm{d}t+\int_{t_{n}}^{t_{n+1}}\boldsymbol{g}\left[\boldsymbol{\xi}\right]\mathrm{d}t=\int_{t_{n}}^{t_{n+1}}\boldsymbol{f}\mathrm{d}t\label{eq:nonlinear_ODEs_discretization}
\end{equation}
Combining this with SAV and 
\begin{align}
\boldsymbol{\xi}(t_{n+1})-\boldsymbol{\xi}(t_{n}) & \approx\frac{h}{2}\left[\dot{\boldsymbol{\xi}}(t_{n})+\dot{\boldsymbol{\xi}}(t_{n+1})\right]\nonumber \\
\dot{\boldsymbol{\xi}}(t_{n+1}) & \approx\frac{2}{h}\boldsymbol{\xi}(t_{n+1})-\frac{2}{h}\boldsymbol{\xi}(t_{n})-\dot{\boldsymbol{\xi}}(t_{n}),\label{eq:nonlinear_update_2}
\end{align}
we obtain the major update scheme: 
\begin{align}
 & \boldsymbol{M}\left[\frac{2}{h}\boldsymbol{\xi}(t_{n+1})-\frac{2}{h}\boldsymbol{\xi}(t_{n})-2\dot{\boldsymbol{\xi}}(t_{n})\right]+\boldsymbol{C}\left[\boldsymbol{\xi}(t_{n+1})-\boldsymbol{\xi}(t_{n})\right]+\frac{h}{2}\boldsymbol{K}\left[\boldsymbol{\xi}(t_{n})+\boldsymbol{\xi}(t_{n+1})\right]\nonumber \\
 & +h\psi(t_{n})\boldsymbol{s}+\frac{h}{2}\boldsymbol{s}\boldsymbol{s}^{\top}\left[\boldsymbol{\xi}(t_{n+1})-\boldsymbol{\xi}(t_{n})\right]=h\boldsymbol{f}(t_{n})+\frac{h^{2}}{2}\dot{\boldsymbol{f}}(t_{n})\label{eq:nonlinear_update_1_0}\\
 & \boldsymbol{\xi}(t_{n+1})=\boldsymbol{Z}_{1}^{-1}\left[\boldsymbol{Z}_{0}\boldsymbol{\xi}(t_{n})+2\boldsymbol{M}\dot{\boldsymbol{\xi}}(t_{n})+h\boldsymbol{f}(t_{n})+\frac{h^{2}}{2}\dot{\boldsymbol{f}}(t_{n})\right],\label{eq:nonlinear_update_1}
\end{align}
where 
\begin{align}
\boldsymbol{Z}_{0} & =\frac{2}{h}\boldsymbol{M}+\boldsymbol{C}-\frac{h}{2}\boldsymbol{K}+\frac{h}{2}\boldsymbol{s}\boldsymbol{s}^{\top},\nonumber \\
\boldsymbol{Z}_{1} & =\frac{2}{h}\boldsymbol{M}+\boldsymbol{C}+\frac{h}{2}\boldsymbol{K}+\frac{h}{2}\boldsymbol{s}\boldsymbol{s}^{\top}.\label{eq:nonlinear_update_where}
\end{align}
Therefore, the SAV update scheme consists of three steps: 
\begin{enumerate}
\item Input $\boldsymbol{\xi}(t_{n})$, $\dot{\boldsymbol{\xi}}(t_{n})$,
$\psi(t_{n})$ into (\ref{eq:nonlinear_update_1}) to compute $\boldsymbol{\xi}(t_{n+1})$.
\item Input $\boldsymbol{\xi}(t_{n})$, $\dot{\boldsymbol{\xi}}(t_{n})$,
$\boldsymbol{\xi}(t_{n+1})$ into (\ref{eq:nonlinear_update_2}) to
compute $\dot{\boldsymbol{\xi}}(t_{n+1})$.
\item Input $\boldsymbol{\xi}(t_{n})$, $\dot{\boldsymbol{\xi}}(t_{n})$,
$\boldsymbol{\xi}(t_{n+1})$, $\psi(t_{n})$ into (\ref{eq:nonlinear_update_3})
to compute $\psi(t_{n+1})$.
\end{enumerate}
It can be seen that the major update scheme (\ref{eq:nonlinear_update_1})
is linear w.r.t. $\boldsymbol{\xi}(t_{n+1})$, allowing for efficient
time stepping. However, we notice that $\boldsymbol{s}\boldsymbol{s}^{\top}$
in (\ref{eq:nonlinear_update_where}) is a fully dense $N\times N$
matrix that may be infeasible to store and solve for large number
DOFs. This problem can be well alleviated by performing eigen decomposition
only once to the time-invariant matrix 
\begin{equation}
\boldsymbol{Z}_{2}=\frac{2}{h}\boldsymbol{M}+\boldsymbol{C}+\frac{h}{2}\boldsymbol{K},\;\boldsymbol{Z}_{2}^{-1}\approx\boldsymbol{\Phi}\boldsymbol{\Lambda}\boldsymbol{\Phi}^{\mathrm{H}},
\end{equation}
where the dimension of $\boldsymbol{\Phi}$ and $\boldsymbol{\Lambda}$
is $M\ll N$, as only the smallest $M$ eigenvalues and corresponding
eigenvectors are needed. This should be regarded as a generalized
eigenvalue problem because it is desired to avoid explicit matrix
inversion. Then by the Shermann--Morrison--Woodbury formula we have
\begin{equation}
\boldsymbol{Z}_{1}^{-1}=\left(\boldsymbol{Z}_{2}+\frac{h}{2}\boldsymbol{s}\boldsymbol{s}^{\top}\right)^{-1}=\boldsymbol{Z}_{2}^{-1}-\left(\frac{2}{h}+\boldsymbol{s}^{\top}\boldsymbol{Z}_{2}^{-1}\boldsymbol{s}\right)^{-1}\left(\boldsymbol{Z}_{2}^{-1}\boldsymbol{s}\right)\left(\boldsymbol{Z}_{2}^{-1}\boldsymbol{s}\right)^{\top},
\end{equation}
where $\boldsymbol{Z}_{2}^{-1}$ can be approximated by the truncated
eigenvalues and eigenvectors. In this way $\boldsymbol{Z}_{1}^{-1}$
multiplying a vector can be efficiently computed without storing and
inverting a $N\times N$ dense matrix. 

To investigate the energy property of this SAV scheme, we multiply
$\frac{1}{h}[\boldsymbol{\xi}(t_{n+1})-\boldsymbol{\xi}(t_{n})]=\frac{1}{2}[\dot{\boldsymbol{\xi}}(t_{n})+\dot{\boldsymbol{\xi}}(t_{n+1})]$
on both sides of (\ref{eq:nonlinear_update_1_0}) to yield\footnote{The reason for the equality $\frac{1}{h}[\boldsymbol{\xi}(t_{n+1})-\boldsymbol{\xi}(t_{n})]=\frac{1}{2}[\dot{\boldsymbol{\xi}}(t_{n})+\dot{\boldsymbol{\xi}}(t_{n+1})]$
to hold here is that this equation is the actual numerical update
scheme we use, no matter how approximate it is to the true continuous
values.} 
\begin{align}
 & \frac{1}{2}\dot{\boldsymbol{\xi}}(t_{n+1})\cdot\boldsymbol{M}\dot{\boldsymbol{\xi}}(t_{n+1})-\frac{1}{2}\dot{\boldsymbol{\xi}}(t_{n})\cdot\boldsymbol{M}\dot{\boldsymbol{\xi}}(t_{n})+\frac{1}{h}\left[\boldsymbol{\xi}(t_{n+1})-\boldsymbol{\xi}(t_{n})\right]\cdot\boldsymbol{C}\left[\boldsymbol{\xi}(t_{n+1})-\boldsymbol{\xi}(t_{n})\right]\nonumber \\
 & +\frac{1}{2}\boldsymbol{\xi}(t_{n+1})\cdot\boldsymbol{K}\boldsymbol{\xi}(t_{n+1})-\frac{1}{2}\boldsymbol{\xi}(t_{n})\cdot\boldsymbol{K}\boldsymbol{\xi}(t_{n})+\frac{1}{2}\psi(t_{n+1})^{2}-\frac{1}{2}\psi(t_{n})^{2}\nonumber \\
 & =\left[\boldsymbol{\xi}(t_{n+1})-\boldsymbol{\xi}(t_{n})\right]\cdot\left[\boldsymbol{f}(t_{n})+\frac{h}{2}\dot{\boldsymbol{f}}(t_{n})\right].
\end{align}
Combining this with the definition of $E$, $E_{k}$, $E_{p}$, $E_{nl}$,
we obtain the evolution of discrete energy as 
\begin{equation}
E\left[\boldsymbol{\xi}(t_{n+1})\right]-E\left[\boldsymbol{\xi}(t_{n})\right]=\left[\boldsymbol{\xi}(t_{n+1})-\boldsymbol{\xi}(t_{n})\right]\cdot\left[\boldsymbol{f}(t_{n})+\frac{h}{2}\dot{\boldsymbol{f}}(t_{n})\right]-\frac{1}{h}\left\Vert \boldsymbol{\xi}(t_{n+1})-\boldsymbol{\xi}(t_{n})\right\Vert _{\boldsymbol{C}}^{2},
\end{equation}
where $\left\Vert \boldsymbol{x}\right\Vert _{\boldsymbol{A}}^{2}=\boldsymbol{x}^{\top}\boldsymbol{A}\boldsymbol{x}$.
Since $\boldsymbol{C}$ is normally positive definite, the discrete
energy is guaranteed to dissipate if $\boldsymbol{g}$ is conservative
and energy injection through $\boldsymbol{f}(t)$ becomes zero. Therefore,
it can be concluded that the SAV scheme applied here is energy stable
at least for kinetic energy, potential energy, nonlinear ``symmetric''
energy, and damping. 

However, it seems energy stablility may fail to satisfy for the non-symmetric
forces $\boldsymbol{f}(t)$, which we should actually write as an
operator $\boldsymbol{f}[\boldsymbol{\xi}(t),\dot{\boldsymbol{\xi}}(t),\ddot{\boldsymbol{\xi}}(t)]$.
These non-symmetric ``endogenous'' forces often relate to the coupling
between systems. For example, for the soundboard system, $\boldsymbol{f}$
is the string's surface force exerted on the soundboard. If $\boldsymbol{f}$
is fully linear with respect to the unknowns, then we can use $t_{n+1}$
values in $\boldsymbol{f}$ to construct linearly implicit schemes
which effectively avoids energy unstability, though the formulas may
be cumbersome to write. But if $\boldsymbol{f}$ exhibits nonlinearity
and $t_{n+1}$ values are used to avoid energy unstability, the scheme
would become nonlinearly implicit and it seems only implicit methods,
e.g. backward differentiation formula (BDF), can solve for the unknown
DOFs. We thus conclude that SAV does not address well the non-symmetric
forces, which is vital for modeling the nonlinear coupling between
the distinct subsystems of the piano. A tradeoff is to keep using
energy-unstabe linear schemes for the nonlinear non-symmetric forces,
but perform several iterations as proposed in section \ref{subsec:Explicit-time-discretization}
to compensate. We hope to see more developments regarding efficient
and energy-stable schemes for nonlinear inter-system coupling.

We now consider an example of nonlinear potential energy arising from
large deformations. For a 3D elastic solid with large deformations,
a major source of nonlinearity is the quadratic term in the Green-Lagrange
strain. The tensor and vector versions of this nonlinear strain (linear
parts omitted) are 
\begin{align}
\boldsymbol{\epsilon}_{nl}(\boldsymbol{x},t) & =\frac{1}{2}\sum_{i=1}^{3}\left(\nabla u_{i}\right)\left(\nabla u_{i}\right)^{\top},\\
\Rightarrow\overrightarrow{\boldsymbol{\epsilon}}_{nl}(\boldsymbol{x},t) & =\mathcal{A}\left(\boldsymbol{u}(\boldsymbol{x},t)\right)=\sum_{i=1}^{3}\left[\begin{array}{c}
\frac{1}{2}\left(\partial_{x}u_{i}\right)^{2}\\
\frac{1}{2}\left(\partial_{y}u_{i}\right)^{2}\\
\frac{1}{2}\left(\partial_{z}u_{i}\right)^{2}\\
\left(\partial_{x}u_{i}\right)\left(\partial_{y}u_{i}\right)\\
\left(\partial_{x}u_{i}\right)\left(\partial_{z}u_{i}\right)\\
\left(\partial_{y}u_{i}\right)\left(\partial_{z}u_{i}\right)
\end{array}\right]
\end{align}
where $\mathcal{A}:\mathbb{R}^{3}\rightarrow\mathbb{R}^{6}$ is a
nonlinear operator. From the nonlinear strain, we find the conservative
part of nonlinear strain energy functional as 
\begin{equation}
E_{nl}\left[\boldsymbol{u}(\boldsymbol{x},t)\right]=\frac{1}{2}\int_{\Omega}\mathcal{A}\left(\boldsymbol{u}(\boldsymbol{x},t)\right)\cdot\boldsymbol{D}\mathcal{A}\left(\boldsymbol{u}(\boldsymbol{x},t)\right)\mathrm{d}V.
\end{equation}
Taking gradient of this functional, the nonlinear stress (body force
version) is derived as 
\begin{equation}
\nabla\cdot\boldsymbol{\sigma}_{nl}=\frac{\delta E_{nl}\left[\boldsymbol{u}\right]}{\delta\boldsymbol{u}}.
\end{equation}
Besides $E_{\mathrm{nl}}$, there may exist damping proportional to
the nonlinear stress, leading to the non-conservative part of nonlinear
strain energy functional as 
\begin{equation}
E_{nlnc}\left[\boldsymbol{u}(\boldsymbol{x},t'),t\right]=\int_{0}^{t}\int_{\Omega}\dot{\mathcal{A}}\left(\boldsymbol{u}(\boldsymbol{x},t')\right)\cdot2\mu\boldsymbol{D}\dot{\mathcal{A}}\left(\boldsymbol{u}(\boldsymbol{x},t')\right)\mathrm{d}V\mathrm{d}t',\label{eq:nonlinear_nonconservative_energy}
\end{equation}
where $2\mu$ is the viscous damping coefficient. The dynamics of
energy dissipation can then be revealed as 
\begin{equation}
\partial_{t}E_{nlnc}\left[\boldsymbol{u}(\boldsymbol{x},t'),t\right]=\int_{\Omega}\dot{\mathcal{A}}\left(\boldsymbol{u}(\boldsymbol{x},t)\right)\cdot2\mu\boldsymbol{D}\dot{\mathcal{A}}\left(\boldsymbol{u}(\boldsymbol{x},t)\right)\mathrm{d}V.
\end{equation}
However, consideration of nonlinear damping seems lacking in existing
IEQ and SAV research. This may stem from that energy dissipation is
path-dependent thus the integration over $[0,t]$ can not be eliminated.
This would require some additional but still acceptable computation
costs, and has been covered by our nonlinear energy model \ref{eq:nonlinear_energy}.
In existing research applying IEQ or SAV to musical instruments simulation,
there can be no damping \cite{van2024implicit} or linear damping
\cite{ducceschi2022simulation}. Now we apply space discretization
to the nonlinear strain energy functional. Based on (\ref{eq:displacement_space_discretization})
(\ref{eq:space_discretization_map}) we should be able to find 
\begin{equation}
\overrightarrow{\boldsymbol{\epsilon}}_{nl}(\boldsymbol{x},t)=\boldsymbol{A}\left(\boldsymbol{x},\boldsymbol{\xi}(t)\right)+\boldsymbol{A}_{0}\left(\boldsymbol{x},\boldsymbol{\xi}(t),\boldsymbol{\xi}_{0}(t)\right),
\end{equation}
where $\boldsymbol{A}$ and $\boldsymbol{A}_{0}$ are $(\mathbb{R}^{3},\mathbb{R}^{N})\rightarrow\mathbb{R}^{6}$
and $(\mathbb{R}^{3},\mathbb{R}^{N},\mathbb{R}^{N_{0}})\rightarrow\mathbb{R}^{6}$
nonlinear functions (a narrower concept than operator) respectively;
$\boldsymbol{A}$ accounts for the conservative part and $\boldsymbol{A}_{0}$
accounts for the non-conservative part when $\boldsymbol{\xi}_{0}$
is nonzero. For simplicity, the detailed expression of $\boldsymbol{A}$
and $\boldsymbol{A}_{0}$ are currently not presented here. The nonlinear
strain energy functional is then 
\begin{equation}
E_{nl}\left[\boldsymbol{\xi}(t)\right]=\frac{1}{2}\int_{\Omega}\boldsymbol{A}\left(\boldsymbol{x},\boldsymbol{\xi}(t)\right)\cdot\boldsymbol{D}\boldsymbol{A}\left(\boldsymbol{x},\boldsymbol{\xi}(t)\right)\mathrm{d}V,\label{eq:nonlinear_energy_space_discretization}
\end{equation}
and its variation and first derivative are 
\begin{align}
\delta E_{nl}\left[\boldsymbol{\xi}(t)\right] & =\delta\boldsymbol{\xi}\cdot\boldsymbol{B}\left(\boldsymbol{\xi}(t)\right),\nonumber \\
\partial_{t}E_{nl}\left[\boldsymbol{\xi}(t)\right] & =\dot{\boldsymbol{\xi}}(t)\cdot\boldsymbol{B}\left(\boldsymbol{\xi}(t)\right),\nonumber \\
\boldsymbol{B}\left(\boldsymbol{\xi}\right) & =\int_{\Omega}\left[\nabla_{\boldsymbol{\xi}}\boldsymbol{A}\left(\boldsymbol{x},\boldsymbol{\xi}\right)\right]^{\top}\boldsymbol{D}\boldsymbol{A}\left(\boldsymbol{x},\boldsymbol{\xi}\right)\mathrm{d}V
\end{align}
Since $\boldsymbol{A}_{0}$ relates to $\boldsymbol{\xi}_{0}(t)$,
i.e. nonzero Dirichlet boundary conditions, it does not satisfy the
symmetry (self-adjoint) condition previously highlighted. We thereby
seek to compute the nonlinear virtual strain energy as 
\begin{align}
\delta E_{nl,0}\left[\boldsymbol{\xi}(t)\right] & =\delta\boldsymbol{\xi}\cdot\boldsymbol{B}_{0}\left(\boldsymbol{\xi}(t),\boldsymbol{\xi}_{0}(t)\right)\nonumber \\
\boldsymbol{B}_{0}\left(\boldsymbol{\xi},\boldsymbol{\xi}_{0}\right) & =\int_{\Omega}\left[\nabla_{\boldsymbol{\xi}}\boldsymbol{A}\left(\boldsymbol{x},\boldsymbol{\xi}\right)\right]^{\top}\boldsymbol{D}\boldsymbol{A}_{0}\left(\boldsymbol{x},\boldsymbol{\xi},\boldsymbol{\xi}_{0}\right)\mathrm{d}V\nonumber \\
 & +\int_{\Omega}\left[\nabla_{\boldsymbol{\xi}}\boldsymbol{A}_{0}\left(\boldsymbol{x},\boldsymbol{\xi},\boldsymbol{\xi}_{0}\right)\right]^{\top}\boldsymbol{D}\left[\boldsymbol{A}\left(\boldsymbol{x},\boldsymbol{\xi}\right)+\boldsymbol{A}_{0}\left(\boldsymbol{x},\boldsymbol{\xi},\boldsymbol{\xi}_{0}\right)\right]\mathrm{d}V
\end{align}
Finally, we add $\boldsymbol{B}$ to $\boldsymbol{g}$ and add $-\boldsymbol{B}_{0}$
to $\boldsymbol{f}$ in the nonlinear ODEs (\ref{eq:nonlinear_ODEs}).
Note that since the nonlinear stress belongs to conservative forces,
its corresponding nonlinear strain energy functional (\ref{eq:nonlinear_energy_space_discretization})
has eliminated the integration over $[0,t]$. This makes it simple
to compute the $E_{nl}[\boldsymbol{\xi}(t_{n})]$ in (\ref{eq:nonlinear_where}).
\clearpage
\end{CJK}

\end{document}